\documentclass[11pt]{article}

\usepackage{bm}
\usepackage[normalem]{ulem}
\usepackage{amsmath,graphics,epsfig,float,braket,multirow,url}
\usepackage{caption}
\usepackage{subcaption}
\usepackage[toc,page]{appendix}
\usepackage{amsmath}
\usepackage{microtype}
\usepackage{comment}
\usepackage{authblk}
\usepackage{hyperref}
\usepackage[capitalize]{cleveref}
\usepackage[normalem]{ulem}
\newcommand{\ith}{\ensuremath{i\textsuperscript{th}~}}

\newcommand{\s}[1]{{\textsf{\textbf{#1}}}}

\begin{document}

%%%% Article title to be placed here
\title{\s{Bistability of midpoint-fused arches with pinned-pinned boundary conditions}}
\author{ \textsf{Rajat Goswami $^\dagger$,  and Safvan Palathingal $^\dagger$}}
\date{
{\it $^{\dagger}$Department of Mechanical and Aerospace Engineering,\\ Indian Institute of Technology Hyderabad, Telangana, India}\\[2ex]
\today
}

 \maketitle
\hrule\vskip 6pt

%%%% Abstract text to be placed here %%%%%%%%%%%%
\begin{abstract}
Arranging multiple arches in a circular pattern and fusing them at their midpoint yields a three-dimensional configuration that we refer to as midpoint-fused arches (MFA). This study investigates the structural bistability of MFA, i.e., their ability to admit two distinct, force-free stable equilibrium states. Starting from an as-fabricated, stress-free configuration, MFA can invert into a stressed, toggled state reminiscent of an umbrella’s ribs. We develop an analytical model for the response of a pinned-pinned MFA subjected to a concentrated mid-span load by minimizing the total potential energy. Individual arches are treated as spatially-deforming, and kinematic compatibility relations are derived at the fusion point to couple their deformations. Various deformation symmetries are then exploited to simplify the problem.

We demonstrate the model’s utility by characterizing the force-displacement response of a two-arch MFA, identifying distinct deformation pathways and discussing the pathway transitions that occur during toggling. In particular, we show how the structure switches between symmetric and asymmetric deformation modes as it moves between stable configurations. The generality of the framework is further established through analysis of a three-arch MFA, which exhibits richer coupled deformation behaviour. Nonlinear finite-element simulations and table-top experiments corroborate the analytical predictions, showing close agreement in both equilibrium states and the associated transition responses.
\end{abstract}

\vskip 6pt
\hrule
\vskip 6pt
\section{Introduction}
Bistable elastic structures have  two stable equilibrium states separated by an energy barrier, and they have  been extensively utilized across a wide range of engineering applications\cite{Hu2015}. They exhibit distinct geometric configurations, a nonlinear force-displacement relationship involving a region of negative stiffness, and the ability to transition between states through multiple actuation modes\cite{palathingal2020statics}. Their ability to maintain stable configurations passively and repeatable snap-through transitions makes them advantageous for applications such as switches, actuators, grippers, microelectromechanical systems (MEMS), programmable Braille displays, etc. \cite{Qiu2003,Balakuntala2020,Abbasi.2024,Yadav.2019,Pal2023}. Their nonlinear force characteristics are particularly beneficial in energy harvesting \cite{nan2021bistable} and zero-stiffness structures\cite{Gilmore2023}, while their geometrically distinct states enable shape-morphing and reconfigurable designs\cite{Balakuntala2020,Faber.2020}. Additionally, their multimodal actuation capability has facilitated technologies such as assistive-chair for the elderly\cite{Kushwaha2021}, RF-MEMS switches\cite{Yadav.2019}, smart actuator systems.

These diverse applications have resulted in a renewed interest in understanding the mechanics of bistable structures in the recent past. Furthermore, this growing interest is widely considered part of a broader paradigm shift toward leveraging elastic instability for functionality rather than avoiding it \cite{Holmes.2019,Reis.2015}. At the same time, seminal works that have extensively examined post-buckling behaviour in slender structures remain relevant in analysing bistability. These, along with some recent studies, include  work on bistable buckled beams\cite{Vangbo.1998,10.1115/1.3179003}, bistable arches\cite{Hoff.1953,fung1952buckling,Palathingal.2019w9j,QiuJMems}, and bistable shells\cite{10.1098/rspa.2017.0230}.

\begin{figure}[!htb]
     \centering
     \begin{subfigure}[b]{0.6\textwidth}
         \centering
         \includegraphics[width=\textwidth]{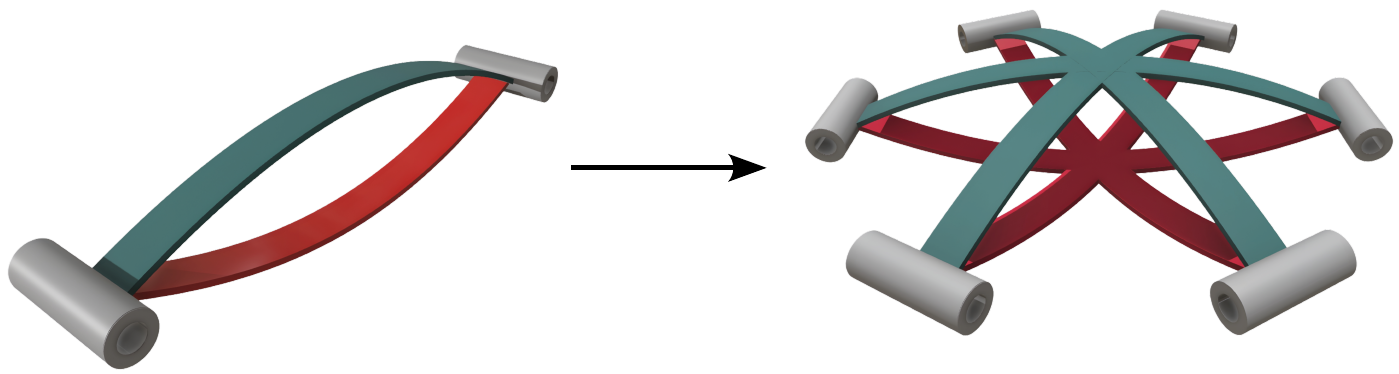}
         \caption{$ $}
         \label{fig:Planar_to_Connected_flip}
     \end{subfigure}
     \hfill
     \begin{subfigure}[b]{0.30\textwidth}
         \centering
         \includegraphics[width=\textwidth]{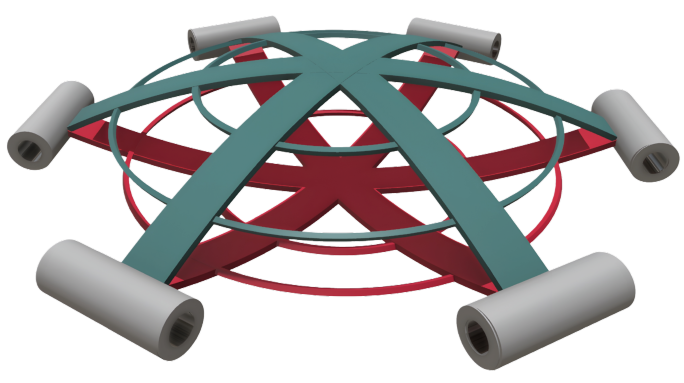}
         \caption{$ $}
         \label{fig:Connected_arch_net}
     \end{subfigure}
\caption{(a) A planar bistable arch (left) and an MFA formed by fusing three arches (right). (b) Interconnected bistable arches, showing the as-fabricated (first) state in green and the second stable state in red.}
\label{fig:Introduction}
\end{figure}

In this study, we investigate midpoint-fused arches (MFA), a class of three-dimensional bistable structures obtained by fusing planar arches at their midpoint, as shown in \cref{fig:Planar_to_Connected_flip}. \Cref{fig:Planar_to_Connected_flip} depicts three arches fused at a common midpoint, whereas the simplest MFA configuration arises when two planar arches are joined at their midpoint. These fused-arch configurations present an attractive alternative to bistable shells, retaining structural robustness while remaining more amenable to modelling. Unlike bistable shells, which often require expensive numerical treatments, MFA, owing to their comparatively simpler geometry, enable a greater degree of analytical tractability. While the present work focuses on MFA, the concept naturally extends to more intricate topologies in which multiple arches are connected at locations beyond the common centre to form interconnected bistable arches, as illustrated in \cref{fig:Connected_arch_net}. Such elastic structures have already motivated functional designs; for example, MFA-like soft bistable grippers\cite{thuruthel2020bistable} have been explored for space-debris clearance\cite{Liu2023}.

We present a simplified analytical framework for studying MFA that builds upon established models for planar arches. Unlike planar arches, each constituent arch in an MFA can experience coupled in-plane bending, out-of-plane bending, and torsion, and the deformations of the constituent arches need not be identical. Accordingly, we model each arch as a spatial arch\cite{Palathingal.2019w9j} capable of out-of-plane and torsional deformations (in addition to in-plane bending) in \cref{sec:modelling}, and introduce an appropriate non-dimensionalization. The formulation is developed by deriving the total potential energy of the MFA and enforcing kinematic compatibility at the fusion point to relate the deformations of the individual arches (\cref{sec:Kinematicsatmidpoint}). The corresponding equilibrium equations are obtained in \cref{Sec:EQEquations}. We then analyse the two-arch MFA (2-MFA) in \cref{Sec:2arches} and the three-arch MFA (3-MFA) in \cref{Sec:3arches}, respectively, by utilizing appropriate deformation symmetries. Furthermore, in \cref{sec:FEAandEXP} we assess the analytical predictions against nonlinear finite-element analysis (FEA) and table-top experiments on 3D-printed MFA. Finally, \cref{sec:summary} summarizes the main findings of this work.

\section{Modelling}
\label{sec:modelling}
\begin{figure}[b!]
\centering
\includegraphics[width=0.92\textwidth]{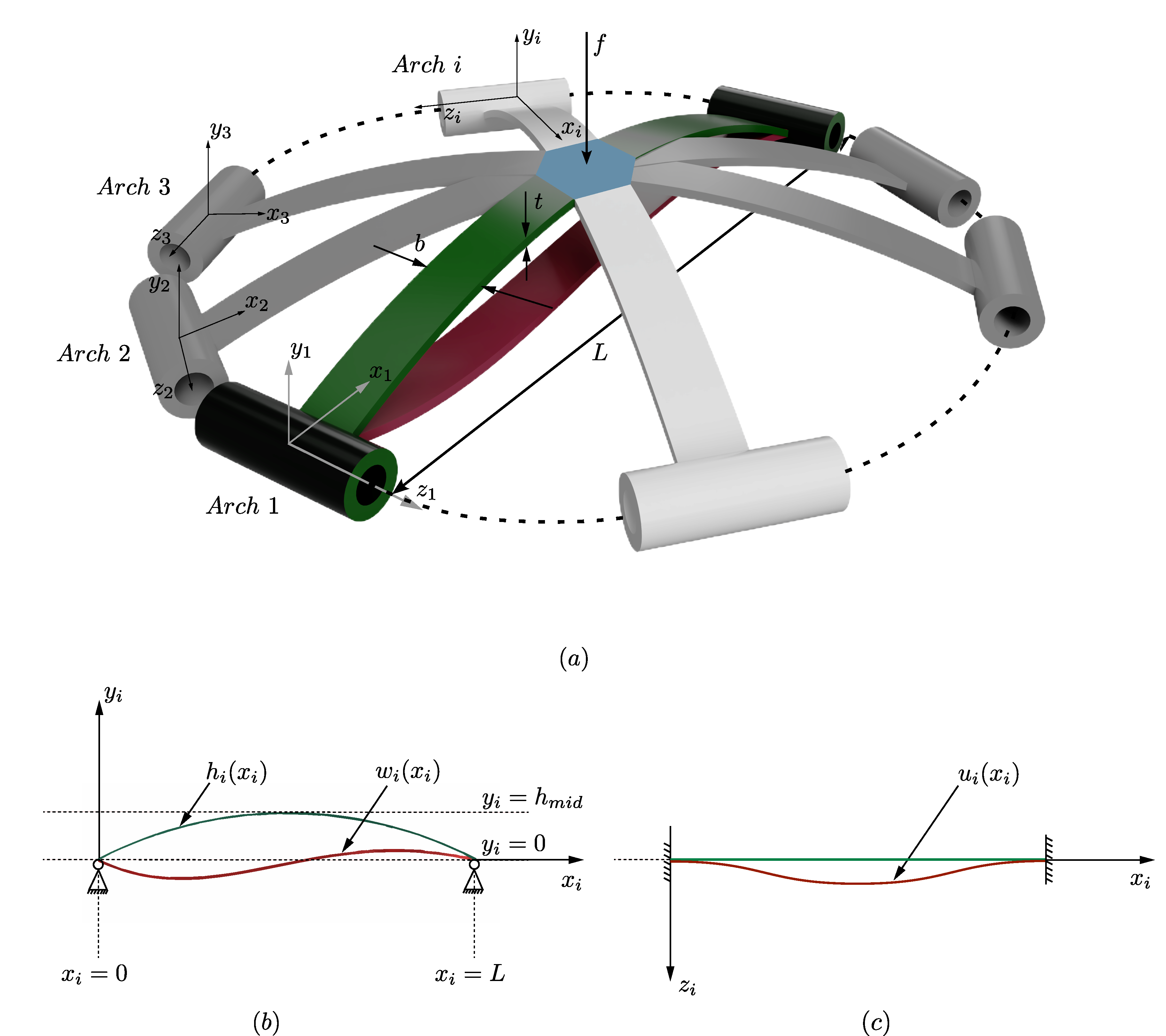}
    \hfill
    \caption{(a) MFA comprising \(i\)  arches with hinged boundary conditions. (b) Projection of the \ith arch in the \(x_i\)-\(y_i\) plane, showing the stress-free in-plane profile, \(h_i(x_i)\) and the deformed in-plane profile \(w_i(x_i)\). (b) Projection onto the \(x_i\)-\(z_i\) plane depicting out-of-plane profile \(u_i(x_i)\).}
    \label{fig:A6_Schematic}
\end{figure}

We model a bistable MFA with pinned-pinned boundary conditions and planform diameter \( L \), as illustrated in \cref{fig:A6_Schematic}a. 
Each arch has a rectangular cross-section of depth \( t \) and width \( b \). The stress-free in-plane centreline of the  \ith arch is denoted by \( h_i(x_i) \), where \( x_i \) is the coordinate along its span (see \cref{fig:A6_Schematic}a). In the stress-free configuration, the out-of-plane profile is taken to be zero. Throughout, we assume  shallow arches, i.e., $\left(\frac{dh_i}{dx_i}\right)^2 \ll 1$. As these arches deform spatially, the in-plane projection of the centreline is 
\( w_i(x_i) \), the out-of-plane displacement is \( u_i(x_i) \), and the twist of the cross-section is represented by \( \phi_i(x_i) \).  We next state the dimensional form of the strain energies due to bending, compression, and torsion. 

\subsection{Strain Energy}
As an arch flips from its natural state to the inverted state, the bending strain energy of the \ith arch \cite{love2013treatise,Palathingal.2019w9j} is 
\(
\frac{E I_{z_i}}{2} \int_0^L\left(\frac{d^2 w_i}{d x_i^2}-\frac{d^2 h_i}{d x_i^2} \right)^2dx_i+\frac{E I_{y_i}}{2} \int_0^L\left(\frac{d^2 u_i}{d x_i^2}-\phi_i \frac{d^2 h_i}{d x_i^2}\right)^2 d x_i,
\)
where \( E \) is the Young's modulus, \( I_{z_i} \) and \( I_{y_i} \) are the second moments of area about the \(z_i\)-  and \(y_i\)-axes, respectively.  
The strain energy associated with axial compression of the \ith arch is approximated as  
\(\frac{p_i}{2}(s^0_{i}-s_i),\) 
where \( s^0_{i} \) given by  
\(
\int_{0}^{L}\left[1+\frac{1}{2}\left(\frac{d h_i}{d x_i}\right)^{2} \right] {d} x_i 
\)
is the stress-free arc length, and \( s_i \) given by 
\(
\int_{0}^{L}\left[1+\frac{1}{2}\left(\frac{{d} w_i}{{d} x_i}\right)^{2}+\frac{1}{2}\left(\frac{{d} u_i}{{d} x_i}\right)^{2}\right] {d} x_i  
\)
is the compressed arc length. The axial force, \( p_i \), is taken as  
\(
E b t\left(\frac{s^0_{i}-s_i}{L}\right)\).  The torsional strain energy is 
\(
\frac{G J}{2} \int_{0}^{L}\left(\frac{{d} \phi_i}{{d} x_i}\right)^{2} d x_i,  
\)  
where \( G \) is the shear modulus, and \( J \) is the polar moment of inertia.  The work potential corresponding to the downward concentrated load \( f \) applied at the fusion point is  
\(
f\left(w_i(L/2)-h_i(L/2)\right).  
\)  Before summing the potential energies of individual arches and discussing their kinematic coupling at the midpoint, we present the appropriate non-dimensionalization.

\subsection{Non-Dimensionalization}  \label{nondimensionalisation}
We introduce the following dimensionless variables:  
\begin{align}
    X_i=\frac{x_i}{L}, \quad 
    H_i=  \frac{h_i}{h_{mid}}, \quad 
    W_i=  \frac{w_i}{h_{mid}}, \quad 
    U_i=  \frac{u_i}{h_{mid}}, \quad 
    \Phi_i=  \phi_i, \quad  
    S_{i}=\frac{s_{i} L}{h_{mid}^2},  \nonumber \\ 
    S^0_{i}=\frac{s^0_{i} L}{h_{mid}^2}, \quad 
    Q=  \frac{h_{mid}}{t}, \quad 
    \beta =\sqrt{ \frac{I_{y_i}}{I_{z_i}}}, \quad 
    \lambda =\frac{L}{h_{mid}}, \quad 
    \mathrm{and} \quad 
    K =\frac{G J \lambda^2}{2 E I_{z_i}}, \label{Eq:NDvariables}
\end{align}  
where \( h_{mid} = h(L/2) \), and since we assume that all the arches have identical cross-sections, we take \(I_{z_i} = I_z\) for all \(i\). Likewise the parameters \(Q\), \(\beta\), \(\lambda\) and \(K\) are identical for all  arches.  The dimensionless form of the strain energy can be obtained by scaling the dimensional strain energy using the non-dimensionalizing factor \( \frac{L^3}{E I_z h_{mid}^2} \).   Thus, the non-dimensional strain energy of an individual arch is given by:  
\begin{align}
SE_i =&  \frac{1}{2} \int_0^1\left(\frac{d^2 W_i}{d X_i^2}-\frac{d^2 H_i}{d X_i^2}\right)^2 d X_i  
+ \frac{\beta^2}{2} \int_0^1\left(\frac{d^2 U_i}{d X_i^2}-\Phi_i \frac{d^2 H_i}{d X_i^2}\right)^2 d X_i \nonumber \\  
& + \frac{3}{2} Q^2\left\{\int_0^1\left[\left(\frac{d H_i}{d X_i}\right)^2-\left(\frac{d W_i}{d X_i}\right)^2-\left(\frac{d U_i}{d X_i}\right)^2\right] d X_i\right\}^2  
+ K \int_0^1\left(\frac{d \Phi_i}{d X_i}\right)^2 d X_i. \label{Eq:NDSE}
\end{align}  
While deriving \cref{Eq:NDSE}, we used \cref{Eq:NDvariables} and expressed \( S_i^0 \) and \( S_i \) as:  
\begin{align}
S_i^0 = \int_0^1\left[\lambda^2+\frac{1}{2}\left(\frac{d H_i}{d X_i}\right)^2\right] d X_i, \quad 
S_i = \int_0^1\left[\lambda^2+\frac{1}{2}\left(\frac{d W_i}{d X_i}\right)^2\right] d X_i.
\end{align}  
Additionally, the non-dimensional axial force is given by  
\begin{align}
P = \frac{p L^2}{E I_z} = 12 Q^2 (S_i^0 - S_i).
\end{align}
By using the relationship between \( E \), \( G \), and Poisson's ratio \( \nu \), along with a standard approximation \cite{young2002roark} for \( J \), \( K \) can be expressed as:  
\begin{align}
K &=\frac{3 \lambda^2}{16(1+\nu)}\left[\frac{16}{3}-\frac{3.36}{\beta}\left(1-\frac{1}{12 \beta^4}\right)\right].
\end{align}  
Finally, the non-dimensional form of the work potential is given by:  
\begin{align}
WP = F \delta_i, \label{Eq:NDWP}
\end{align}  
where \( F = \frac{f L^3}{E I_z h_{mid}} \) and 
\begin{equation}
    \delta_i=H_i(1/2) - W_i(1/2).\label{Eq:delta}
\end{equation}  

The total potential energy of the MFA is the summation of the strain energies of the individual arches (\cref{Eq:NDSE}) and the work potential (\cref{Eq:NDWP}), i.e.,  
\begin{align}
PE = \sum_{i} SE_i - WP. \label{Eq:NDPE}
\end{align}  
In the next section, we derive the kinematic relations required to capture the fusion of arches at the midpoint.

\section{Kinematics at the Midpoint} 
\label{sec:Kinematicsatmidpoint}
The fusion of the arches at the midpoint ensures that their deformations are kinematically related at that point. Consequently, the displacement at the midpoint is the same for all arches, i.e.,
\begin{equation}
      W_1 \big|_{X_1 = 1/2} = W_2 \big|_{X_2 = 1/2}=\dots=W_i \big|_{X_i = 1/2} \label{Eq:midpoint_disp}.
\end{equation}
 Furthermore, the rotation of the region surrounding the fusion point establishes a kinematic relationship between \( \Phi_i \big|_{X_i = 1/2} \) and \( \frac{1}{\lambda}\frac{d W_i}{d X_i} \big|_{X_i = 1/2} \) of the individual arches. For instance, \( \Phi \big|_{X = 1/2} \) of one arch will be equal to \( \frac{1}{\lambda}\frac{d W}{d X} \big|_{X = 1/2} \) of an arch fused with it at an angle of \( \frac{\pi}{2} \) radians. This is because the portion where all the arches are fused together displaces like a rigid body. 
 
\begin{figure}[!b]
     \centering
     \begin{subfigure}[b]{0.463\textwidth}
         \centering
         \includegraphics[width=\textwidth]{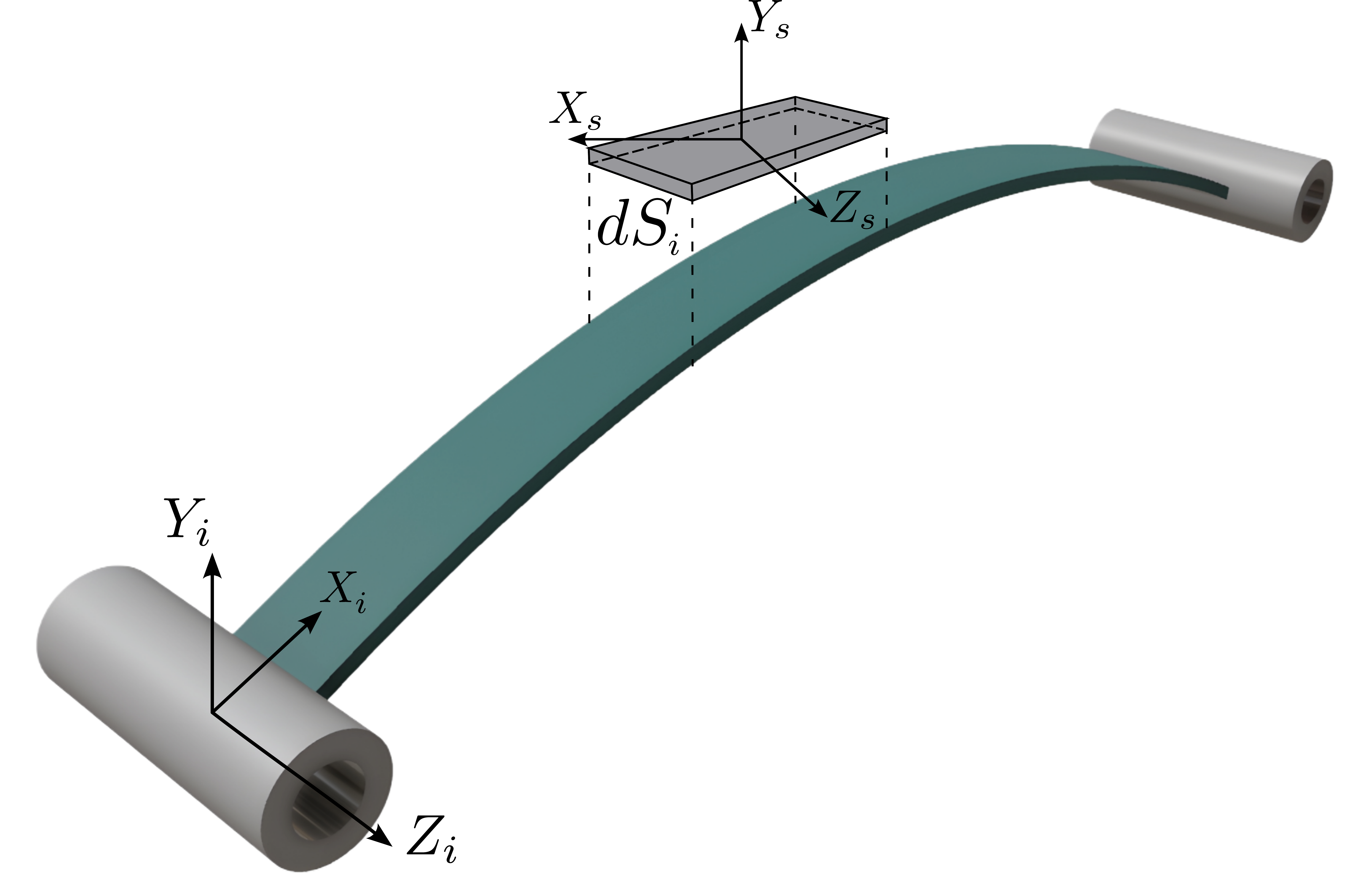}
         \caption{$ $}
         \label{fig:A6_midpatch1}
     \end{subfigure}
     \hfill
     \begin{subfigure}[b]{0.525\textwidth}
         \centering
         \includegraphics[width=\textwidth]{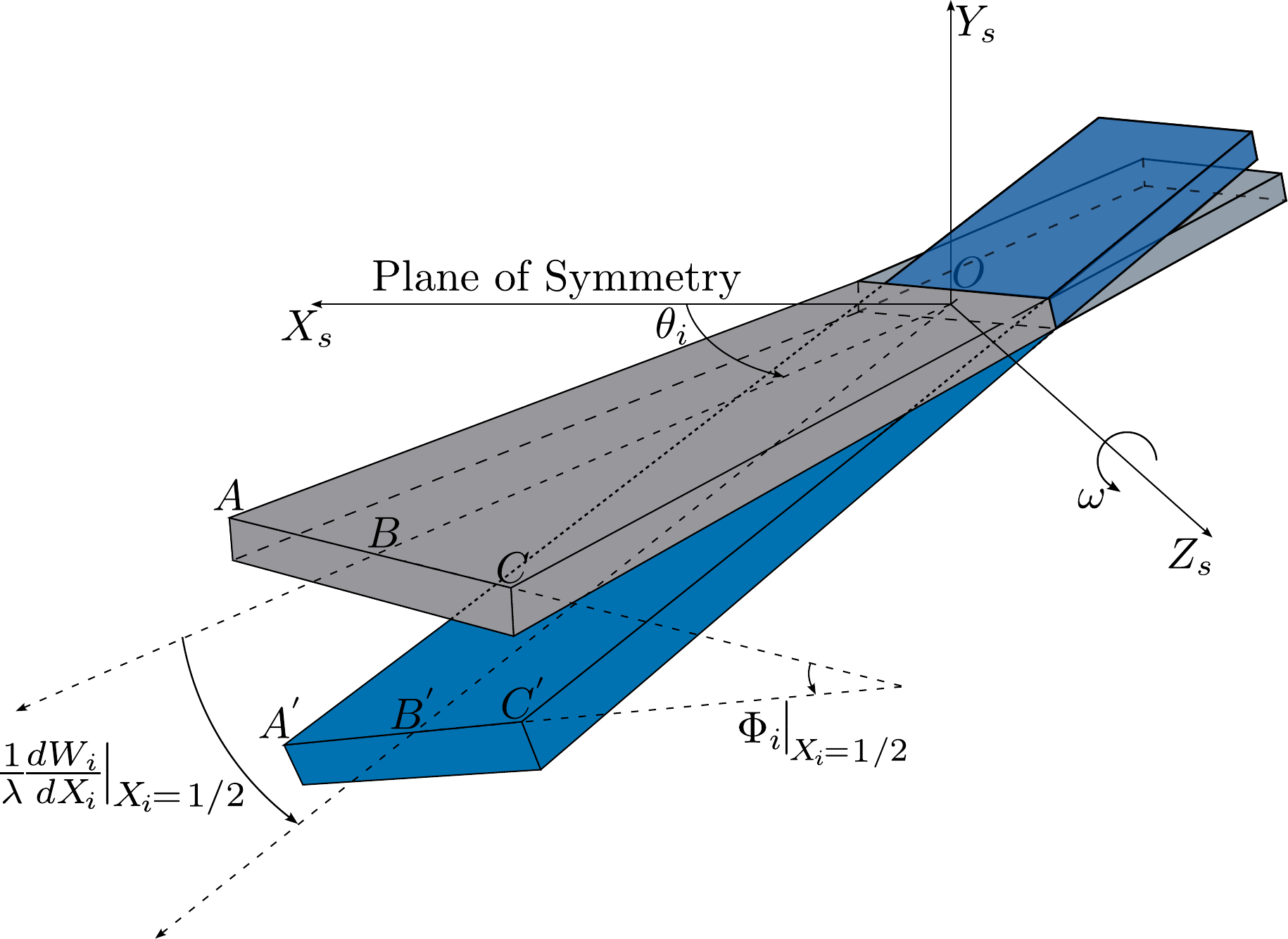}
         \caption{$ $}
         \label{fig:A6_midpatch2}
     \end{subfigure}
\caption{(a) A small element $dS_i$ of an arch. (b) Schematic of the element $dS_i$ before rotation (grey) and after rotation (blue) about the $Z_{s}$-axis.}
        \label{fig:A6_midpatch}
\end{figure}

 Generalizing this, we next derive the kinematic relations between \( \Phi_i \big|_{X_i = 1/2} \) and \( \frac{1}{\lambda}\frac{d W_i}{d X_i} \big|_{X_i = 1/2} \) for the \ith  arch. Consider a small element $dS_i$ (grey colour in \cref{fig:A6_midpatch1})  at its midpoint. 
The element $dS_i$ is inclined at an angle $\theta_i$ with the $X_s$ axis in the $X_sZ_s$ plane as shown in \cref{fig:A6_midpatch2}. The coordinate system $X_sY_sZ_s$  is defined such that the $X_sY_s$ plane represents the plane of symmetry of deformation.

 As shown in \cref{fig:A6_midpatch2}, as $dS_i$ rotates by a small angle $\omega$ about the $Z_{s}$-axis, the edge $A B C$ rotates and translates to $A^{\prime} B^{\prime} C^{\prime}$. The angle between $A C$ and $A^{\prime} C^{\prime}$ is $\Phi_i \big|_{X_i = 1/2}$. The angle between $OB$ and $OB^{\prime}$ is $\frac{1}{\lambda}\frac{d W_i}{d X_i} \big|_{X_i = 1/2}$. 
We can show that $\Phi_i \big|_{X_i = 1/2}$ is related to $\omega$ and $\theta_i$ as (details in \cref{Appendix:kinematics}):
\begin{align}
 \Phi_i \big|_{X_i = 1/2}=\cos ^{-1}\left(\sin ^2 \theta_i \cos \omega+\cos ^2 \theta_i\right), 
\end{align}
which for small angles of $\Phi_i \big|_{X_i = 1/2}$ and $\omega$ simplifies to,
\begin{align}
\Phi_i \big|_{X_i = 1/2}  =\omega \sin \theta_i. \label{Eq:phi_omega_relation}
\end{align}
Similarly, we obtain (details in \cref{Appendix:kinematics}),
\begin{align}
\frac{1}{\lambda}\frac{d W_i}{d X_i} \big|_{X_i = 1/2} & =\omega \cos \theta_i. \label{Eq:slope_omega_relation}
\end{align}
For each arch in the MFA (by choosing appropriate $\theta_i$ values), we use \cref{Eq:phi_omega_relation,Eq:slope_omega_relation} to express its slope and rotation in terms of $\omega$. Then we eliminate $\omega$ to get direct relationships between slopes and rotations of the arches at the midspan. We will illustrate this for 2-MFA and 3-MFA in the later sections. 

\section{Equilibrium Equations}  
\label{Sec:EQEquations}
We take the as-fabricated shape of the $i$th arch, \( H_i(X_i) = a_1 \sin(\pi X_i) \), and approximate \( W_i(X_i), U_i(X_i), \Phi_i(X_i) \) as:  
\begin{align}
W_i(X_i) &= \sum_{j=1,2,3,\dots,n} A_{ij} \sin(j \pi X_i), \nonumber \\  
U_i(X_i) &= \sum_{j=1,3,5,\dots,m} B_{ij} \bar{U}_j(X_i) + \sum_{j=2,4,6,\dots,m} B_{ij} \tilde{U}_j(X_i), \quad \mathrm{and} \nonumber \\  
\Phi_i(X_i) &= \sum_{j=1,2,3,\dots,l} C_{ij} \sin(j \pi X_i), \label{Eq:modeshapes}  
\end{align}  
where,
\begin{align}  
\bar{U}_j &= \frac{1}{2} - \frac{1}{2} \cos((j+1) \pi X), \\  
\tilde{U}_j &= \frac{1}{2}\left[1 - 2 X - \cos(N_j X) + \frac{2}{N_j} \sin(N_j X)\right], \\  
\tan(N_j/2)&= N_j/2,
\end{align}  
and \( A_{ij}, B_{ij}, C_{ij} \) are the unknown mode weights corresponding to \( W(X_i) \), \( U(X_i) \), and \( \phi(X_i) \), respectively. The subscript \( i \) denotes the arch, while the subscripts \( j \) and \( k \) refer to the mode numbers. The parameter \( r \) represents the total number of arches in the MFA. The parameters \( n \), \( m \), and \( l \) denote the maximum numbers of modes used to approximate the in-plane, out-of-plane, and torsional deformations, respectively. By substituting \cref{Eq:modeshapes} into \cref{Eq:NDPE} and simplifying, we obtain:  
\begin{align}
P E=& \sum_{i=1,2,3, \ldots r} \frac{\pi^4}{4}\left(a_1-A_{i 1}\right)^2+\sum_{\substack{i=1,2,3, \ldots r \nonumber\\
j=2,3,4, \ldots n}} \frac{\pi^4 j^4 A_{i j}^2}{4}+\sum_{\substack{i=1,2,3, \ldots r \\
j=1,3,5, \ldots m}} \frac{\beta^2 B_{i j}^2(j+1)^4 \pi^4}{16} \\ \nonumber
& +\sum_{\substack{i=1,2,3, \ldots r \\
j=2,4,6, \ldots m}} \frac{\beta^2 B_{i j}^2}{16} N_j^4+\sum_{i=1,2,3, \ldots r} \frac{3 a_1^2 \pi^4 \beta^2 C_{i 1}^2}{16}+\sum_{\substack{i=1,2,3, \ldots r \\
j=2,3,4, \ldots l}} \frac{\beta^2 a_1^2 \pi^4}{8} C_{i j}^2 \\ \nonumber
& +\sum_{\substack{i=1,2,3, \ldots r \\
j=1,3,5, \ldots m \\
k=1,3,5, \ldots l}} 2 \beta^2 a_1 \pi^2 B_{i j} C_{i k}  M_{j k}^*+\sum_{\substack{i=1,2,3, \ldots r \\
j=2,4,6, \ldots m \\
k=2,4,6, \ldots l}} 2 \beta^2 a_1 \pi^2 B_{i j} C_{i k}  M_{j k}^* \\ \nonumber
& +\frac{3 Q^2}{2}\left(\frac{\pi^2 a_1^2}{2}-\sum_{\substack{i=1,2,3, \ldots r \\
j=1,2,3, \ldots n}}\frac{\pi^2 j^2 A_{i j}^2}{2}-\sum_{\substack{i=1,2,3, \ldots r \\
j=1,3,5, \ldots m}}\frac{(j+1)^2 \pi^2 B_{i j}^2}{8}-\sum_{\substack{i=1,2,3, \ldots r \\
j=2,4,6, \ldots m}}\frac{N_j^2 B_{i j}^2}{8}\right)^2 \\  &+\sum_{\substack{i=1,2,3, \ldots r \\
j=1,2,3, \ldots l}} \frac{\pi^2 K j^2 C_{i j}^2}{2} - F\left(1 - W(1/2)\right)
\label{Eq:PEintermsofmodeweights}
\end{align}
where,
\begin{align}
M_{j k}^*= \begin{cases}\int_0^1 \left[\sum_{k=1,3,5, \ldots l} \bar{U}_j^{\prime \prime} \sin (k \pi X) \sin (\pi X) \right]d X &  \quad \text{if} \quad k=1,3,5, \ldots l \\ \int_0^1 \left[ \sum_{k=2,4,6, \ldots l} \tilde{U}_j^{\prime \prime} \sin (k \pi X) \sin (\pi X) \right]d X & \quad \text{if} \quad  k=2,4,6, \ldots l\end{cases},
\end{align}
and some computed values of $M_{j k}^*$  are:
\begin{align}
\begin{array}{c|c|cccccc}
M_{j k}^* & k & 1 & 2 & 3 & 4 & 5 & 6 \\
\hline j & & & & & & \\
\hline 1 && -\pi^2 / 2 & 0 & 0 & 0 & 0 & 0 \\
2 && 0 & -11.155 & 0 & 0.549477 & 0 & 0.20847\\
3 && \pi^2 / 2 & 0 & -2 \pi^2 & 0 & 0 & 0 \\
4 && 0 & 9.52924 & 0 & -31.3504&0&0.84847 \\
5 && 0 & 0 & 2 \pi^2 & 0 & -9 \pi^2 / 2 & 0 \\
6 && 0 & 0.28611 & 0 & 28.783 & 0 & -61.4454
\end{array}
\end{align}

A detailed derivation of \cref{Eq:PEintermsofmodeweights} is included in \cref{appendix:PEintermsofmodeweights}. Next, we use \cref{Eq:PEintermsofmodeweights} to derive the equilibrium equations for the MFA and determine their stability.  

\section{Two-Arch MFA (2-MFA)}  \label{Sec:2arches}
\begin{figure}[!htb]
    \centering
    \includegraphics[width=1.0 \linewidth]{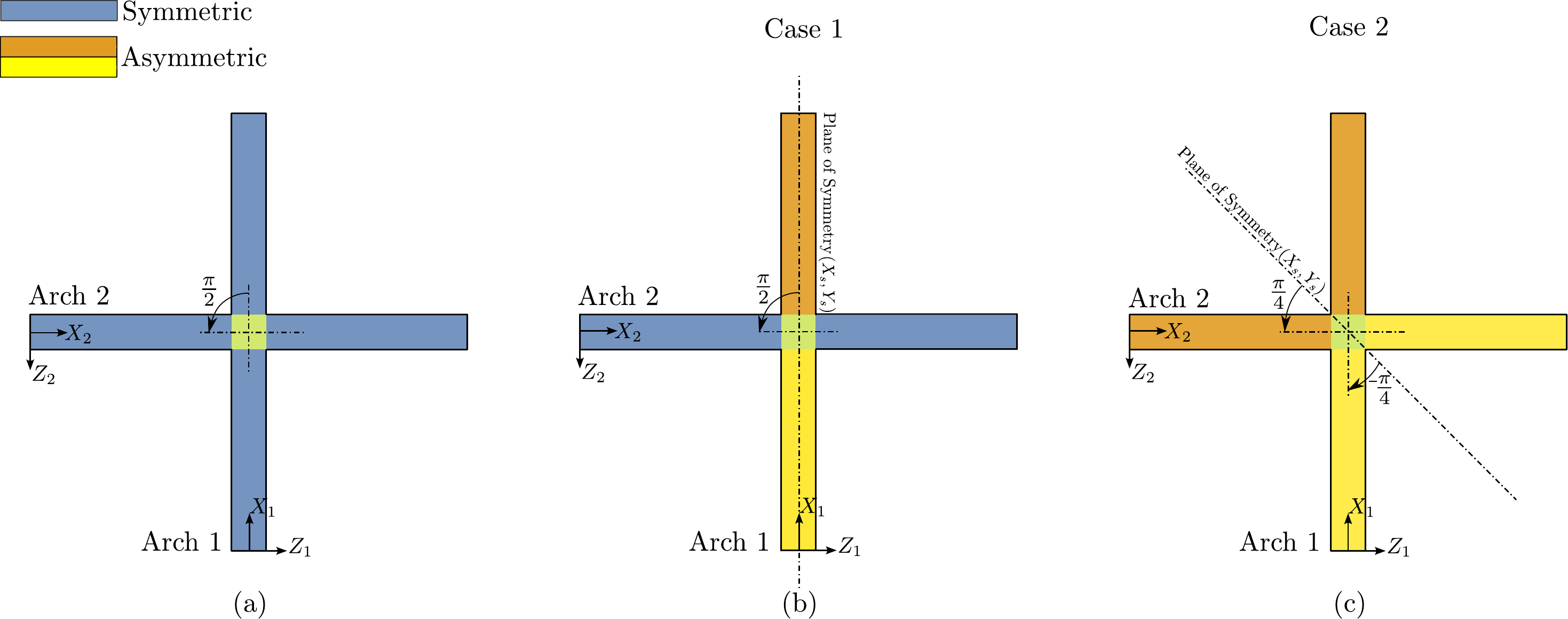}
\caption{Deformation patterns of a 2-MFA: (a) symmetric-symmetric deformation of Arch~1 and Arch~2; (b) Case~1: Arch~1 deforms asymmetrically while Arch~2 remains symmetric; (c) Case~2: both arches deform asymmetrically.}
    \label{fig:DP_doublearch}
\end{figure}

We consider a 2-MFA with two planar arches fused at an angle of \(\frac{\pi}{2}\) between them, as shown in \cref{fig:DP_doublearch}a. Thus, the total potential energy of the 2-MFA  as given by \cref{Eq:NDPE}, is \( SE_1 + SE_2 - WP \). However, note that when we express this potential energy by using \cref{Eq:PEintermsofmodeweights}, some of the mode weights would become dependent on the others due to the kinematics described in \cref{sec:Kinematicsatmidpoint}.

The 2-MFA may follow three deformation pathways while switching from its initial profile to the inverted profile. The first pathway involves both arches switching symmetrically, as schematically depicted in \cref{fig:DP_doublearch}a. The second pathway is when one arch
deforms asymmetrically (Arch 1 in \cref{fig:DP_doublearch}b),  while the other remains symmetric (Arch 2 in \cref{fig:DP_doublearch}b). The third pathway results in both arches becoming identically asymmetric, as shown in \cref{fig:DP_doublearch}c. 

To simplify modelling these distinct cases, we categorize the problem into two cases based on the plane of symmetry, which defines the symmetry of the deformation. In the first case (\cref{fig:DP_doublearch}b)), the deformation is such that the asymmetric arch lies on the plane of symmetry, while the other arch deforms symmetrically.  In the second case, when both arches deform asymmetrically, the plane of symmetry bisects the angle between them, as shown in \cref{fig:DP_doublearch}c. Note that the deformation pathway in which both arches remain symmetric can be captured using either of these planes of symmetry.

\subsection{Case 1} \label{Sec:doublearchescase1}
To analyse Case 1, we start by taking $n=2$, $m=0$, and $l=1$ in \cref{Eq:PEintermsofmodeweights} for both arches. We assume that Arch~1 (\(i=1\)) undergoes asymmetric in-plane deformation with neither torsion (\(C_{11}=0\)) nor out-of-plane deformation (\(B_{11}=0\)), whereas Arch~2 (\(i=2\)) deforms symmetrically in-plane and does not exhibit out-of-plane deformation (\(B_{21}=0\)).  The asymmetric deformation of Arch~1 induces torsion in Arch~2. Accordingly, the unknown coefficients are \(A_{11}\), \(A_{12}\), \(A_{21}\), \(A_{22}\), and \(C_{21}\), and the potential energy in \cref{Eq:PEintermsofmodeweights} reduces to:
\begin{align}
PE &=-F \left(a_1-A_{11}\right)+\frac{3}{2} Q^2 \left[\frac{1}{2} \pi ^2 a_1^2-\pi ^2 \left(\frac{A_{11}^2}{2}+2
   A_{12}^2\right)\right]^2+\nonumber\\ &\frac{3}{2} Q^2 \left[\frac{1}{2} \pi ^2 a_1^2-\pi ^2 \left(\frac{A_{21}^2}{2}+2
   A_{22}^2\right)\right]^2+\frac{1}{4} \pi ^4 \left(a_1-A_{11}\right)^2+\frac{1}{4} \pi ^4
   \left(a_1-A_{21}\right)^2+\nonumber\\ &\frac{3}{16} \pi ^4 a_1^2 \beta ^2 C_{21}^2+4 \pi ^4 A_{12}^2+4 \pi ^4 A_{22}^2+\frac{1}{2} \pi ^2
   C_{21}^2 K. \label{Eq:PEn2m1l1case1}
\end{align}

From the conditions at the midpoint  (\cref{Eq:midpoint_disp,Eq:slope_omega_relation}), the compatibility constraints are:
\begin{align}
     {W_1} \big|_{X_1 = 1/2}&={W_2} \big|_{X_2 = 1/2}  &\implies \qquad A_{21}&=A_{11} \label{i2n2m0l1c1:deltacondition}\\
     \Phi_1 \big|_{X_2 = 1/2} &=  0  &\implies \qquad C_{11}&=0 \label{i2n2m0l1c1:phicondition}\\
     \Phi_2 \big|_{X_2 = 1/2} \cos(0) &=  \frac{1}{\lambda}\frac{d W_1}{d X_1} \bigg|_{X_1 = 1/2} \sin(\pi/2)  &\implies \qquad C_{21}&=-\frac{2 \pi  \text{A}_{12}}{\lambda } \label{i2n2m0l1c1:slope&phicondition}\\
        \frac{d W_2}{d X_2} \bigg|_{X_2 = 1/2} &=0  &\implies \qquad A_{22}&=0 \label{i2n2m0l1c1:slopecondition},
\end{align}
where \cref{i2n2m0l1c1:slopecondition} follows from the assumed symmetric deformation of Arch 2. Eliminating  $A_{21}$, $A_{22}$ and $C_{21}$ using \cref{i2n2m0l1c1:deltacondition,i2n2m0l1c1:phicondition,i2n2m0l1c1:slope&phicondition,i2n2m0l1c1:slopecondition}, we solve for the remaining unknowns $A_{11}$ and $A_{12}$ from the equilibrium equations:
\begin{align}
       \frac{\partial PE}{\partial A_{11}}&=3 \pi ^2 A_{11} Q^2 \left( \Delta+\frac{1}{2} \pi ^2 a_1^2-\frac{1}{2} \pi ^2 A_{11}^2 \right)+\pi ^4 \left(a_1-A_{11}\right)-F=0 \label{Eq:EE_i2n2m1l1_A11}\\
     \frac{\partial PE}{\partial A_{12}}&=   A_{12}\left(\frac{3 \pi ^6 a_1^2 \beta ^2}{2 \lambda ^2}+\frac{4 \pi ^4 K}{\lambda ^2}-12 \pi ^2 \Delta Q^2+8 \pi ^4\right)=0 \label{Eq:EE_i2n2m1l1_A12},
\end{align}
where $ \Delta= \frac{1}{2} \pi ^2 a_1^2-\pi ^2\left(\frac{A_{11}^2}{2}+2 A_{12}^2\right)$.
\Cref{Eq:EE_i2n2m1l1_A12} admits a symmetric solution for \(A_{12}=0\), and an asymmetric solution when the corresponding \(\Delta\), denoted \(\Delta_{sa}\), satisfies
\begin{equation}
   \frac{3 \pi ^6 a_1^2 \beta ^2}{2 \lambda ^2}+\frac{4 \pi ^4 K}{\lambda ^2}-12 \pi ^2 \Delta_{sa} Q^2+8 \pi ^4=0.
   \label{Eq:Delta1Eq_i2n2m1l1}
\end{equation}

\begin{figure}[!hb]
     \centering
         \includegraphics[width=0.65\textwidth]{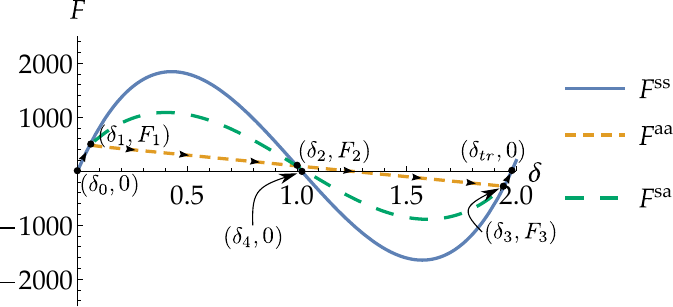}
         \caption{ Force-displacement characteristics of a 2-MFA with $n=2$, $m=0$, $l=1$. $F^{ss}$ (solid blue curve) is the symmetric solution, whereas \(F^{sa}\) (large dashed green curve) and \(F^{aa}\) (small dashed orange curve) correspond to asymmetric solutions for Case~1 and Case~2, respectively for \(Q=4\), \(\beta=3\), and \(\lambda=25\).}
        \label{fig:Fdelta_i2n2m0l1_c1_c2}
\end{figure}

The force, $F^{ss}$ \footnote{Here, “ss” denotes symmetric-symmetric deformation, “sa” denotes symmetric-asymmetric deformation, and “aa” denotes asymmetric-asymmetric deformation.}, corresponding to the symmetric solution ($A_{12}=0$), is 
\begin{equation}
    F^{ss}=\pi ^4 \delta  \left[3 a_1 Q^2 \left(2 a_1-3 \delta \right)+3 \delta ^2 Q^2+1\right], \label{Eq:Fss_i2n2m1l1}
\end{equation}
where $\delta=a_1-A_{11}$ is the midpoint deflection along the direction of the applied force (see \cref{Eq:delta}). The asymmetric force branch $F^{sa}$ is obtained as
\begin{equation}
    F^{sa}=  \frac{1}{2} \left[-2 \pi ^4 \left(a_1-\delta \right)+6 \pi ^2 \Delta_{sa}  Q^2 \left(a_1-\delta \right)-3 \pi ^4 Q^2
   \left(a_1-\delta \right){}^3+3 \pi ^4 a_1^2 Q^2 \left(a_1-\delta \right)+2 \pi ^4 a_1\right],  \label{Eq:Fas_i2n2m1l1}
\end{equation}
where $A_{11}=a_1-\delta$, and
\begin{equation}
    A_{12}= \pm\frac{\sqrt{2 \pi ^2 a_1 \delta -\pi ^2 \delta ^2-2 \Delta_{sa}}}{2 \pi }. \label{Eq:A12as_i2n2m1l1}
\end{equation}
\Cref{Eq:Fss_i2n2m1l1,Eq:Fas_i2n2m1l1} show that   \( F^{ss} \) and  \( F^{sa} \) exhibit  a cubic dependence on the displacement \( \delta \), as illustrated in \cref{fig:Fdelta_i2n2m0l1_c1_c2}. 

\subsection{Case 2} \label{Sec:doublearchescase2}
 In Case 2, both Arch 1 and Arch 2 deform asymmetrically, with no out-of-plane deformation (\(B_{11}=B_{21}=0\)). The potential energy is computed as \(SE_1 + SE_2 - WP \), where \(SE_1\) and \(SE_2\) are the total strain energies of Arch~1 and Arch~2, respectively. The unknown coefficients are \(A_{11}\), \(A_{12}\), \(A_{21}\), \(A_{22}\), \(C_{11}\), and \(C_{21}\). Using \cref{Eq:PEintermsofmodeweights}, the potential energy for this case is 
\begin{align}
PE &=-F \left(a_1-A_{11}\right)+\frac{3}{2} Q^2 \left[\frac{1}{2} \pi ^2 a_1^2-\pi ^2 \left(\frac{A_{11}^2}{2}+2
   A_{12}^2\right)\right]^2\nonumber\\&+\frac{3}{2} Q^2 \left[\frac{1}{2} \pi ^2 a_1^2-\pi ^2 \left(\frac{A_{21}^2}{2}+2
   A_{22}^2\right)\right]^2+\frac{1}{4} \pi ^4 \left(a_1-A_{11}\right)^2+\frac{1}{4} \pi ^4
   \left(a_1-A_{21}\right)^2\nonumber\\&+\frac{3}{16} \pi ^4 a_1^2 \beta ^2 C_{11}^2+\frac{3}{16} \pi ^4 a_1^2 \beta ^2 C_{21}^2+4 \pi
   ^4 A_{12}^2+4 \pi ^4 A_{22}^2+\frac{1}{2} \pi ^2 C_{11}^2 K+\frac{1}{2} \pi ^2 C_{21}^2 K.\label{Eq:PEn2m1l1case2}
\end{align}
Since both arches deform asymmetrically, the plane of symmetry lies between them. The resulting compatibility conditions are:
\begin{align}
     {W_1} \big|_{X_1 = 1/2}&={W_2} \big|_{X_2 = 1/2}  &\implies \qquad A_{21}&=A_{11} \label{i2n2m0l1c2:deltacondition},\\
      \frac{d W_2}{d X_2} \bigg|_{X_2 = 1/2}\cos(-\pi/4) &=  \frac{d W_1}{d X_1} \bigg|_{X_1 = 1/2}\cos(\pi/4)  &\implies \qquad A_{22}&=A_{12} \label{i2n2m0l1c2:slope&slopecondition},\\
     \Phi_2 \big|_{X_2 = 1/2}\sin(-\pi/4) &=  \Phi_1 \big|_{X_1 = 1/2}\sin(\pi/4)  &\implies \qquad C_{21}&=-C_{11} \label{i2n2m0l1c2:phi&phicondition},\\
         \Phi_1 \big|_{X_1 = 1/2}\cos(-\pi/4) &=\frac{1}{\lambda}\frac{d W_1}{d X_1} \bigg|_{X_1 = 1/2}\sin(-\pi/4) &\implies \qquad C_{11}&=\frac{2 \pi  \text{A}_{12}}{\lambda } \label{i2n2m0l1c2:slope&phicondition}.
\end{align}
Substituting these relations into \cref{Eq:PEn2m1l1case2} and enforcing equilibrium yields the governing equations for \(A_{11}\) and \(A_{12}\):
\begin{align}
    \frac{\partial PE}{\partial A_{11}}&=6 \pi ^2 A_{11} Q^2 \Delta+\pi ^4
   \left(a_1-A_{11}\right)-F=0, \label{Eq:EEi2n2m1l1_1_c2}\\
    \frac{\partial PE}{\partial A_{12}}&=2 \left(\frac{3 \pi ^6 a_1^2 A_{12} \beta ^2}{2 \lambda ^2}+8 \pi ^4 A_{12}\right)-24 \pi ^2 A_{12} Q^2 \Delta+\frac{8 \pi ^4 A_{12} K}{\lambda ^2}=0,\label{Eq:EEi2n2m1l1_2_c2}
\end{align}
where $ \Delta= \frac{1}{2} \pi ^2 a_1^2-\pi ^2 \left(\frac{A_{11}^2}{2}+2 A_{12}^2\right)$.
\Cref{Eq:EEi2n2m1l1_1_c2,Eq:EEi2n2m1l1_2_c2} admit  the same symmetric solution as in Case~1 when $A_{12}=0$.  In addition, an asymmetric solution (with both arches deforming asymmetrically) exists when \(\Delta=\Delta_{aa}\), where
\begin{align}
\frac{3 \pi ^6 a_1^2 \beta ^2}{\lambda ^2}+\frac{8 \pi ^4 K}{\lambda ^2}-24 \pi ^2 \Delta_{aa} Q^2+16 \pi ^4=0.
   \label{Eq:Delta1Eq_i2n2m1l1c2}
\end{align}
The corresponding asymmetric force branch \(F^{aa}\) is linear and is given by
\begin{equation}
    F^{aa}=  -\pi ^4 \left(a_1-\delta \right)+6 \pi ^2 \Delta_{aa}  Q^2 \left(a_1-\delta \right)+\pi ^4 a_1,  \label{Eq:Faa_i2n2m1l1}
\end{equation}
as shown in \cref{fig:Fdelta_i2n2m0l1_c1_c2}. As expected, this branch has a substantially smaller magnitude than \(F^{sa}\). Here, \(A_{11}=a_1-\delta\) and $A_{12}=\pm\frac{\sqrt{2 \pi ^2 a_1 \delta -\pi ^2 \delta ^2-2 \Delta_{aa}}}{2 \pi }$.
\subsection{Critical Points on the Force-Displacement Curves}
We next summarize expressions for the key points on the force-displacement curve, as illustrated in \cref{fig:Fdelta_i2n2m0l1_c1_c2}. 
 The force branches \(F^{aa}\), \(F^{sa}\), and \(F^{ss}\) depend on the non-dimensional geometric parameter \(Q\). Since \(F^{ss}\) is cubic in \(\delta\), it can attain three distinct zeros for appropriate values of \(Q\). Specifically, \(F^{ss}=0\) at \[
\delta_0 = 0, \quad \delta_4 = \frac{1}{6} \left[9 a_1-\frac{ \sqrt{3 Q^2 \left(3 a_1^2 Q^2-4\right)}}{Q^2}\right], \, \& \quad \delta_{tr} = \frac{1}{6} \left[9 a_1 + \frac{ \sqrt{3 Q^2 \left(3 a_1^2 Q^2-4\right)}}{Q^2}\right],
\]where the last point corresponds to the $\delta$ in the second stable state, typically referred to as the travel of midpoint, \(\delta_{tr}\) \cite{Palathingal.2017}. While \(F^{ss}=0\) always at \(\delta=0\), the remaining two roots depend on \(Q\). When \(Q = \frac{2}{\sqrt{3}a_1}\), these two roots coincide; for \(Q\) below this value, no real solutions exist. Therefore, for a 2-MFA with uniform cross-sections to exhibit bistability, the condition
\begin{equation}
    Q > \frac{2}{\sqrt{3}a_1},\label{Eq:bistabilityCondition}
\end{equation} must be satisfied. The maximum value of \( F^{ss} \) is
\(F^{ss}_{max}=\frac{2}{3} \pi ^4 a_1^2 \sqrt{Q^2 \left(3 a_1^2 Q^2-1\right)}-\frac{2 \pi ^4 \sqrt{Q^2 \left(3 a_1^2 Q^2-1\right)}}{9 Q^2}+\pi ^4
   a_1,\) which occurs at  $\delta=a_1-\frac{\sqrt{Q^2 \left(3 a_1^2 Q^2-1\right)}}{3 Q^2}$.

The symmetric and asymmetric branches intersect at three points \((\delta_1, F_1)\), \((\delta_2, F_2)\), and \((\delta_3, F_3)\) on the force-displacement curve, as indicated in \cref{fig:Fdelta_i2n2m0l1_c1_c2}. These intersection points can be expressed in terms of $\Delta{aa}$ (see \cref{Eq:Faa_i2n2m1l1}), $Q$, and $a_1$ as:
\begin{align}
& \delta_1= \frac{\pi  a_1-\sqrt{\pi ^2 a_1^2-2 \Delta_{aa} }}{\pi }, \qquad &F_1=\pi ^4 a_1-  \pi  \sqrt{\pi ^2 a_1^2-2 \Delta_{aa} } \left(\pi ^2-6 \Delta_{aa}  Q^2\right),\\
   & \delta_2=a_1, \qquad    &F_2= \pi ^4 a_1,\\
     & \delta_3\frac{\pi  a_1+\sqrt{\pi ^2 a_1^2-2 \Delta_{aa} }}{\pi },      \qquad  &F_3=\pi ^4 a_1+   \pi  \sqrt{\pi ^2 a_1^2-2 \Delta_{aa} } \left(\pi ^2-6 \Delta_{aa}  Q^2\right).\label{Eq:switchbackforce}
\end{align}
The curves  intersect at three points when \(\frac{d F^{ss}}{d \delta} < \frac{d F^{aa}}{d \delta}\) at \(\delta=\delta_2\). From \cref{Eq:Fss_i2n2m1l1,Eq:Faa_i2n2m1l1}, this condition is satisfied when
\begin{equation}
    \Delta_{aa}<\frac{1}{2} \pi ^2 a_1^2 .\label{Eq:intersectionCondition}
\end{equation}

The effective force-displacement curve is obtained by combining the \(F^{ss}\) and \(F^{aa}\) branches, with transitions between them occurring at the intersection points above. This hybrid response is indicated by the black arrows in \cref{fig:Fdelta_i2n2m0l1_c1_c2} and is discussed in the next section.

\subsection{Force-Displacement Response}
The force-displacement response is governed by the interaction between the symmetric and asymmetric solution branches. This behaviour is analogous to that reported for bistable buckled beams \cite{Camescasse.2013,Vangbo.1998}, planar arches \cite{palathingal2019analysis,QiuJMems}, and spatial arches \cite{Palathingal.2019w9j}. Here, we employ an energy-based argument to identify the effective path followed by the arch.

\begin{figure}[!htb]
    \centering
    \includegraphics[width=1\linewidth]{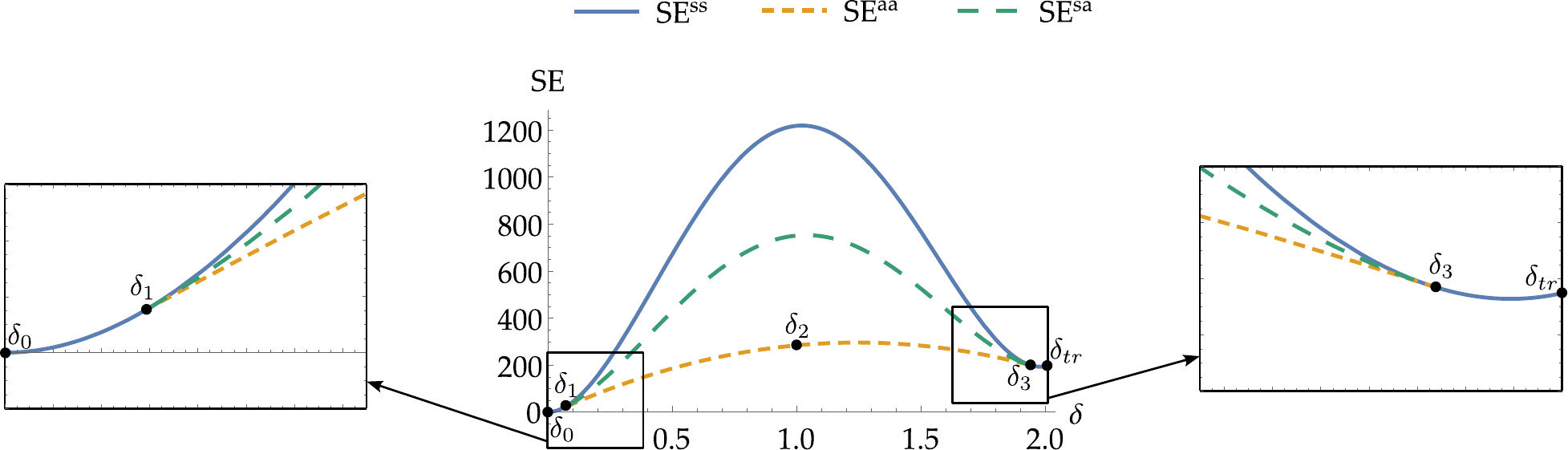}
    \caption{Strain energy variation for different force pathways for \(Q=4\), \(\beta=3\), and \(\lambda=25\).}
    \label{fig:SEdelta_i2n2m0l1_c1_c2}
\end{figure}

Starting from \(\delta=0\), the arch initially deforms along the \(F^{ss}\) branch up to \(\delta=\delta_1\). At \(\delta=\delta_1\), the response may either continue along \(F^{ss}\) or transition to the \(F^{aa}\) or \(F^{sa}\) branches. The energy associated with the \(F^{aa}\) pathway is the lowest, provided that the \(F^{ss}\), \(F^{aa}\), and \(F^{sa}\) branches intersect at three points, as specified by \cref{Eq:intersectionCondition}. The strain energies corresponding to these force pathways are plotted in \cref{fig:SEdelta_i2n2m0l1_c1_c2}.
Using \cref{Eq:Delta1Eq_i2n2m1l1c2}, the switching from the symmetric to the asymmetric pathway can alternatively be expressed as the condition
\begin{equation}
\frac{\pi ^2 \left(3 \pi ^2 a_1^2 \beta ^2+16 \lambda ^2+8 K\right)}{24 \lambda ^2 Q^2} <\frac{1}{2} \pi ^2 a_1^2 .\label{Eq:switchingCondition}
\end{equation} At \(\delta=\delta_3\), the strain energy associated with the symmetric deformation pathway becomes smaller again, and the arches return to this pathway and reach the second stable state at \(\delta_{tr}\).

\begin{figure}[!htbp]
    \centering
\includegraphics[width=0.5\linewidth]{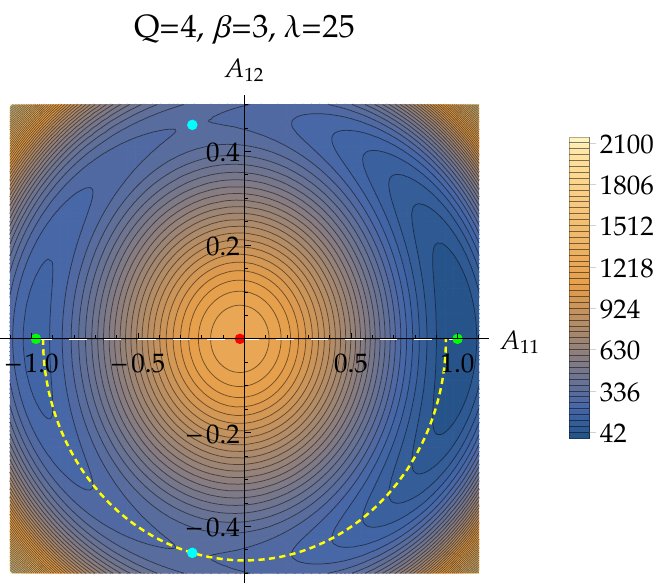}
\caption{Strain-energy contours in  \(A_{11}-A_{12}\) plane for \(Q=4\), \(\beta=3\), and \(\lambda=25\). Stable equilibria are shown in green, the unstable equilibrium in red, and saddle points in cyan. The white dashed line denotes the symmetric branch, and the yellow dotted curve denotes an asymmetric branch.}
    \label{fig:EnergyContour_n2m1l1}
\end{figure}

Alternatively, if only the \(F^{ss}\) and \(F^{sa}\) branches are considered, the strain energy of the 2-MFA can be visualised as contours, as shown in \cref{fig:EnergyContour_n2m1l1}. Here, the strain energy is plotted as a function of \(A_{11}\) and \(A_{12}\) for \(Q=4\), \(\beta=3\), and \(\lambda=25\). The energy attains a global minimum at the stress-free configuration, i.e., \(A_{11}=a_1=1\) and \(A_{12}=0\). The contours are symmetric about the \(A_{11}\)-axis because the arch may switch asymmetrically with either half deflecting downward (i.e., \(\pm A_{12}\)). A second local minimum occurs at the inverted (second) stable state, again along the \(A_{11}\)-axis. These two stable equilibria are indicated by the green points in \cref{fig:EnergyContour_n2m1l1}. The red point located between them corresponds to an unstable equilibrium. In addition, two saddle points with nonzero \(A_{12}\) appear in symmetric pairs corresponding to \(\pm A_{12}\), and are marked in cyan. The white dashed curve connecting the stable equilibria along the \(A_{11}\)-axis (and passing through the unstable point) corresponds to the symmetric solution branch, whereas the yellow dotted curve corresponds to one asymmetric solution branch. A second asymmetric branch exists that passes through the upper two quadrants, although it is not shown in the figure. The intersection of the yellow curve with the white dashed curve corresponds to \(\delta=\delta_1\). This representation also highlights that the asymmetric branch follows a substantially lower-energy path.

While the results for \(n=2\), \(m=0\), and \(l=1\) provide useful qualitative insight, quantitative accuracy can be improved by increasing the number of modes used to approximate the deformations. In this context, we first revisit the simplifying assumption \(m=0\).

\subsection{Out-of-Plane Deformation} \label{Sec:2archesn2}
In this section, we allow non-zero out-of-plane deformation by setting \(m=1\), while keeping the remaining mode weights unchanged from the previous sections, i.e., \(n=2\) and \(l=1\). With these choices, \cref{Eq:PEintermsofmodeweights} becomes
\begin{align}
PE &=-F \left(a_1-A_{11}\right)+\frac{3}{2} Q^2 \left[\frac{1}{2} \pi ^2 a_1^2-\pi ^2 \left(\frac{A_{11}^2}{2}+2 A_{12}^2\right)-\frac{1}{2} \pi ^2
   B_{11}^2\right]^2\nonumber\\&+\frac{3}{2} Q^2 \left[\frac{1}{2} \pi ^2 a_1^2-\pi ^2 \left(\frac{A_{21}^2}{2}+2
   A_{22}^2\right)-\frac{1}{2} \pi ^2 B_{21}^2\right]^2+\frac{1}{4} \pi ^4
   \left(a_1-A_{11}\right){}^2+\frac{1}{4} \pi ^4 \left(a_1-A_{21}\right){}^2\nonumber\\&+\beta ^2 \left(-\frac{1}{2} \pi ^4 a_1 B_{11}
   C_{11}+\frac{3}{16} \pi ^4 a_1^2 C_{11}^2+\pi ^4 B_{11}^2\right)+4 \pi ^4 A_{12}^2+4 \pi ^4 A_{22}^2\nonumber\\&+\beta ^2 \left(-\frac{1}{2} \pi ^4 a_1 B_{21}
   C_{21}+\frac{3}{16} \pi ^4 a_1^2 C_{21}^2+\pi ^4 B_{21}^2\right)+\frac{1}{2} \pi ^2
   C_{11}^2 K+\frac{1}{2} \pi ^2 C_{21}^2 K \label{Eq:PEn2m1l1}
\end{align}

\begin{figure}[!htb]
     \begin{subfigure}[b]{0.49\textwidth}
         \centering\includegraphics[width=\textwidth]{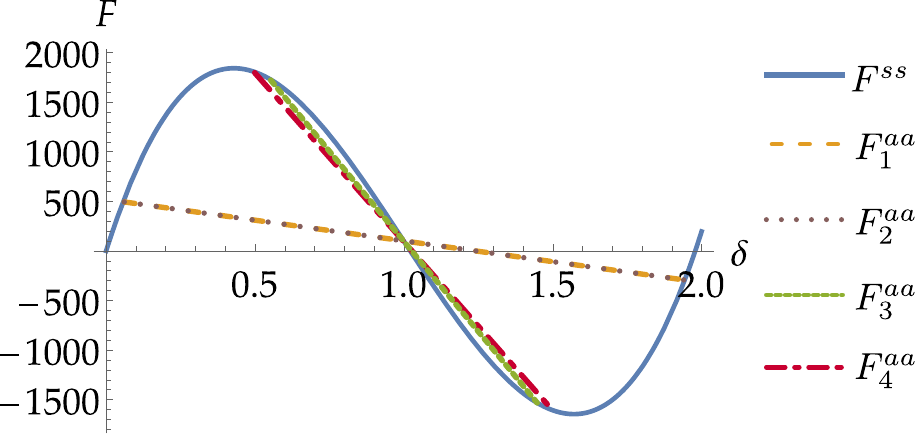}
         \caption{$ $}
         \label{fig:Fd_i2n2m1l1}
     \end{subfigure}
     \hfill
     \begin{subfigure}[b]{0.49\textwidth}
         \centering   \includegraphics[width=\textwidth]{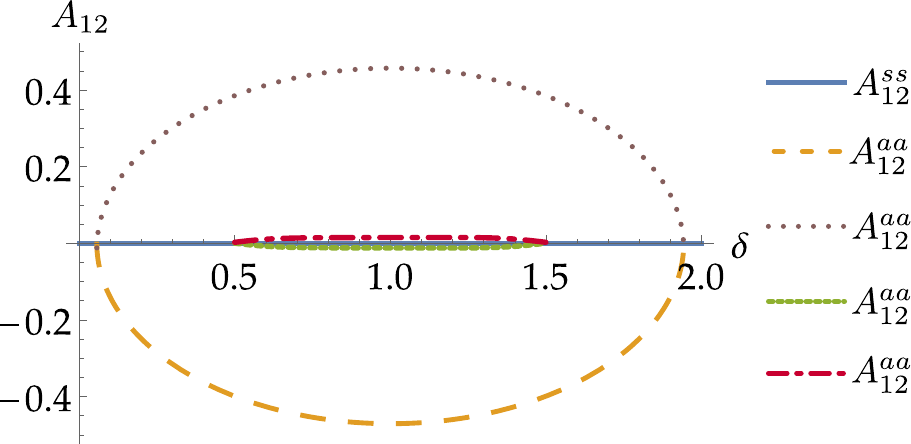}
         \caption{$ $}
         \label{fig:A12_i2n2m1l1}
     \end{subfigure}
     \hfill
     \begin{subfigure}[b]{0.49\textwidth}
         \centering
         \includegraphics[width=\textwidth]{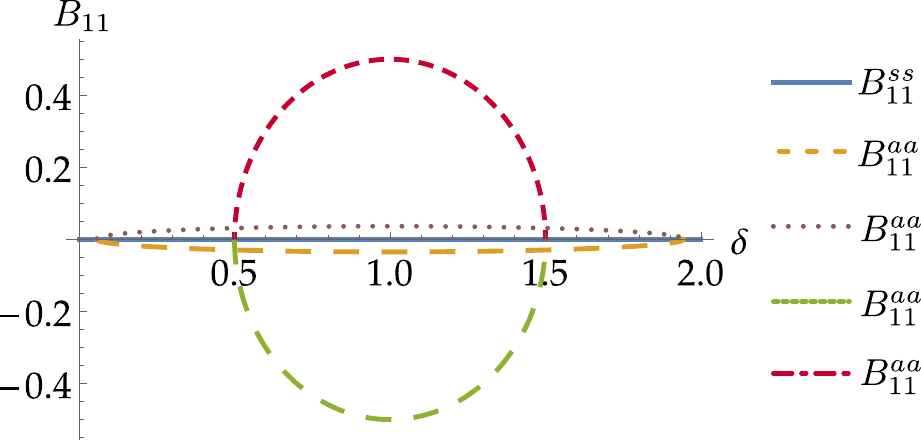}
         \caption{$ $}
         \label{fig:B11_i2n2m1l1}
     \end{subfigure}
     \hfill
     \begin{subfigure}[b]{0.49\textwidth}
         \centering\includegraphics[width=\textwidth]{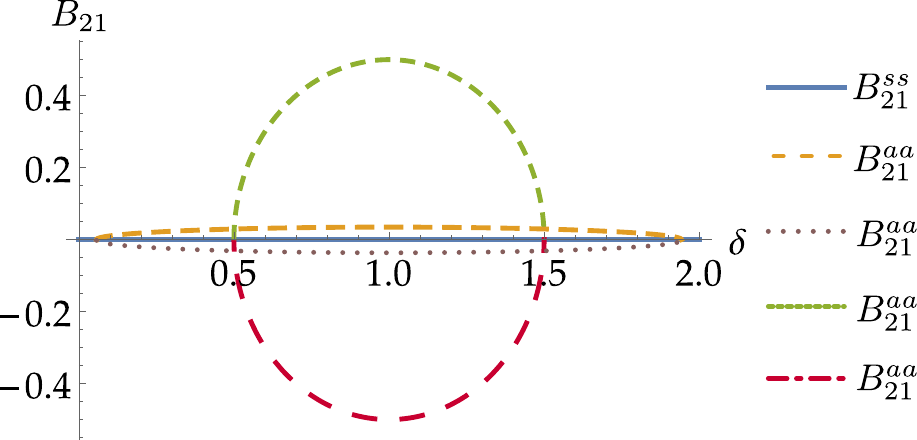}
         \caption{$ $}
         \label{fig:B21_i2n2m1l1}
     \end{subfigure}
 \caption{(a) Force-displacement response of a 2-MFA with \(n=2\), \(m=1\), and \(l=1\). \(F^{ss}\) is the symmetric branch; \(F^{aa}_{1,2}\) are asymmetric branches with \(A_{12}\)-dominated response, and \(F^{aa}_{3,4}\) are asymmetric branches with \(B_{11}\)-dominated response. (b-d) Evolution of \(A_{12}\), \(B_{11}\), and \(B_{21}\). Parameters: \(Q=4\), \(\beta=3\), \(\lambda=25\).}
        \label{fig:$n=2$,$m=1$,$l=1$}
\end{figure}

Since the effective deformation characteristics can be obtained solely from Case~2, we do not consider Case~1 here.  We assume that both arches exhibit out-of-plane deformation and therefore retain \(B_{11}\) and \(B_{21}\). As in Case~2, we substitute for \(A_{21}\), \(A_{22}\), \(C_{21}\), and \(C_{11}\) in \cref{Eq:PEn2m1l1} using the conditions in \cref{i2n2m0l1c2:deltacondition,i2n2m0l1c2:slope&slopecondition,i2n2m0l1c2:phi&phicondition,i2n2m0l1c2:slope&phicondition}, and solve for the remaining unknowns \(A_{11}\), \(A_{12}\), \(B_{11}\), and \(B_{21}\) from the equilibrium equations:
\begin{align}
    \frac{\partial PE}{\partial A_{11}}&=-3 \pi ^2 Q^2 A_{11}\Delta_1-3 \pi ^2 Q^2 A_{11}\Delta_2-\pi ^4 \left(a_1-A_{11}\right)+F=0 \label{Eq:EEn2m1l1_1},\\
       \frac{\partial PE}{\partial A_{12}}&=\beta ^2 \left(\frac{3 \pi ^6 a_1^2 A_{12}}{2 \lambda ^2}-\frac{\pi ^5 a_1 B_{11}}{\lambda }\right)+\beta ^2 \left(\frac{3 \pi
   ^6 a_1^2 A_{12}}{2 \lambda ^2}+\frac{\pi ^5 a_1 B_{21}}{\lambda }\right)-12 \pi ^2 A_{12} Q^2 \Delta_1\nonumber\\&-12 \pi ^2 A_{12} Q^2
   \Delta_2+\frac{8
   \pi ^4 A_{12} K}{\lambda ^2}+16 \pi ^4 A_{12}=0\label{Eq:EEn2m1l1_2},\\
      \frac{\partial PE}{\partial B_{11}    }&=\beta ^2 \left(2 \pi ^4 B_{11}-\frac{\pi ^5 a_1 A_{12}}{\lambda }\right)-3 \pi ^2 B_{11} Q^2 \Delta_1=0\label{Eq:EEn2m1l1_3},\\
   \frac{\partial PE}{\partial B_{21}    }&=\beta ^2 \left(2 \pi ^4 B_{21}+\frac{\pi ^5 a_1 A_{12}}{\lambda }\right)-3 \pi ^2 B_{21} Q^2 \Delta_2=0,\label{Eq:EEn2m1l1_4}
\end{align}
where $\Delta_1 = \frac{1}{2} \pi ^2 a_1^2-\pi ^2 \left(\frac{A_{11}^2}{2}+2 A_{12}^2\right)-\frac{1}{2} \pi ^2 B_{11}^2$, and $\Delta_2 = \frac{1}{2} \pi ^2 a_1^2-\pi ^2 \left(\frac{A_{11}^2}{2}+2 A_{12}^2\right)-\frac{1}{2} \pi ^2 B_{21}^2$. Note that  $\Delta_1$ and \(\Delta_2\) now include \(B_{11}\) and \(B_{21}\) terms, respectively. Solving \cref{Eq:EEn2m1l1_1,Eq:EEn2m1l1_2,Eq:EEn2m1l1_3,Eq:EEn2m1l1_4} yields multiple solution branches, as shown in \cref{fig:Fd_i2n2m1l1}. 
The expressions for \(F^{ss}\) (noting that \(A_{12}=B_{11}=B_{21}=0\) on this branch; see \cref{fig:A12_i2n2m1l1,fig:B11_i2n2m1l1,fig:B21_i2n2m1l1}) remain identical to \cref{Eq:Fss_i2n2m1l1}. In contrast, the four asymmetric force branches \(F^{aa}\) (\(F^{aa}_{1}\), \(F^{aa}_{2}\), \(F^{aa}_{3}\), \(F^{aa}_{4}\)) exhibit non-zero \(A_{12}\), \(B_{11}\), and \(B_{21}\), as seen in \cref{fig:A12_i2n2m1l1,fig:B11_i2n2m1l1,fig:B21_i2n2m1l1}.

The branches \(F^{aa}_{1}\) and \(F^{aa}_{2}\) correspond to solutions in which in-plane deformation dominates over out-of-plane deformation, whereas \(F^{aa}_{3}\) and \(F^{aa}_{4}\) correspond to solutions in which out-of-plane deformation dominates. Our interest is primarily in arches with width much greater than depth, i.e., \(\beta \gg 1\). In this regime, the switching force associated with \(F^{aa}_{3}\) and \(F^{aa}_{4}\) is substantially higher, and the arch follows the \(F^{aa}_{1}\) and \(F^{aa}_{2}\) pathway, making the \(F^{aa}_{3}\) and \(F^{aa}_{4}\) less practically relevant.

\begin{figure}[!htbp]
    \centering
\begin{subfigure}[b]{0.47\textwidth}
         \centering
    \includegraphics[width=\textwidth]{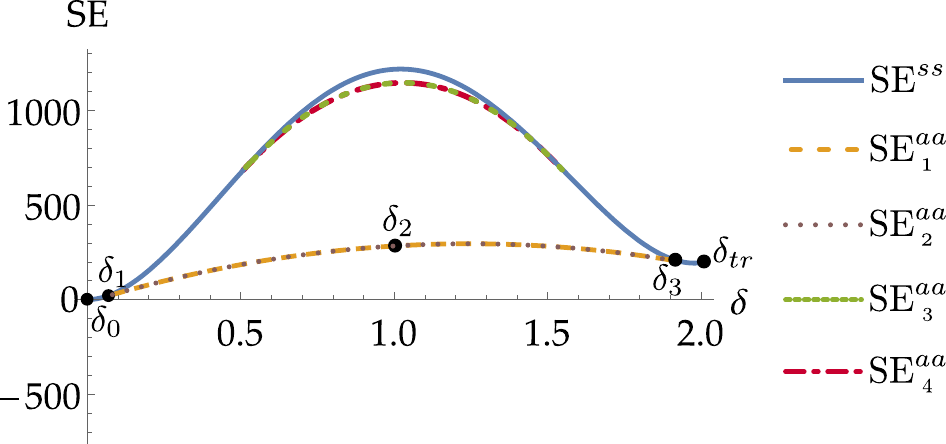}
    \caption{$ $}
    \label{fig:SE_n2m1l1}
     \end{subfigure}
     \hfill
     \begin{subfigure}[b]{0.52\textwidth}
         \centering
    \includegraphics[width=\textwidth]{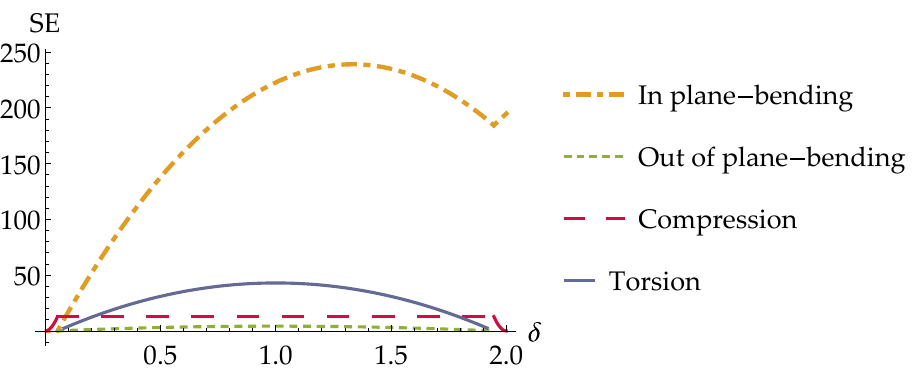}
    \caption{$ $}
    \label{fig:SEseparate_n2m1l1}
     \end{subfigure}
\caption{(a) Total strain energy for the symmetric (solid) and two asymmetric (dashed) solution branches. (b) Strain-energy evolution along the hybrid deformation pathway indicated by the black arrows in \cref{fig:Fdelta_i2n2m0l1_c1_c2}. Parameters: \(Q=4\), \(\beta=3\), and \(\lambda=25\).}
\end{figure}

This energy difference between \(F^{aa}_{1}\), \(F^{aa}_{2}\), \(F^{aa}_{3}\), and \(F^{aa}_{4}\) is further evident from the strain-energy plots in \cref{fig:SE_n2m1l1}. 
The arch initially deforms symmetrically and switches to an asymmetric pathway at \(\delta_1\). It then follows the asymmetric pathway until \(\delta_3\), after which it switches back to symmetric deformation and reaches the second stable state at \(\delta_{tr}\). Notably, by including out-of-plane deformation modes (\(m\neq 0\)), \(B_{11}\) becomes non-zero along \(F^{aa}_{1}\) and \(F^{aa}_{2}\), although its magnitude remains small. \Cref{fig:SEseparate_n2m1l1} shows the evolution of the strain-energy components along the hybrid pathway indicated by the black arrows in \cref{fig:Fdelta_i2n2m0l1_c1_c2}. The in-plane bending energy clearly dominates the total strain energy, while the out-of-plane bending energy remains negligible by comparison. Consequently, the asymmetric branches \(F^{aa}_{1}\) and \(F^{aa}_{2}\) closely resemble the asymmetric branch \(F^{aa}\) obtained in \cref{Sec:doublearchescase2}.

\subsection{Increased Number of Modeshapes}
\begin{figure}[!htbp]
     \begin{subfigure}[b]{0.49\textwidth}
         \centering\includegraphics[width=\textwidth]{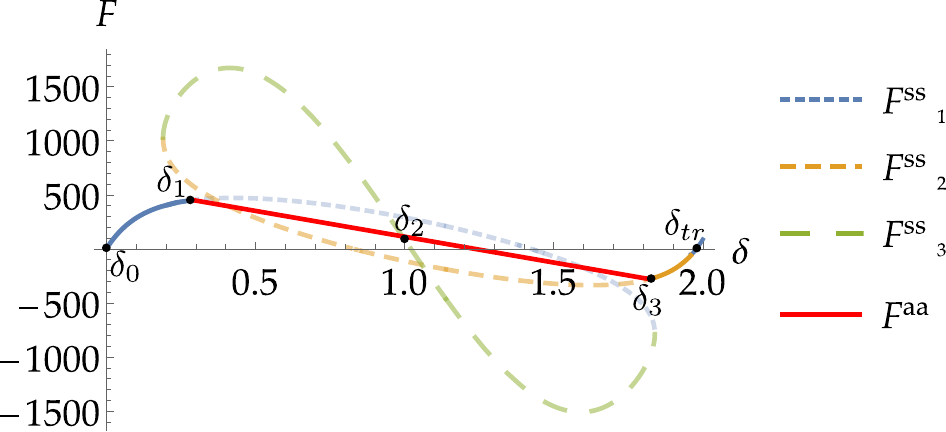}
         \caption{$ $}
         \label{fig:Fdelta_n3m0l1}
     \end{subfigure}
     \hfill
     \begin{subfigure}[b]{0.49\textwidth}
         \centering   \includegraphics[width=\textwidth]{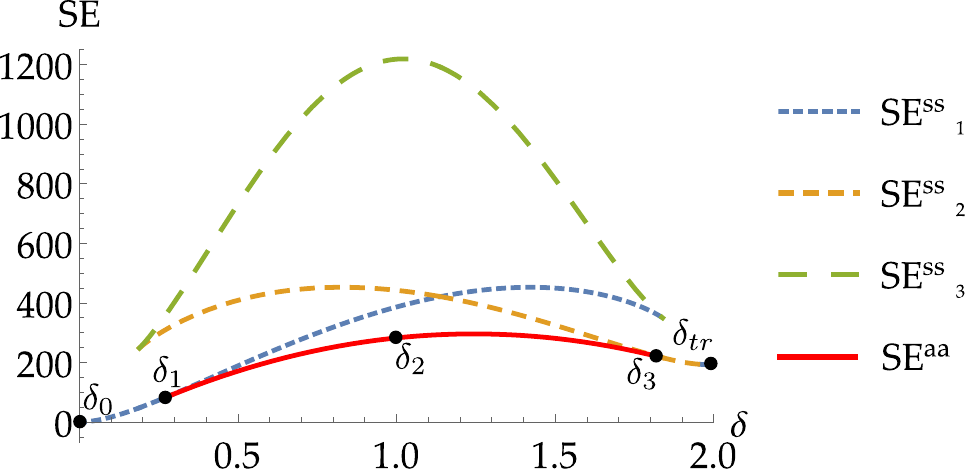}
         \caption{$ $}
         \label{fig:SE_i2n3m0l1_combined}
     \end{subfigure}
     \hfill
     \begin{subfigure}[b]{0.49\textwidth}
         \centering
         \includegraphics[width=\textwidth]{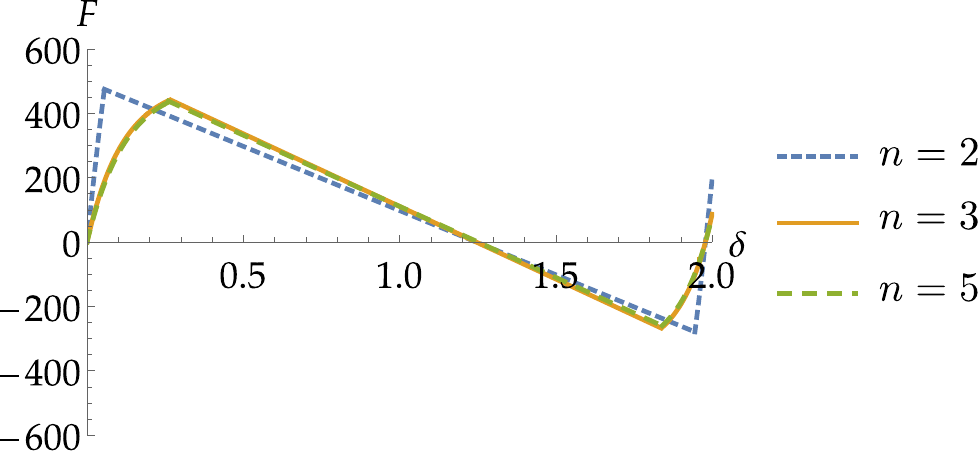}
         \caption{$ $}
         \label{fig:Fdelta_n2n3n5m0l1}
     \end{subfigure}
     \hfill
     \begin{subfigure}[b]{0.49\textwidth}
         \centering\includegraphics[width=\textwidth]{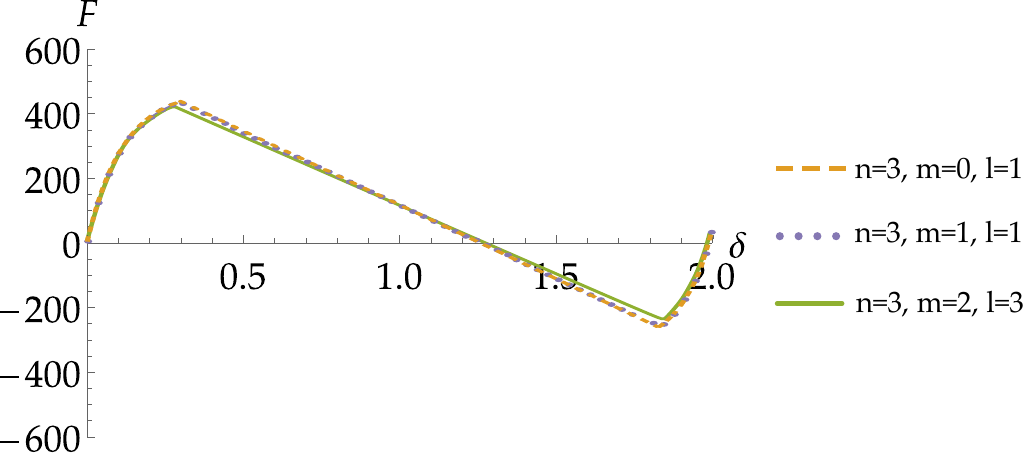}
         \caption{$ $}
         \label{fig:A4L100T1B3H4_analytical_compare}
     \end{subfigure}
 \caption{(a) Force-displacement response for \(n=3\), \(m=0\), and \(l=1\). (b) Total strain energy of the 2-MFA for the symmetric (dashed) and asymmetric (solid) solution branches. (c) Effect of \(n\) on the force-displacement curve of a 2-MFA (\(m=0\), \(l=1\)). (d) Comparison of force-displacement plot for \(n=3\) with varying \(m\) and \(l\) for \(Q=4\), \(\beta=3\), and \(\lambda=25\).}
        \label{fig:$n=3$,$m=0$,$l=1$}
\end{figure}

In this section, we seek analytical predictions that are comparable with the experimental and FEA results. Since the in-plane bending energy dominates the other strain-energy components, we first increase the number of in-plane modes by increasing \(n\). We consider Case~2 with \(n=3\), \(m=0\), and \(l=1\), for which the potential energy becomes  
\begin{align}  
    PE &= -F \left(a_1-A_{11}+A_{13}\right)+\frac{3}{2} Q^2 \left[\frac{1}{2} \pi ^2 a_1^2-\pi ^2 \left(\frac{A_{11}^2}{2}+2
   A_{12}^2+\frac{9 A_{13}^2}{2}\right)\right]^2+\frac{1}{4} \pi ^4
   \left(a_1-A_{11}\right){}^2\nonumber\\&+\frac{3}{2} Q^2 \left[\frac{1}{2} \pi ^2 a_1^2-\pi ^2
   \left(\frac{A_{21}^2}{2}+2 A_{22}^2+\frac{9 A_{23}^2}{2}\right)\right]^2+\frac{1}{4} \pi ^4 \left(a_1-A_{21}\right){}^2+\frac{3}{16} \pi ^4 a_1^2 \beta ^2
   C_{11}^2\nonumber\\&+\frac{3}{16} \pi ^4 a_1^2 \beta ^2 C_{21}^2+4 \pi ^4 A_{12}^2+\frac{81}{4} \pi ^4 A_{13}^2+4 \pi ^4
   A_{22}^2+\frac{81}{4} \pi ^4 A_{23}^2+\frac{1}{2} \pi ^2 C_{11}^2 K+\frac{1}{2} \pi ^2 C_{21}^2 K. \label{Eq:PEn3m0l1}  
\end{align}
As in \cref{Sec:doublearchescase2}, we use the midpoint conditions to substitute for \(A_{21}\), \(A_{22}\), \(C_{21}\), and \(C_{11}\) in \cref{Eq:PEn3m0l1}. We then solve for \(A_{11}\), \(A_{12}\), \(A_{13}\), and \(A_{23}\) using the corresponding four equilibrium equations. The resulting force-displacement characteristics are shown in \cref{fig:Fdelta_n3m0l1}; note that \(\delta=a_1-A_{11}+A_{13}\) (see \cref{Eq:delta}).

The additional symmetric pathways associated with a dominant \(A_{13}\) contribution are shown as dashed curves, while the asymmetric pathway is shown as a solid curve. Despite these additional branches, the hybrid deformation pathway (opaque superimposed curve in \cref{fig:Fdelta_n3m0l1}) remains qualitatively consistent with the results in the previous sections.

To further improve accuracy, we systematically increase the number of mode weights used in the formulation. We first examine the effect of increasing the number of in-plane modes \(n\). The prediction obtained with \(n=2\) improves substantially when \(n=3\) is used, as shown in \cref{fig:Fdelta_n2n3n5m0l1}. Increasing to \(n=5\) yields a response similar to \(n=3\); therefore, we set \(n=3\) and then improve the approximation by increasing the mode weights in \(m\) and \(l\). Increasing \(m\) and \(l\) show slight improvements in the force-displacement response, as illustrated in \cref{fig:A4L100T1B3H4_analytical_compare}. A detailed comparison with the FEA and experimental results is presented in \cref{sec:FEAandEXP}.
 
\section{Three-Arch MFA (3-MFA)}\label{Sec:3arches}
\begin{figure}[!htb]
    \centering
    \includegraphics[width=1.0 \linewidth]{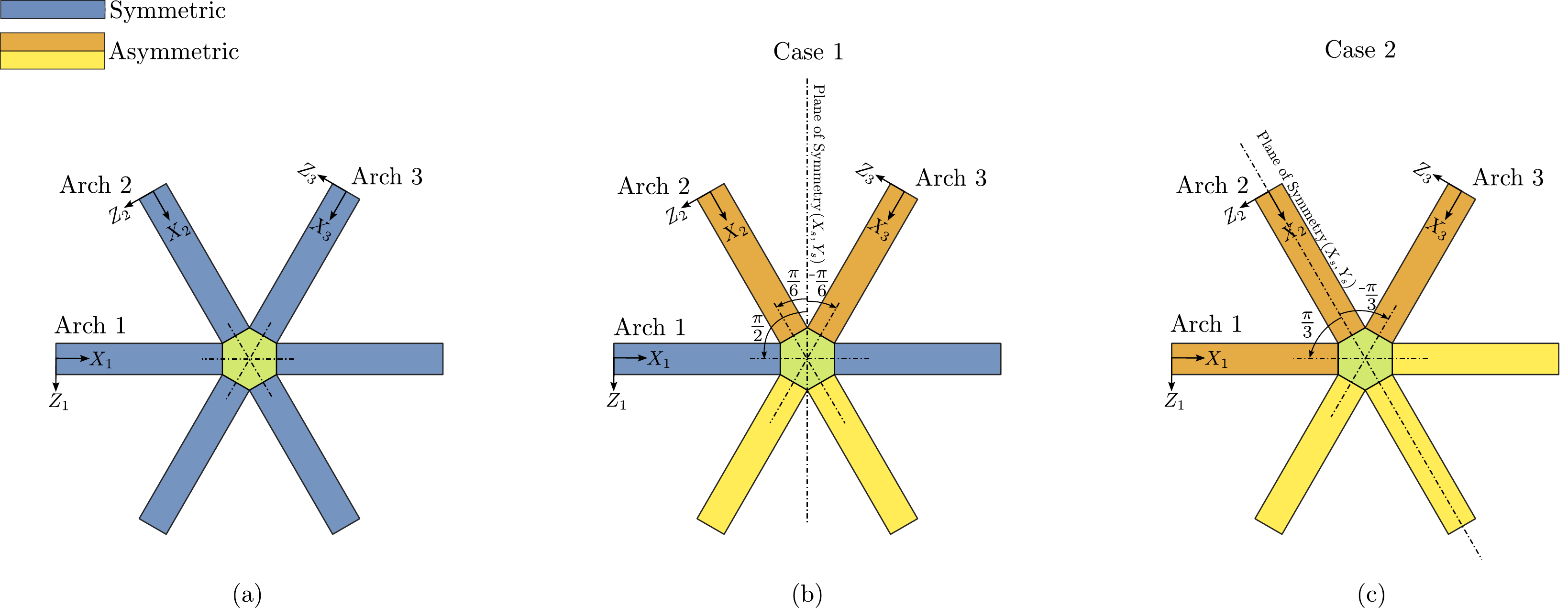}
\caption{Deformation pathways of a 3-MFA: (a) symmetric deformation of all arches; (b) Case~1: Arch~2 and Arch~3 deform identically while Arch~1 remains symmetric; (c) Case~2: all three arches deform asymmetrically.}
    \label{fig:DP_TripleArch}
\end{figure}

To illustrate  the generality of the modelling approach, we now consider a 3-MFA as shown in \cref{fig:DP_TripleArch}a. To simplify the analysis, its deformations are classified into two distinct cases, each associated with a different plane of symmetry, as depicted in \cref{fig:DP_TripleArch}b-c. In Case 1,  Arch 1 deforms symmetrically, while Arch 2 and Arch 3 exhibit similar types of deformation, either symmetric or asymmetric. Note that Arches 1 to 3 make angles \(\pi/2\), \(\pi/6\), and \(-\pi/6\), respectively with respect to the plane of symmetry. In Case 2, all three arches deform in the same manner, either all asymmetric or all symmetric. Here, the angles with respect to the plane of symmetry are \(\pi/3\), \(0\), and \(-\pi/3\). Similar to the case of 2-MFA, the scenario where all arches deform symmetrically is captured in both cases.
\subsection{Case 1} \label{Sec:triplearchescase1}
We consider \(n = 2\) for Arches~2 and~3, and \(n = 3\) for Arch 1, while keeping \(m = 0\) and \(l = 1\) for all arches. The additional mode ensures that the symmetric deformation of Arch~1 is accurately captured. Thus, from \cref{Eq:PEintermsofmodeweights}, the total potential energy is given by,
\begin{align}
    PE &=-F \left(a_1-A_{11}+A_{13}\right)+\frac{3}{2} Q^2 \left[\frac{1}{2} \pi ^2 a_1^2-\pi ^2 \left(\frac{A_{11}^2}{2}+2
   A_{12}^2+\frac{9 A_{13}^2}{2}\right)\right]^2\nonumber\\&+\frac{3}{2} Q^2 \left[\frac{1}{2} \pi ^2 a_1^2-\pi ^2
   \left(\frac{A_{21}^2}{2}+2 A_{22}^2\right)\right]^2+\frac{3}{2} Q^2 \left[\frac{1}{2} \pi ^2 a_1^2-\pi ^2
   \left(\frac{A_{31}^2}{2}+2 A_{32}^2\right)\right]^2\nonumber\\&+\frac{1}{4} \pi ^4 \left(a_1-A_{11}\right){}^2+\frac{1}{4} \pi ^4
   \left(a_1-A_{21}\right){}^2+\frac{1}{4} \pi ^4 \left(a_1-A_{31}\right){}^2+\frac{3}{16} \pi ^4 a_1^2 \beta ^2
   C_{11}^2\nonumber\\&+\frac{3}{16} \pi ^4 a_1^2 \beta ^2 C_{21}^2+\frac{3}{16} \pi ^4 a_1^2 \beta ^2 C_{31}^2+4 \pi ^4
   A_{12}^2+\frac{81}{4} \pi ^4 A_{13}^2+4 \pi ^4 A_{22}^2+4 \pi ^4 A_{32}^2\nonumber\\&+\frac{1}{2} \pi ^2 C_{11}^2 K+\frac{1}{2} \pi ^2
   C_{21}^2 K+\frac{1}{2} \pi ^2 C_{31}^2 K \label{PEi3n3m0l1case1},
\end{align}
where $A_{11}$, $A_{12}$, $A_{13}$, $A_{21}$, $A_{22}$, $A_{31}$, $A_{32}$, $C_{11}$, $C_{21}$, and $C_{31}$ are the unknown mode weights. By applying the midpoint conditions from \cref{Eq:midpoint_disp,Eq:slope_omega_relation} and condition due to the  symmetric deformation of Arch 2, $A_{21}$, $A_{12}$, $A_{31}$, $A_{32}$, $C_{11}$, $C_{21}$ and $C_{31}$ are eliminated as:
\begin{align}
     {W_1} \big|_{X_1 = 1/2}&={W_2} \big|_{X_2 = 1/2}  &\implies \qquad A_{21}&=A_{11}-A_{13} \label{deltacondition1}\\
     \frac{d W_1}{d X_1} \bigg|_{X_1 = 1/2}  &=  0 &\implies \qquad A_{12}&=0 \label{zeroslopecondition}\\
     {W_1} \big|_{X_1 = 1/2}&={W_3} \big|_{X_3 = 1/2}  &\implies \qquad A_{31}&=A_{11}-A_{13}\label{deltacondition2}\\
     \frac{d W_2}{d X_2} \bigg|_{X_2 = 1/2}&=\frac{d W_3}{d X_3} \bigg|_{X_3 = 1/2}  &\implies \qquad A_{32}&=A_{22} \label{slope1&slope3}\\
     \Phi_3 \big|_{X_3 = 1/2} &= - \Phi_2 \big|_{X_2 = 1/2}  &\implies \qquad C_{31}&= -C_{21} \label{phi1&phi3condition}\\
     \Phi_1 \big|_{X_1 = 1/2} \sin(\pi/6) &=  \Phi_2 \big|_{X_2 = 1/2}  &\implies \qquad C_{21}&=\frac{\text{C}_{11}}{2 } \label{phi&phicondition}\\
         \Phi_1 \big|_{X_1 = 1/2} \cos(\pi/6)  &= \frac{1}{\lambda}\frac{d W_2}{d X_2} \bigg|_{X_2 = 1/2}  &\implies \qquad C_{11}&=-\frac{4 \pi  \text{A}_{22}}{\sqrt{3}\lambda } \label{slope&phicondition}.
\end{align}
The remaining unknown mode weights $A_{11}$, $A_{22}$, and $A_{13}$ are determined using the following equilibrium equations: 
\begin{align}
    \frac{\partial PE}{\partial A_{11}}&=-\frac{1}{2} \pi ^4 \left(a_1-A_{11}\right)-\pi ^4 \left(a_1-A_{11}+A_{13}\right)-3 \pi ^2 A_{11} \Delta _1 Q^2\nonumber\\&-6 \pi ^2
   \left(A_{11}-A_{13}\right) \Delta _2 Q^2+F=0 \label{Eq:EEi3n2&3m0l1_1_c1_asym}\\
    \frac{\partial PE}{\partial A_{22}}&=\frac{3 \pi ^6 a_1^2 A_{22} \beta ^2}{\lambda ^2}+\frac{8 \pi ^4 A_{22} K}{\lambda ^2}-24 \pi ^2 A_{22} \Delta _2 Q^2+16 \pi
   ^4 A_{22}=0\label{Eq:EEi3n2&3m0l1_2_c1_asym}\\
    \frac{\partial PE}{\partial A_{13}}&=\pi ^4 \left(a_1-A_{11}+A_{13}\right)-27 \pi ^2 A_{13} \Delta _1 Q^2-6 \pi ^2 \left(A_{13}-A_{11}\right) \Delta _2
   Q^2+\frac{81 \pi ^4 A_{13}}{2}-F=0 \label{Eq:EEi3n2&3m0l1_3_c1_asym},
\end{align}
where,
\begin{align}
\Delta_1&=\frac{1}{2} \pi ^2 a_1^2-\pi ^2 \left(\frac{A_{11}^2}{2}+\frac{9 A_{13}^2}{2}\right) \label{i3n3&2Delta1},~ \text{and} \\
\Delta_2&=\frac{1}{2} \pi ^2 a_1^2-\pi ^2 \left[\frac{1}{2} \left(A_{11}-A_{13}\right){}^2+2 A_{22}^2\right] \label{i3n3&2Delta2}.
\end{align}
\Cref{Eq:EEi3n2&3m0l1_1_c1_asym,Eq:EEi3n2&3m0l1_2_c1_asym,Eq:EEi3n2&3m0l1_3_c1_asym} yield symmetric solution branches when  $A_{22}=0$. These symmetric branches are given by solid blue curves in \cref{fig:Fdelta_i3n2&3m0l1}. For asymmetric solution ($A_{22} \ne0$), \(\Delta_2\)    satisfies,
\begin{equation}
   \frac{3 \pi ^6 a_1^2 \beta ^2}{\lambda ^2}+\frac{8 \pi ^4 K}{\lambda ^2}-24 \pi ^2 \Delta _2 Q^2+16 \pi ^4=0.
   \label{Eq:Delta1Eq_i3n2m0l1_c1}
\end{equation}

Solving  \Cref{Eq:EEi3n2&3m0l1_1_c1_asym,Eq:EEi3n2&3m0l1_2_c1_asym,Eq:EEi3n2&3m0l1_3_c1_asym} for $A_{22} \ne0$ results in asymmetric solution branches, as shown by large dashed orange curves in \cref{fig:Fdelta_i3n2&3m0l1}.   We omit the force expressions here due to their complexity and length. A detailed explanation of the force pathways taken by the arch during deformation will be presented following the discussion of Case 2. 
\subsection{Case 2} \label{Sec:triplearchescase2} 
In Case 2, we focus on capturing the deformation pathway in which all three arches deform asymmetrically, as shown in \cref{fig:DP_TripleArch}c. In this case, the kinematic conditions at the midpoint are:
\\
\\
\noindent 
\begin{minipage}[b]{.38\textwidth}
    \vspace{-\baselineskip} 
    \begin{equation}
         {W_1} \big|_{X_1 = 1/2}={W_2} \big|_{X_2 = 1/2} \label{Eq:3MFAC2M1}
    \end{equation}
\end{minipage}
\hfill
\begin{minipage}[b]{.55\textwidth}
    \vspace{-\baselineskip} 
    \begin{equation}
         \frac{d W_1}{d X_1} \bigg|_{X_1 = 1/2}   = \frac{d W_2}{d X_2} \bigg|_{X_2 = 1/2} \cos(\pi/3) \label{Eq:3MFAC2M2}
    \end{equation}
\end{minipage}
\\
\\
\\
\begin{minipage}[b]{.38\textwidth}
    \vspace{-\baselineskip} 
    \begin{equation}
         {W_1} \big|_{X_1 = 1/2}={W_3} \big|_{X_3 = 1/2}  \label{Eq:3MFAC2M3}
    \end{equation}
\end{minipage}
\hfill
\begin{minipage}[b]{.55\textwidth}
    \vspace{-\baselineskip} 
    \begin{equation}
         \frac{d W_1}{d X_1} \bigg|_{X_1 = 1/2}  = \frac{d W_3}{d X_3} \bigg|_{X_3 = 1/2} \label{Eq:3MFAC2M4}
    \end{equation}
\end{minipage}
\\
\\
\\
\begin{minipage}[b]{.38\textwidth}
    \vspace{-\baselineskip} 
    \begin{equation}
    \Phi_2 \big|_{X_2 = 1/2} =0  \label{Eq:3MFAC2M6}
    \end{equation}
\end{minipage}
\hfill
\begin{minipage}[b]{.55\textwidth}
    \vspace{-\baselineskip} 
    \begin{equation}
          \Phi_1 \big|_{X_1 = 1/2} =  \frac{1}{\lambda}\frac{d W_2}{d X_2} \bigg|_{X_2 = 1/2}  \sin(\pi/3)  \label{Eq:3MFAC2M5}
    \end{equation}
\end{minipage}
\\
\\
\\
\begin{minipage}[b]{.38\textwidth}
    \vspace{-\baselineskip} 
    \begin{equation}
        \Phi_3 \big|_{X_3 = 1/2} = - \Phi_1 \big|_{X_1 = 1/2}  \label{Eq:3MFAC2M7}
    \end{equation}
\end{minipage}
\\

Here, if only \( n = 2 \) modes are retained for each arch, \cref{Eq:3MFAC2M1} implies \( A_{21} = A_{11} \), and \cref{Eq:3MFAC2M3} gives \( A_{31} = A_{11} \). Furthermore, \cref{Eq:3MFAC2M2} and \cref{Eq:3MFAC2M4} yield \( A_{22} = 2 A_{12} \) and \( A_{32} = 2 A_{12} \), respectively. Consequently, only \( A_{11} \) and \( A_{12} \) remain as independent variables. This constraint enforces identical first-mode amplitudes for all three arches, while the second-mode amplitudes of the second and third arches become twice that of the first. Such restrictions overly constrain the deformation and result in an incorrectly stiff response. Therefore, in Case~2, at least two asymmetric modes must be retained in the approximation, i.e.\ \( n \geq 4 \).

Thus, we consider \(n=4\), \(m=0\), and \(l=1\) but exclude the third in-plane bending mode for all three arches in order to enable a direct comparison with the results obtained for Case 1. Hence, the total potential energy is given by,
\begin{align}
    PE &= -F \left(a_1-A_{11}\right)+\frac{3}{2} Q^2 \left[\frac{1}{2} \pi ^2 a_1^2-\pi ^2 \left(\frac{A_{11}^2}{2}+2 A_{12}^2+8
   A_{14}^2\right)\right]^2+\frac{1}{4} \pi ^4 \left(a_1-A_{11}\right)^2\nonumber\\&+\frac{3}{16} \pi ^4 a_1^2 \beta ^2
   C_{11}^2+\frac{3}{2} Q^2 \left[\frac{1}{2} \pi ^2 a_1^2-\pi ^2 \left(\frac{A_{21}^2}{2}+2 A_{22}^2+8
   A_{24}^2\right)\right]^2+\frac{1}{4} \pi ^4
   \left(a_1-A_{21}\right)^2\nonumber\\&+\frac{3}{2} Q^2 \left[\frac{1}{2} \pi ^2 a_1^2-\pi ^2 \left(\frac{A_{31}^2}{2}+2 A_{32}^2+8
   A_{34}^2\right)\right]^2+\frac{1}{4} \pi ^4 \left(a_1-A_{31}\right)^2+\frac{3}{16} \pi ^4 a_1^2 \beta ^2 C_{31}^2\nonumber\\&+4 \pi ^4 A_{12}^2+64 \pi ^4 A_{14}^2+4 \pi ^4 A_{22}^2+64 \pi ^4
   A_{24}^2+4 \pi ^4 A_{32}^2+64 \pi ^4 A_{34}^2+\frac{1}{2} \pi ^2 C_{11}^2 K+\frac{1}{2} \pi ^2 C_{31}^2 K \label{PEi3n3m0l1case2}
\end{align}

The midpoint conditions \cref{Eq:3MFAC2M1,Eq:3MFAC2M2,Eq:3MFAC2M3,Eq:3MFAC2M4,Eq:3MFAC2M5,Eq:3MFAC2M6,Eq:3MFAC2M7} imply that $A_{21}=A_{11}$,
$A_{22}=2\left(A_{12}-2A_{14}\right)$,
$A_{31}=A_{11}$,
$A_{32}=A_{12}-2A_{14}+2A_{34}$,
$C_{21}=0$,
$C_{11}=-\dfrac{\sqrt{3}\pi A_{22}}{\lambda}$, and 
$C_{31}=-C_{11}$, respectively.
The remaining unknown mode weights $A_{11}$, $A_{12}$, $A_{14}$, $A_{24}$, and $A_{34}$ are determined using the equilibrium conditions given by,
\begin{align}
    \frac{\partial PE}{\partial A_{11}}=&-\frac{3}{2} \pi ^4 \left(a_1-A_{11}\right)-3 \pi ^2 A_{11} \Delta _1 Q^2-3 \pi ^2 A_{11} \Delta _2 Q^2-3 \pi ^2 A_{11} \Delta
   _3 Q^2+F=0 \label{Eq:EEi3n2&3m0l1_c2_1},\\
       \frac{\partial PE}{\partial A_{12}}=&\frac{9 \pi ^6 a_1^2 A_{12} \beta ^2}{\lambda ^2}-\frac{18 \pi ^6 a_1^2 A_{14} \beta ^2}{\lambda ^2}+\frac{24 \pi ^4 A_{12}
   K}{\lambda ^2}-\frac{48 \pi ^4 A_{14} K}{\lambda ^2}-12 \pi ^2 A_{12} \Delta _1 Q^2\nonumber\\&-48 \pi ^2 A_{12} \Delta _2 Q^2+96 \pi
   ^2 A_{14} \Delta _2 Q^2-48 \pi ^2 A_{24} \Delta _2 Q^2-12 \pi ^2 A_{12} \Delta _3 Q^2+24 \pi ^2 A_{14} \Delta _3 Q^2\nonumber\\&-24 \pi
   ^2 A_{34} \Delta _3 Q^2+48 \pi ^4 A_{12}-80 \pi ^4 A_{14}+32 \pi ^4 A_{24}+16 \pi ^4 A_{34}=0\label{Eq:EEi3n2&3m0l1_c2_2},\\
   \frac{\partial PE}{\partial A_{14}}=&-\frac{18 \pi ^6 a_1^2 A_{12} \beta ^2}{\lambda ^2}+\frac{36 \pi ^6 a_1^2 A_{14} \beta ^2}{\lambda ^2}-\frac{48 \pi ^4 A_{12}
   K}{\lambda ^2}+\frac{96 \pi ^4 A_{14} K}{\lambda ^2}-48 \pi ^2 A_{14} \Delta _1 Q^2\nonumber\\&+96 \pi ^2 A_{12} \Delta _2 Q^2-192 \pi
   ^2 A_{14} \Delta _2 Q^2+96 \pi ^2 A_{24} \Delta _2 Q^2+24 \pi ^2 A_{12} \Delta _3 Q^2-32 \pi ^4 A_{34}\nonumber\\&-48 \pi ^2 A_{14} \Delta _3 Q^2+48 \pi
   ^2 A_{34} \Delta _3 Q^2-80 \pi ^4 A_{12}+288 \pi ^4 A_{14}-64 \pi ^4 A_{24}=0 \label{Eq:EEi3n2&3m0l1_c2_3},\\
   \frac{\partial PE}{\partial A_{24}}=&-48 \pi ^2 A_{12} \Delta _2 Q^2+96 \pi ^2 A_{14} \Delta _2 Q^2-96 \pi ^2 A_{24} \Delta _2 Q^2+32 \pi ^4 A_{12}-64 \pi ^4
   A_{14}\nonumber\\&+160 \pi ^4 A_{24}=0 \label{Eq:EEi3n2&3m0l1_c2_4},\\
   \frac{\partial PE}{\partial A_{34}}=&-24 \pi ^2 A_{12} \Delta _3 Q^2+48 \pi ^2 A_{14} \Delta _3 Q^2-96 \pi ^2 A_{34} \Delta _3 Q^2+16 \pi ^4 A_{12}-32 \pi ^4
   A_{14}\nonumber\\&+160 \pi ^4 A_{34}=0 \label{Eq:EEi3n2&3m0l1_c2_5},
\end{align}
where \(\Delta_1\), \(\Delta_2\), and \(\Delta_3\) are given by,
\begin{align}
\Delta_1&=\frac{1}{2} \pi ^2 a_1^2-\pi ^2 \left(\frac{A_{11}^2}{2}+2 A_{12}^2+8 A_{14}^2\right), \nonumber\\
\Delta_2&=\frac{1}{2} \pi ^2 a_1^2-\pi ^2 \left[\frac{A_{11}^2}{2}+8 A_{24}^2+8 \left(A_{12}-2 A_{14}+A_{24}\right)^2\right], \nonumber\\
\Delta_3&=\frac{1}{2} \pi ^2 a_1^2-\pi ^2 \left[\frac{A_{11}^2}{2}+8 A_{34}^2+2 \left(A_{12}-2 A_{14}+2 A_{34}\right)^2\right],
\end{align}

\begin{figure}[!htbp]
    \centering
\begin{subfigure}[b]{0.495\textwidth}
         \centering
    \includegraphics[width=\textwidth]{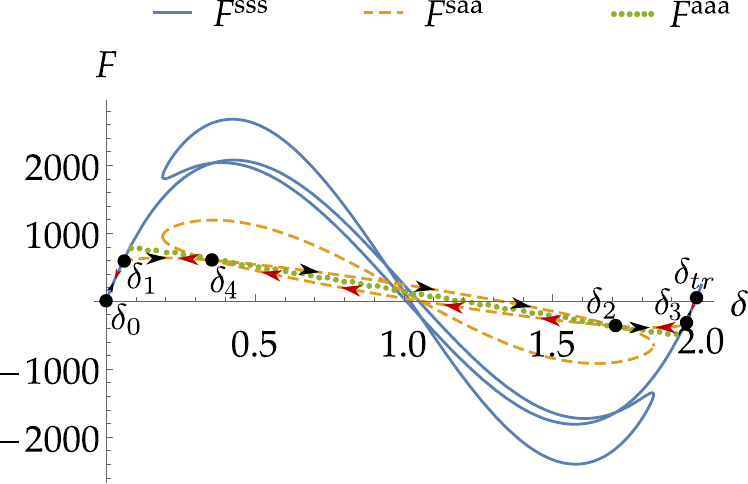}
    \caption{$ $}
    \label{fig:Fdelta_i3n2&3m0l1}
     \end{subfigure}
     \hfill
     \begin{subfigure}[b]{0.45\textwidth}
         \centering
    \includegraphics[width=\textwidth]{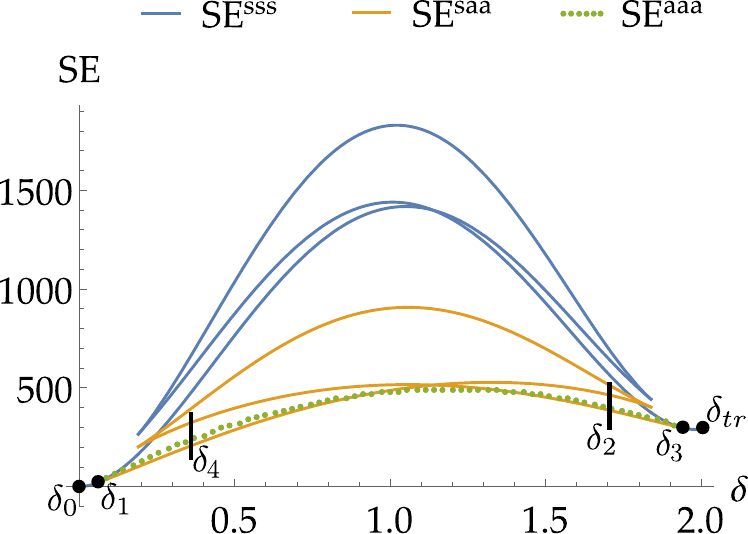}
    \caption{$ $}
    \label{fig:SE_i3n2&3m0l1_combined}
     \end{subfigure}
\caption{(a) Force-displacement response of a 3-MFA for Case~1 (\(n=2\) for Arches~1 and~3, \(n=3\) for Arch~2) and Case~2 (\(n=2\) for all arches), with \(m=0\) and \(l=1\) in both cases. (b) Total strain energy for the symmetric (solid) and asymmetric (dashed) solution branches for \(Q=4\), \(\beta=3\), and \(\lambda=25\).}
\end{figure}

However, solving \cref{Eq:EEi3n2&3m0l1_c2_1,Eq:EEi3n2&3m0l1_c2_2,Eq:EEi3n2&3m0l1_c2_3,Eq:EEi3n2&3m0l1_c2_4,Eq:EEi3n2&3m0l1_c2_5} to obtain a closed-form solution is intractable. Accordingly, the system is solved numerically, and only the solution branch with the least magnitude is retained for Case~2. The corresponding force-displacement curve is shown by the dotted green line in \cref{fig:Fdelta_i3n2&3m0l1}.

As in the case of the 2-MFA, the effective deformation pathway of the 3-MFA is identified by examining the force-displacement and strain-energy curves associated with each admissible branch. Consider a 3-MFA with $Q=4$, $\lambda=3$, $\beta=25$. All three arches initially deform symmetrically from the as-fabricated state, progressing from \(\delta_0\) to \(\delta_1\) along the blue solid curve in \cref{fig:Fdelta_i3n2&3m0l1}. At \(\delta_1\), the response transitions to the Case 1 pathway and follows the orange long-dashed curve. At \(\delta_2\), the structure remains on the Case 1 branch and continues along it up to \(\delta_3\), where it rejoins the symmetric (blue) branch and proceeds with symmetric deformation and reaches second stable state at \(\delta_{tr}\).

During the switching-back, the 3-MFA initially deforms symmetrically between \(\delta_{tr}\) and \(\delta_3\). At \(\delta_3\), Case 1 deformation begins and persists until \(\delta_1\), passing through \(\delta_4\). Beyond \(\delta_1\), all three arches revert to symmetric deformation as the structure approaches the primary stable state at \(\delta_0\). The \(F\)-\(\delta\) response indicates that the switching and switching-back paths do not coincide (see black arrows for switching and red arrows for switching-back in \cref{fig:Fdelta_i3n2&3m0l1}). This asymmetry arises because Arch 1 deforms predominantly with a positive third-mode contribution during switching, whereas the switch-back is dominated by a negative third mode. At each transition point, 3-MFA moves onto the branch with the lowest strain energy, as evident from \cref{fig:SE_i3n2&3m0l1_combined}.

For the present 3-MFA configuration, the analytical model predicts that Case 2 deformation does not occur, since the strain energy of the Case 2 branch (green dotted curve) exceeds that of the competing branches at their intersection points. However, simulations and experiments indicate that the 3-MFA undergoes transitions between Case 1 and Case 2 during both switching and  switching-back (see \cref{sec:FEAandEXP}). This behaviour is not captured by the model because it corresponds to intermediate deformation states that lack a distinct plane of symmetry, representing a transition from the symmetry plane between two arches (Case 1) to a symmetry plane aligned with a single arch (Case 2).  Nevertheless, \cref{fig:SE_i3n2&3m0l1_combined} reveals regions near \(\delta \approx 1\) where the Case 2 strain energy becomes lower, providing insight into the intermediate jumps observed in simulations and experiments.
 
\section{FEA and Experimental Validation}
\label{sec:FEAandEXP}
 In this section, we compare the analytical predictions with nonlinear FEA and tabletop experimental results. Specifically, we examine the deformation pathways observed along with the corresponding force-displacement responses. 

 The MFA model used in the FEA simulations is generated using a Python script that defines the span and the sinusoidal as-fabricated centreline profile of the arches. In ABAQUS/CAE 2018, the script is used to create the part, which is then assigned beam sections with a rectangular cross-section. The details of the cross-section are provided in \cref{tab:1}. 

The resulting geometry is partitioned to obtain a refined mesh using two-noded linear beam elements (B31) along the span of the MFA.

\begin{table}[!h]
\caption{Geometrical and material properties of MFA}
\centering
\label{table_example}
\begin{tabular}{llll}
\hline
Parameter &Value &Property &Value \\
\hline
Length (L)  &100 mm &Elastic Modulus &1.2 GPa \\
Width (b)  &3 mm & & \\
Thickness (t)  &1 mm &Poisson's ratio &0.35 \\
 Midrise (h)  &4 mm & & \\\hline
\end{tabular}
\label{tab:1}
\vspace*{-4pt}
\end{table}
The analysis is carried out using a quasi-static displacement-controlled setup with a dynamic implicit solver. The ends of the MFA are assigned pinned-pinned boundary conditions, and a displacement equal to twice the mid-rise of the arch (\(\delta_{tr} \approx 2\)) is applied at the midpoint.
 
\begin{figure}[!htbp]
\centering

\begin{subfigure}[b]{0.32\textwidth}
\centering
\includegraphics[width=\textwidth]{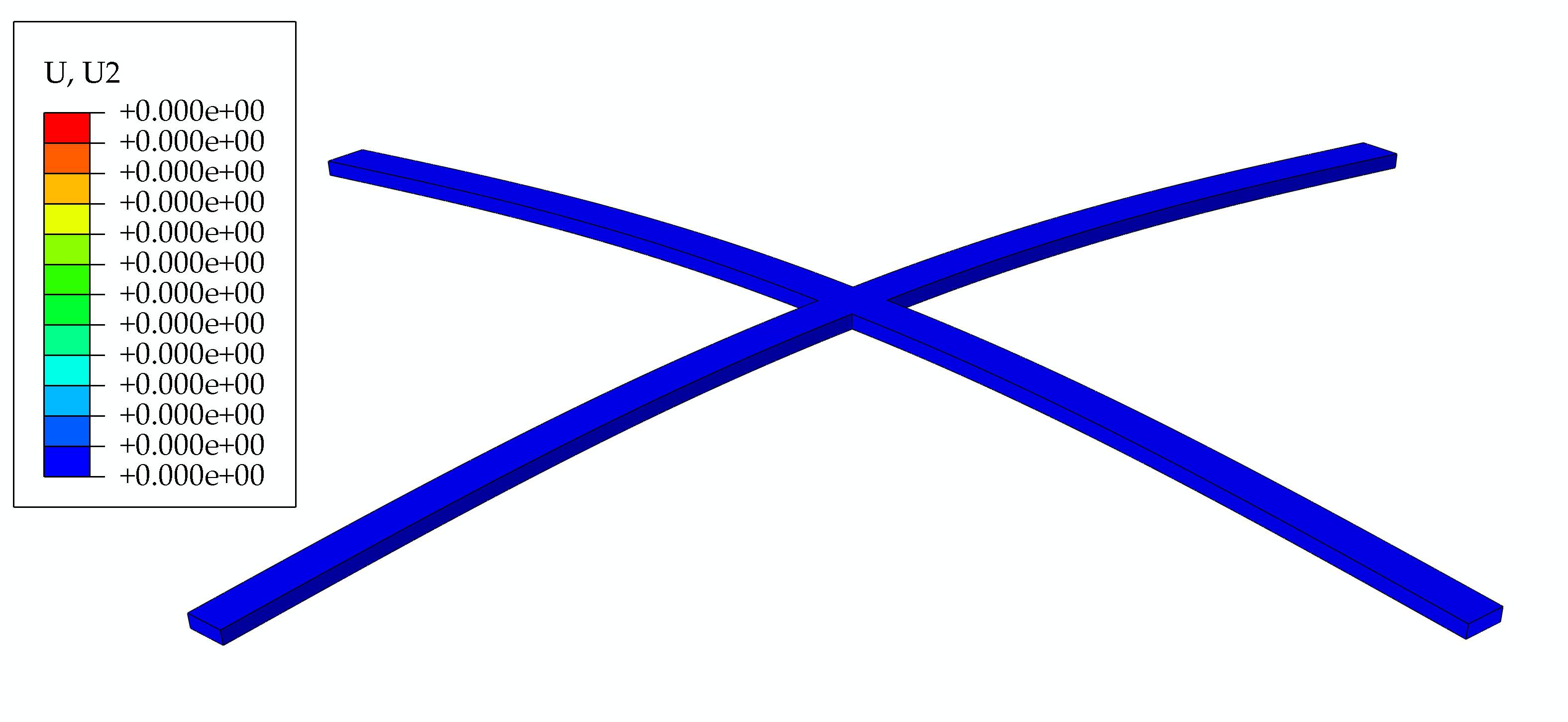}
\caption{}
\label{fig:deformation_doublearcha}
\end{subfigure}
\hfill
\begin{subfigure}[b]{0.32\textwidth}
\centering
\includegraphics[width=\textwidth]{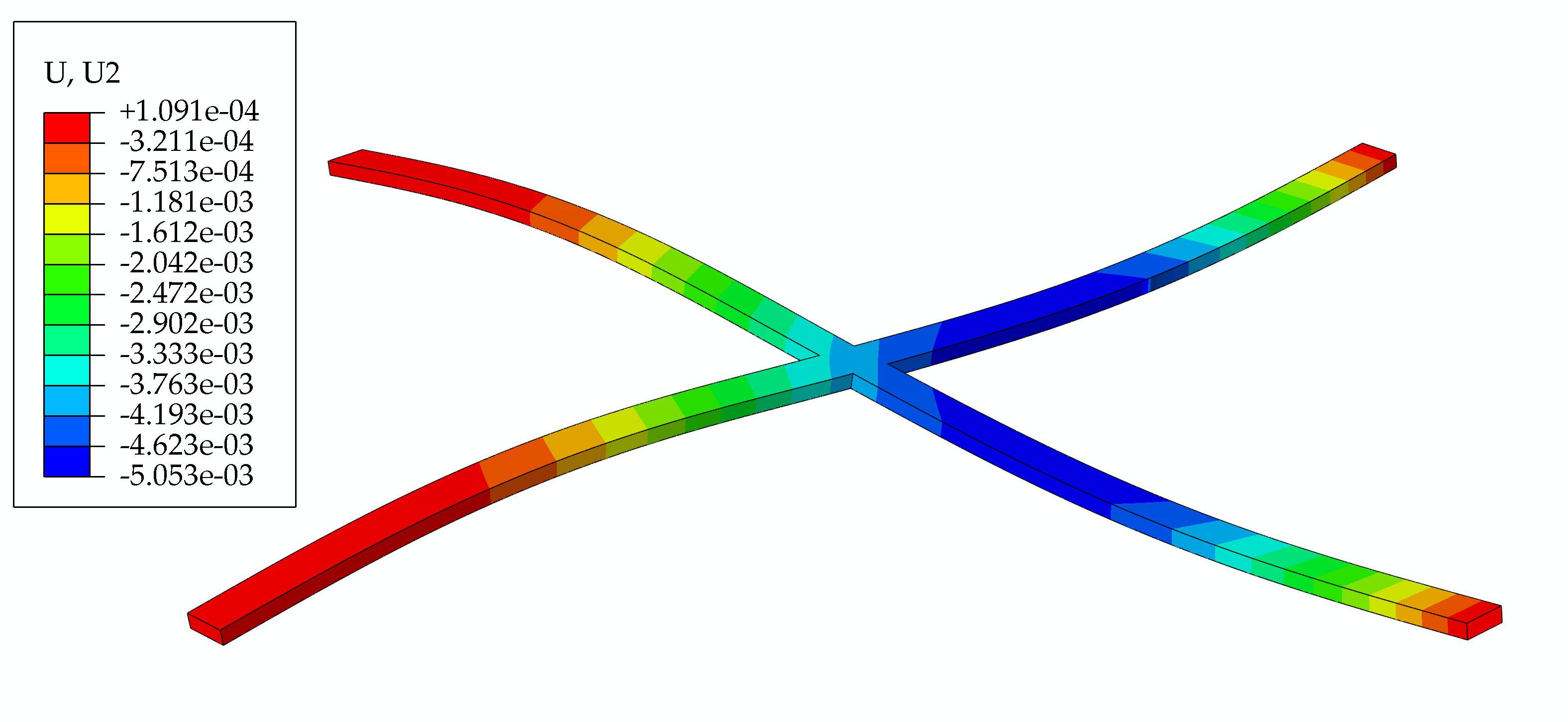}
\caption{}
\label{fig:deformation_doublearchb}
\end{subfigure}
\hfill
\begin{subfigure}[b]{0.32\textwidth}
\centering
\includegraphics[width=\textwidth]{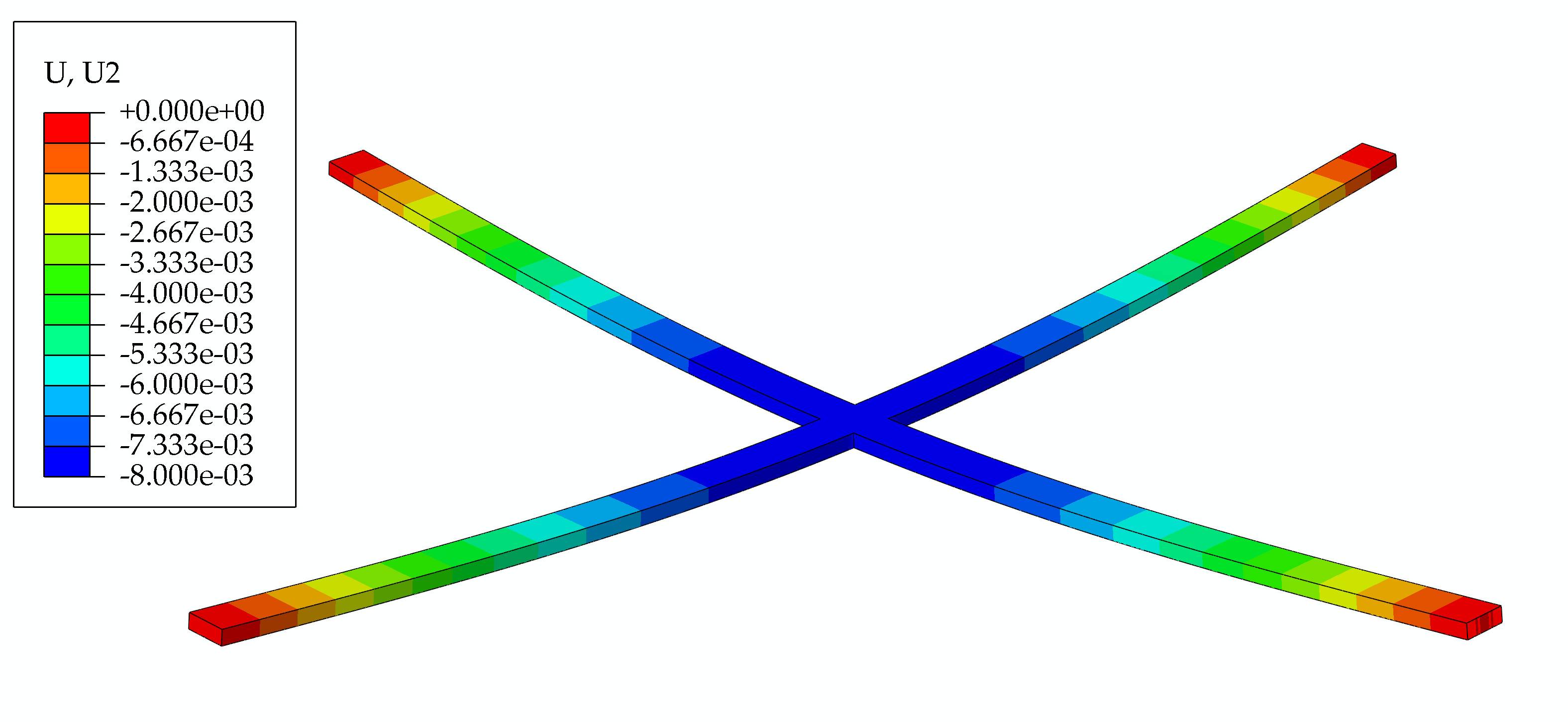}
\caption{}
\label{fig:deformation_doublearchc}
\end{subfigure}

\begin{subfigure}[b]{0.32\textwidth}
\centering
\includegraphics[width=\textwidth]{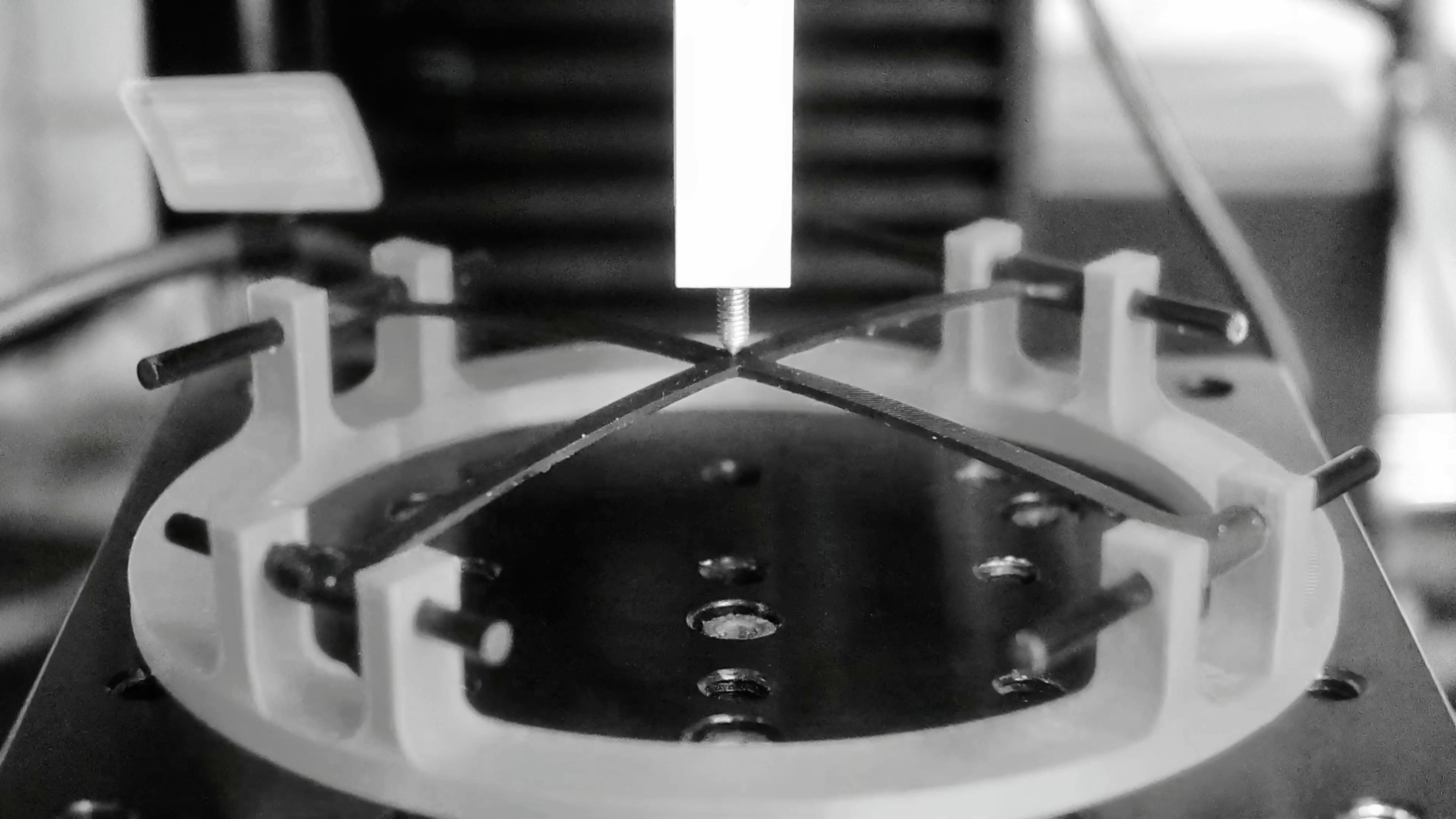}
\caption{}
\label{fig:deformation_doublearchd}
\end{subfigure}
\hfill
\begin{subfigure}[b]{0.32\textwidth}
\centering
\includegraphics[width=\textwidth]{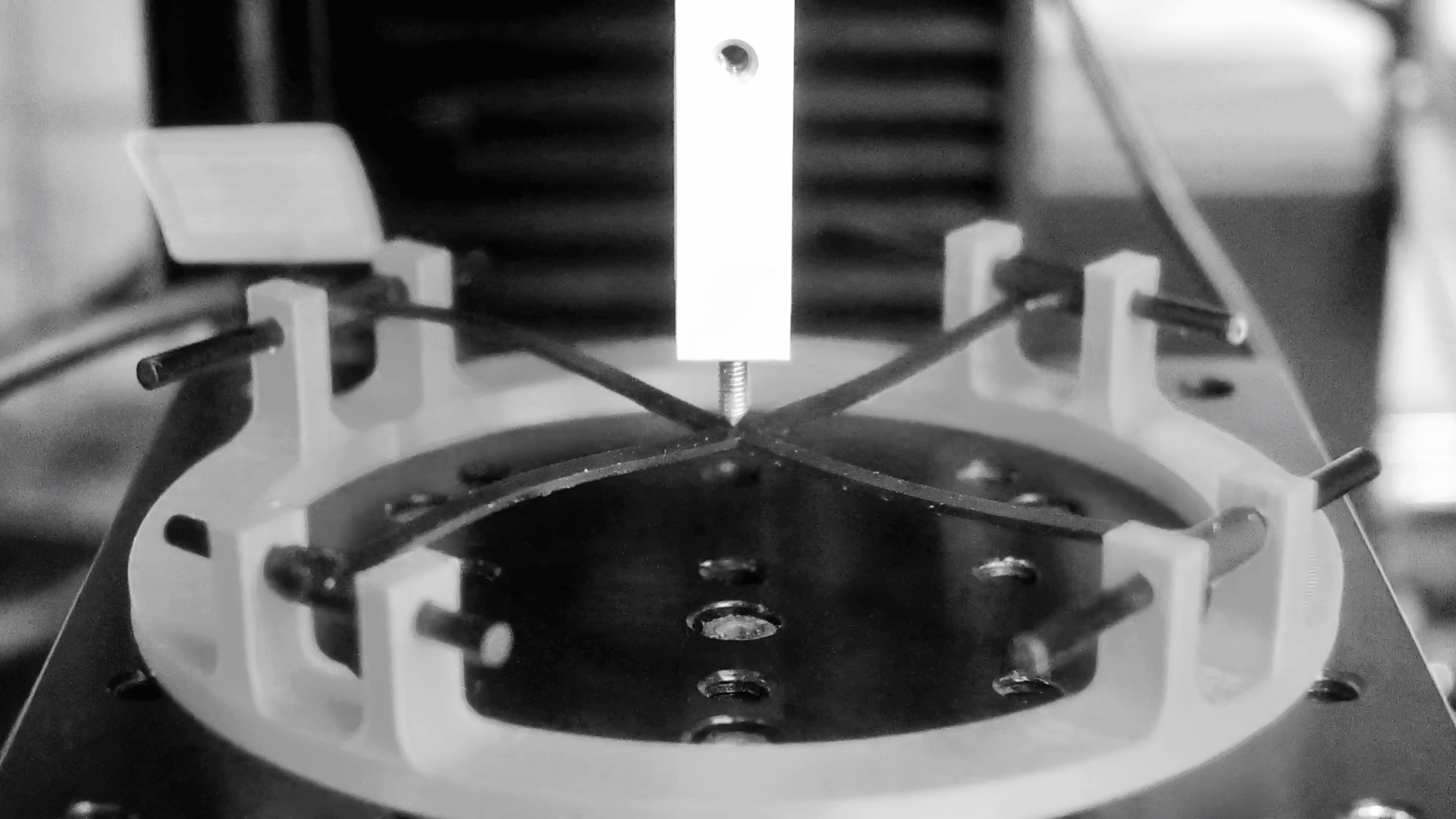}
\caption{}
\label{fig:deformation_doublearche}
\end{subfigure}
\hfill
\begin{subfigure}[b]{0.32\textwidth}
\centering
\includegraphics[width=\textwidth]{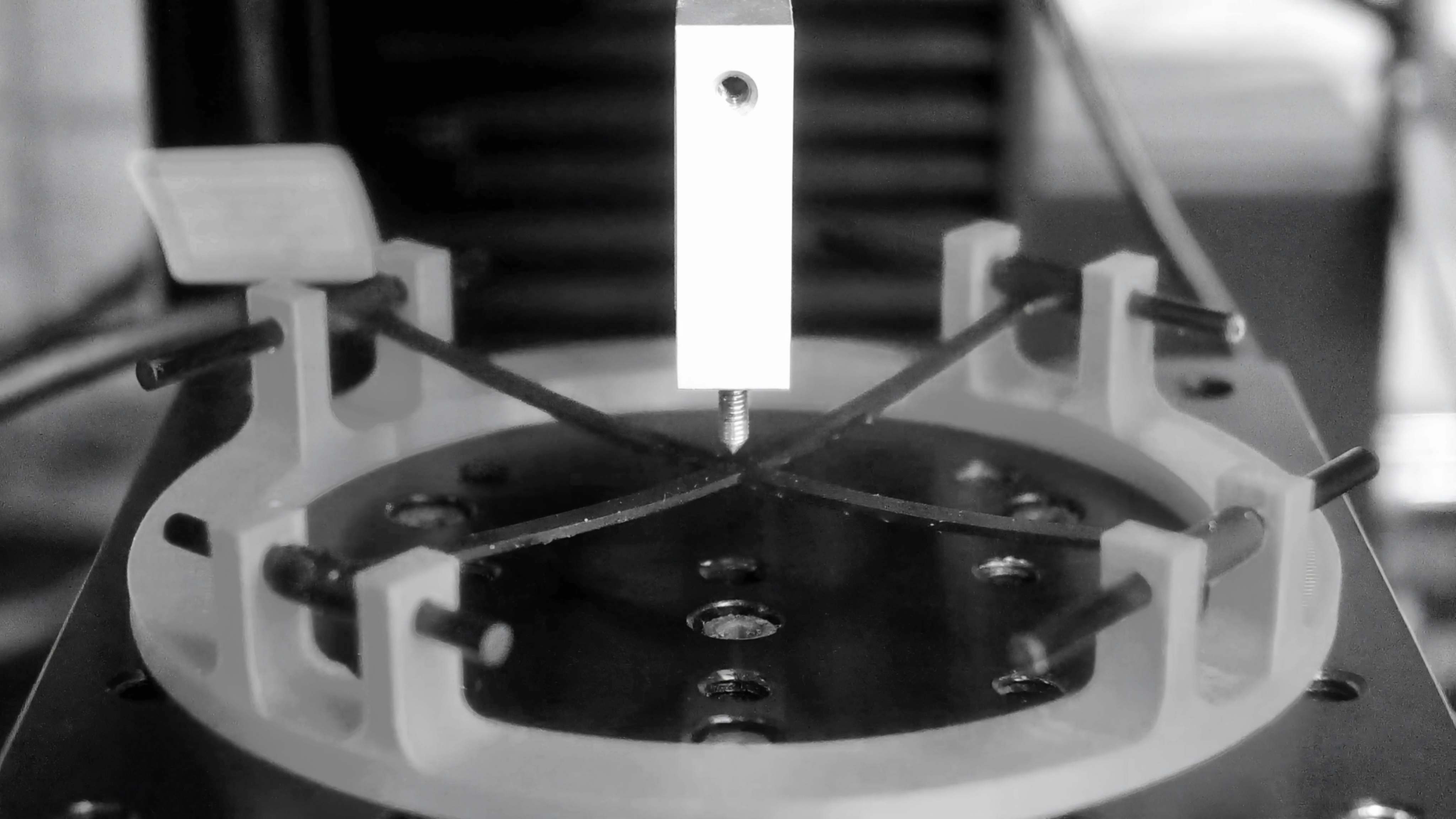}
\caption{}
\label{fig:deformation_doublearchf}
\end{subfigure}
\caption{Deformation of the 2-MFA from FEA (a-c) and experiments (d-f), showing the as-fabricated state, an intermediate asymmetric deformation, and the toggled state.}
\label{fig:deformation_doublearch}
\end{figure}

\begin{figure}[!htbp]
\centering

\begin{subfigure}[b]{0.24\textwidth}
\centering
\includegraphics[width=\textwidth]{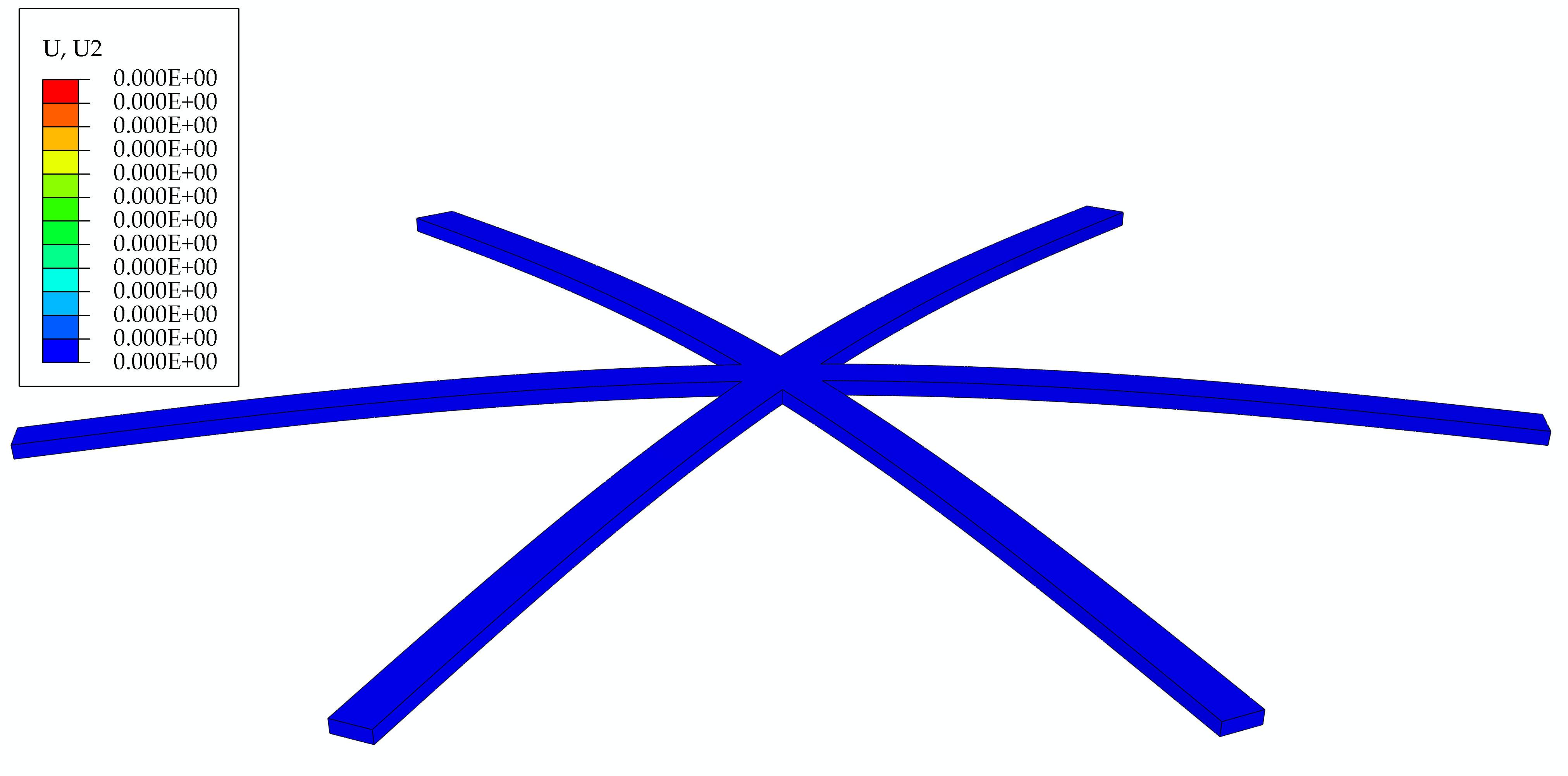}
\caption{}
\label{fig:triplearch_a}
\end{subfigure}
\hfill
\begin{subfigure}[b]{0.24\textwidth}
\centering
\includegraphics[width=\textwidth]{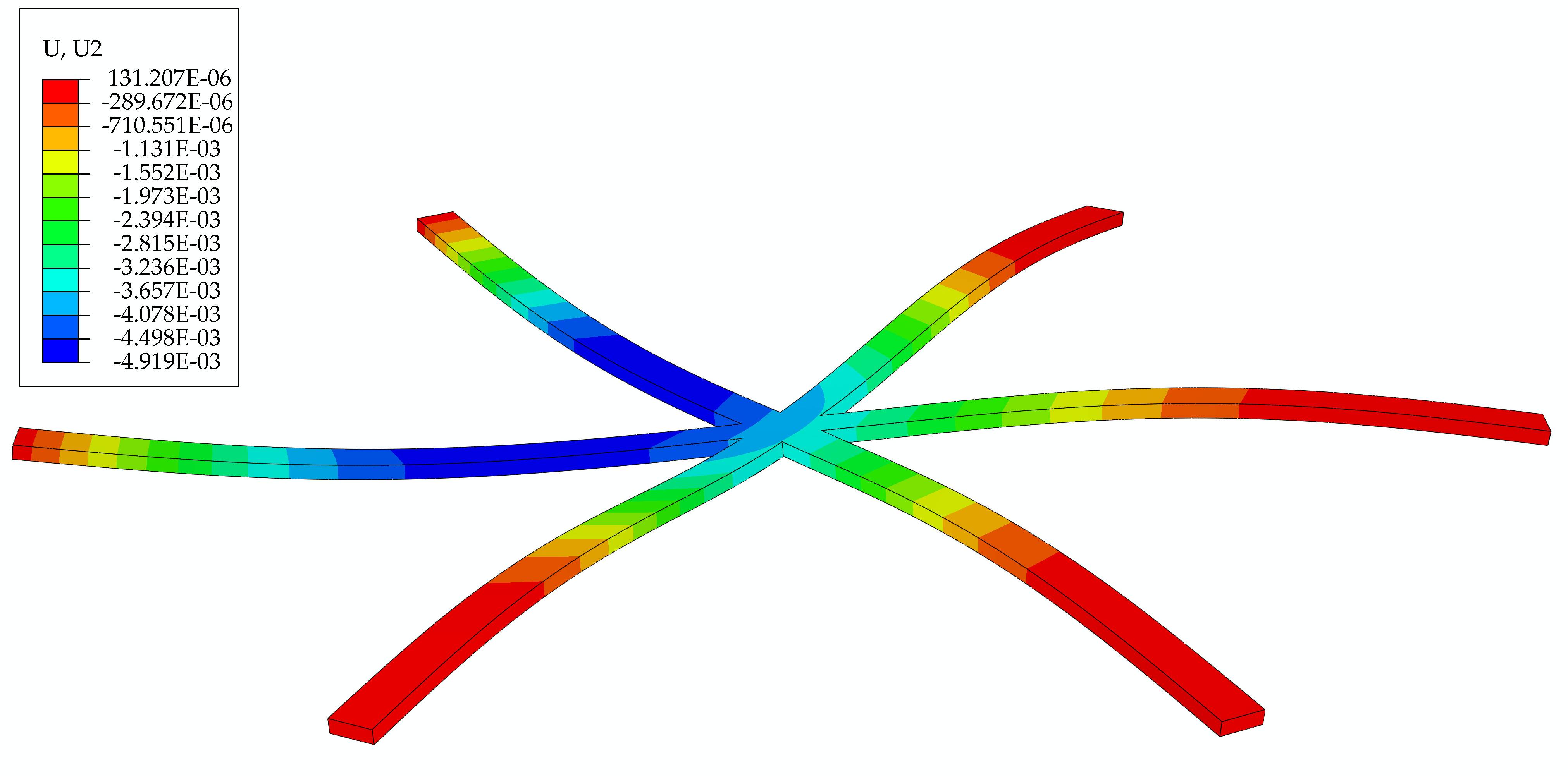}
\caption{}
\label{fig:triplearch_b}
\end{subfigure}
\hfill
\begin{subfigure}[b]{0.24\textwidth}
\centering
\includegraphics[width=\textwidth]{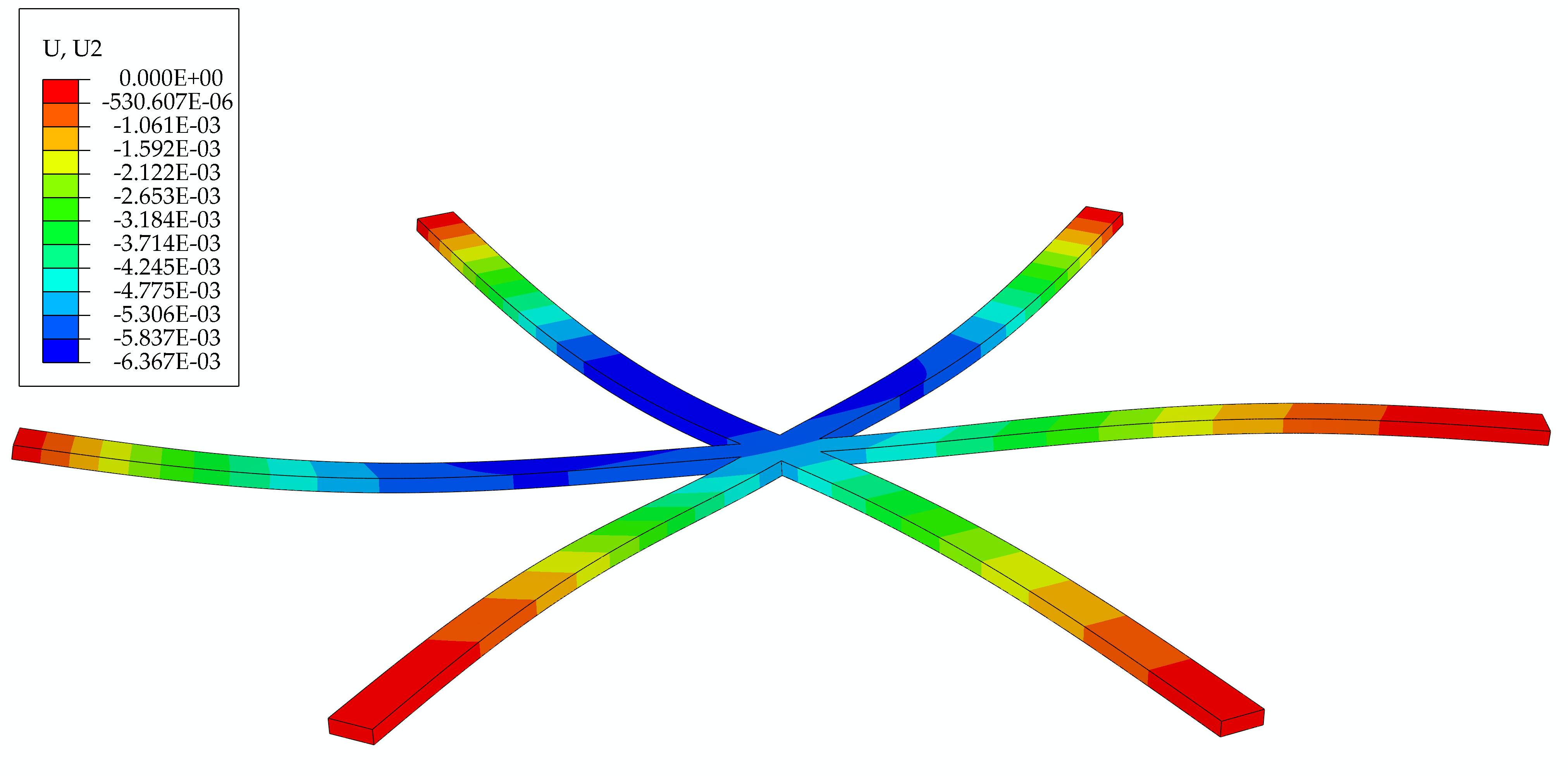}
\caption{}
\label{fig:triplearch_c}
\end{subfigure}
\hfill
\begin{subfigure}[b]{0.24\textwidth}
\centering
\includegraphics[width=\textwidth]{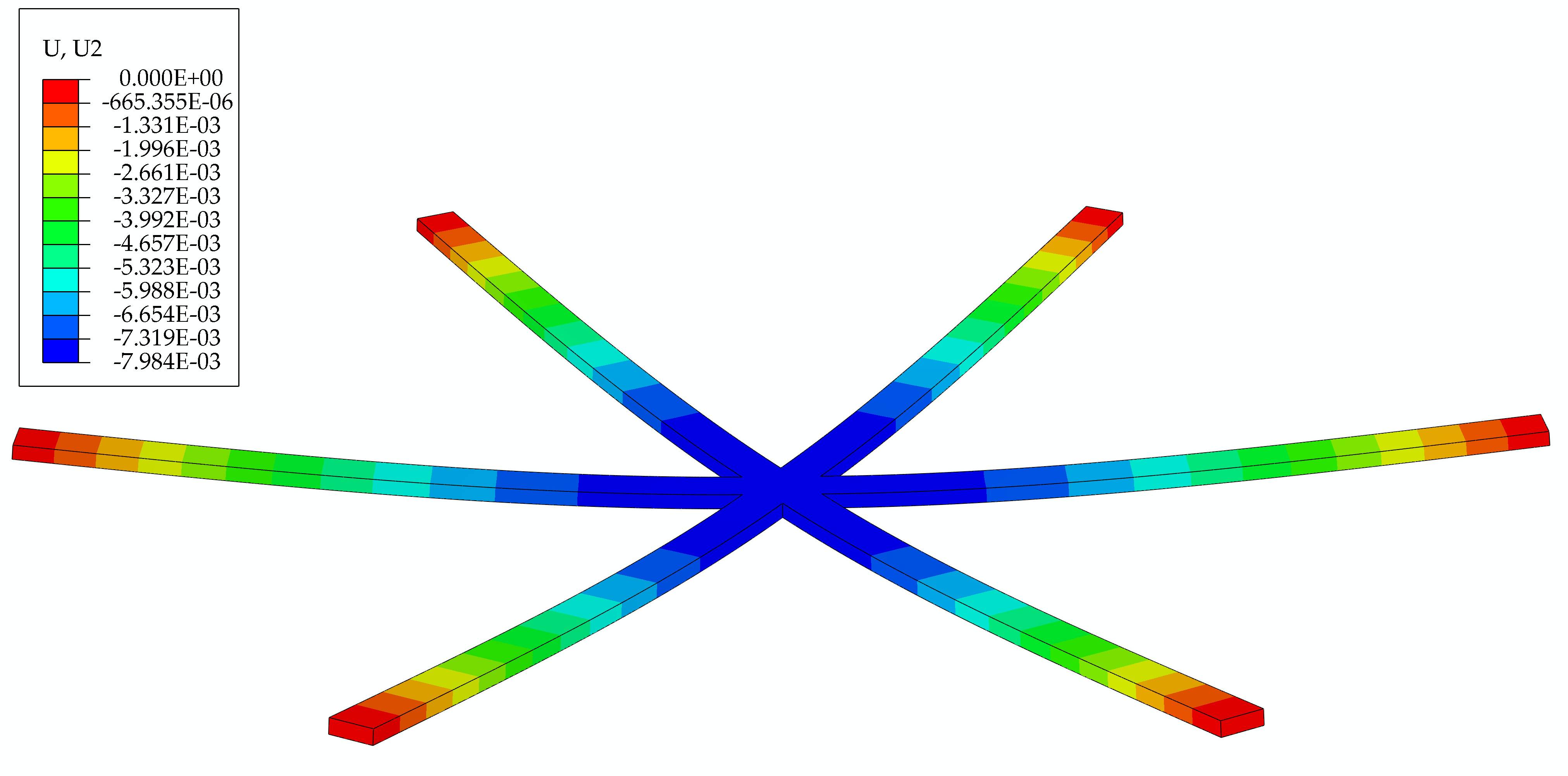}
\caption{}
\label{fig:triplearch_d}
\end{subfigure}

\begin{subfigure}[b]{0.24\textwidth}
\centering
\includegraphics[width=\textwidth]{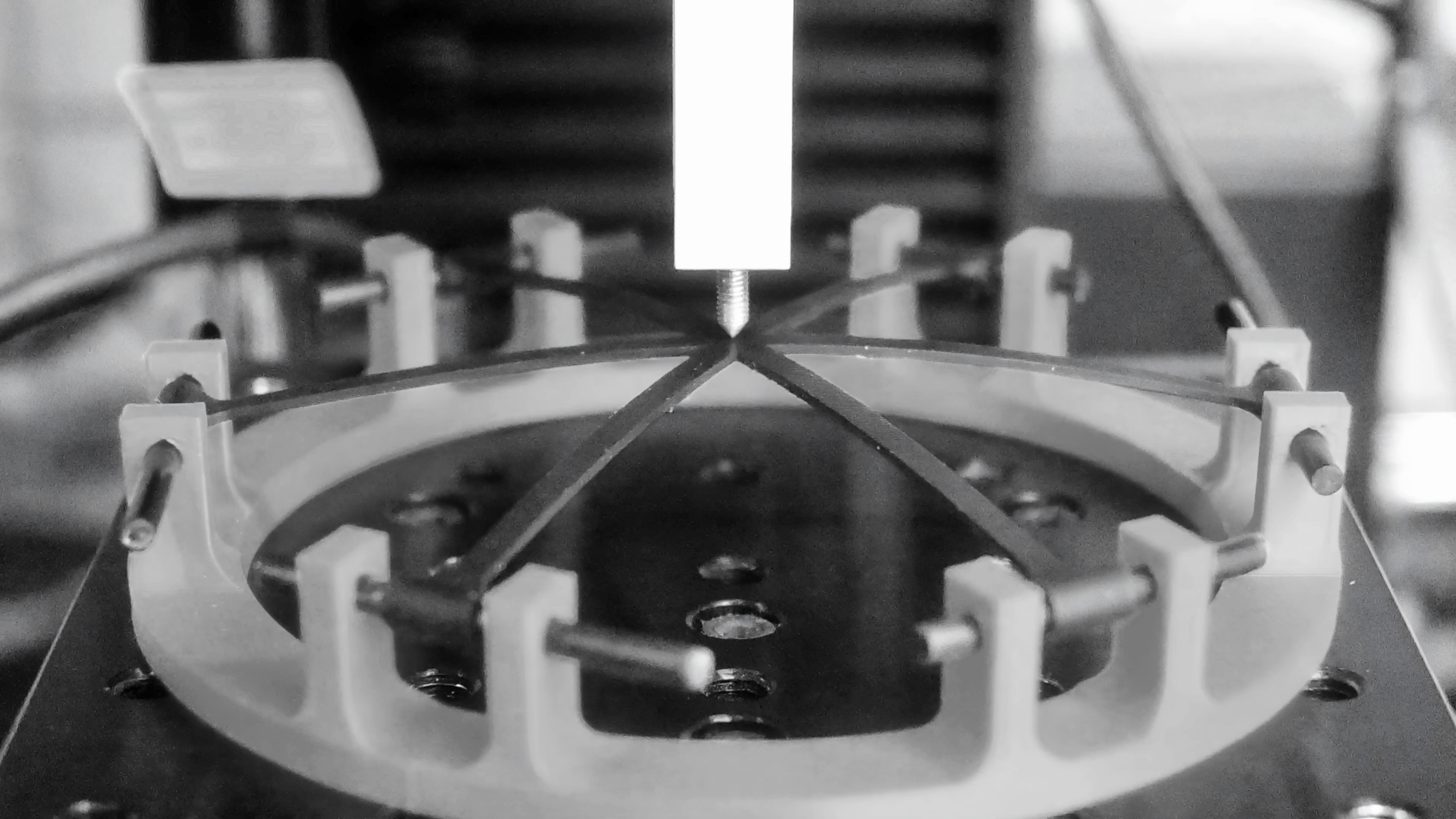}
\caption{}
\label{fig:triplearch_e}
\end{subfigure}
\hfill
\begin{subfigure}[b]{0.24\textwidth}
\centering
\includegraphics[width=\textwidth]{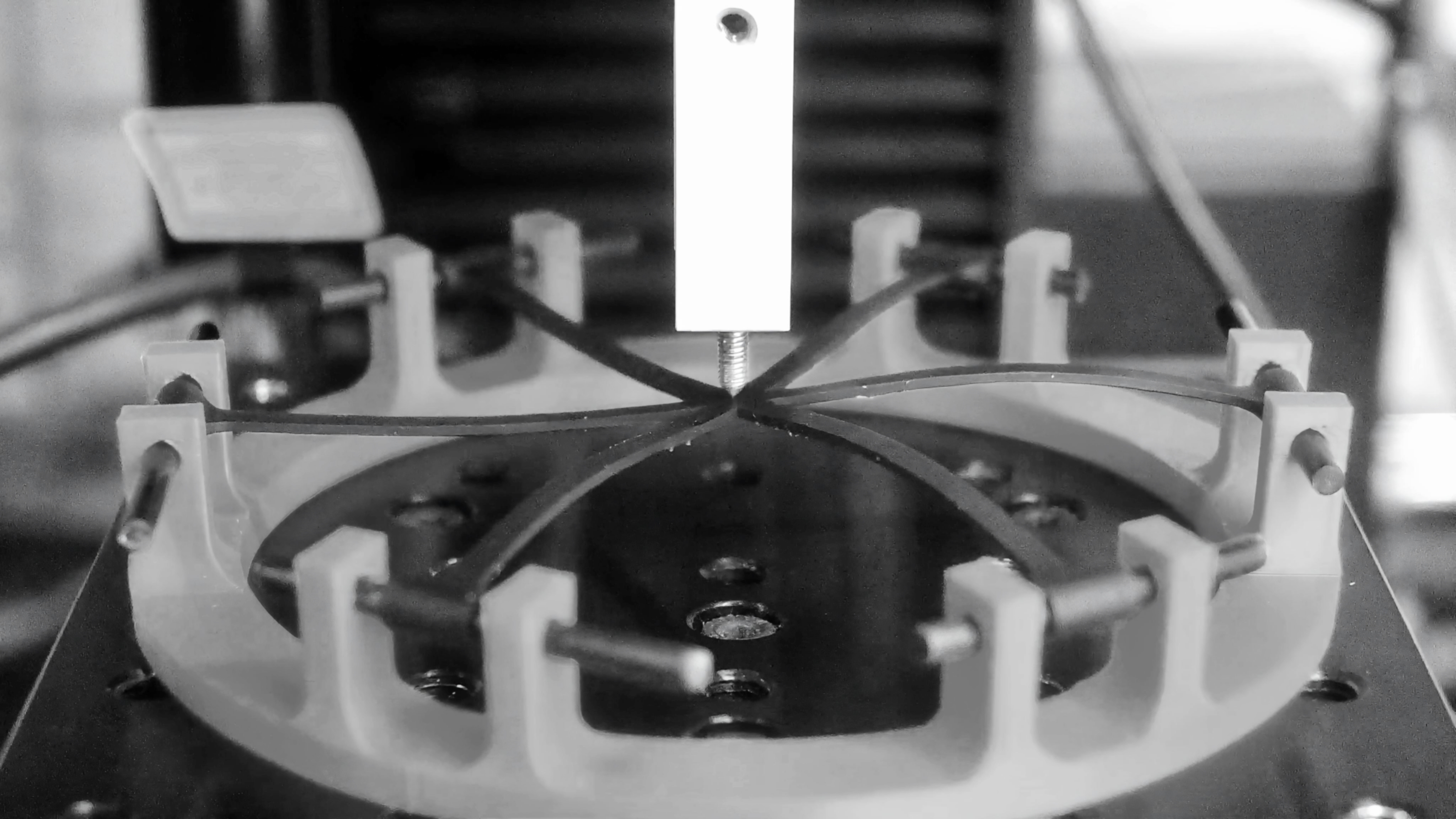}
\caption{}
\label{fig:triplearch_f}
\end{subfigure}
\hfill
\begin{subfigure}[b]{0.24\textwidth}
\centering
\includegraphics[width=\textwidth]{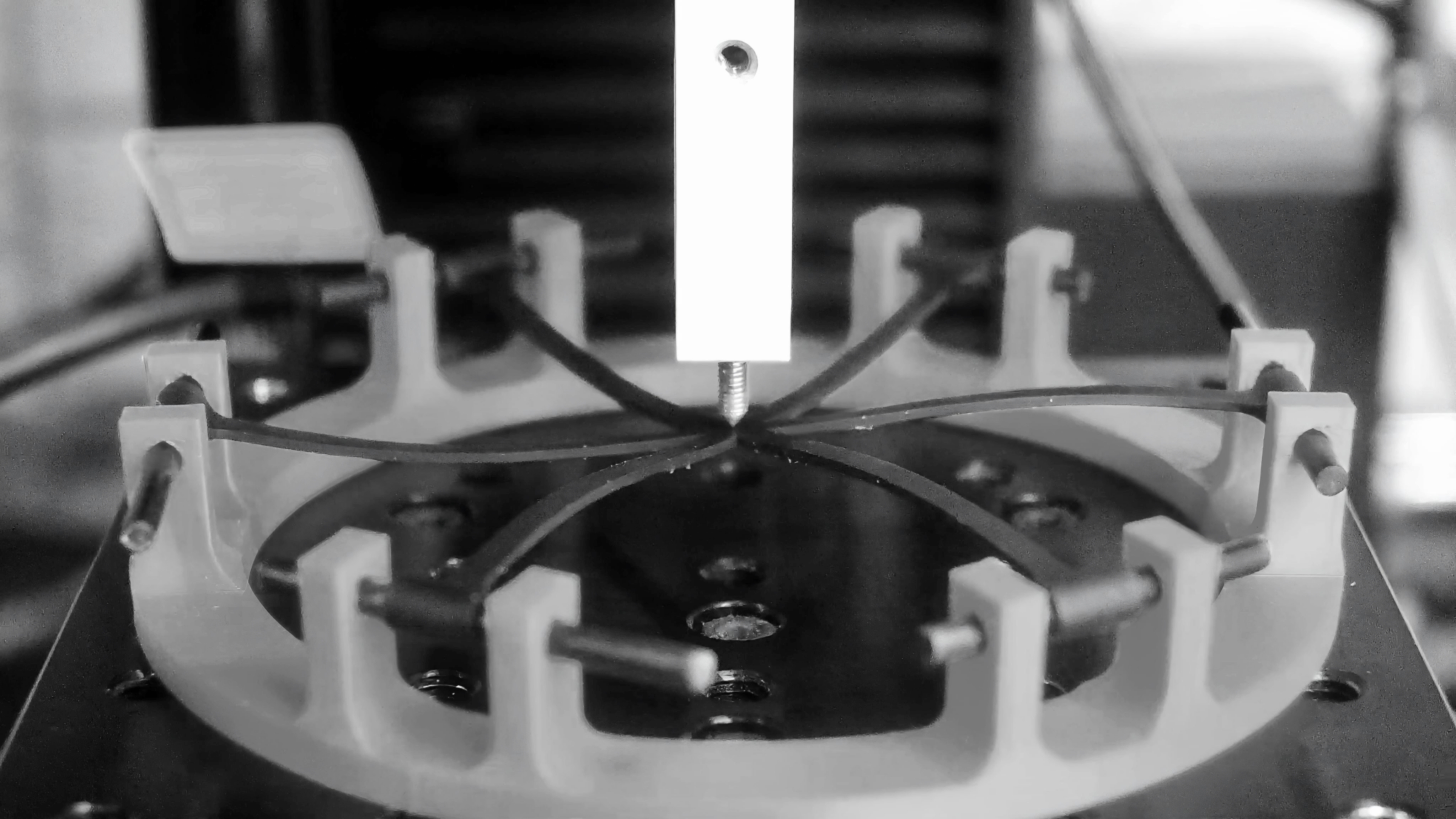}
\caption{}
\label{fig:triplearch_g}
\end{subfigure}
\hfill
\begin{subfigure}[b]{0.24\textwidth}
\centering
\includegraphics[width=\textwidth]{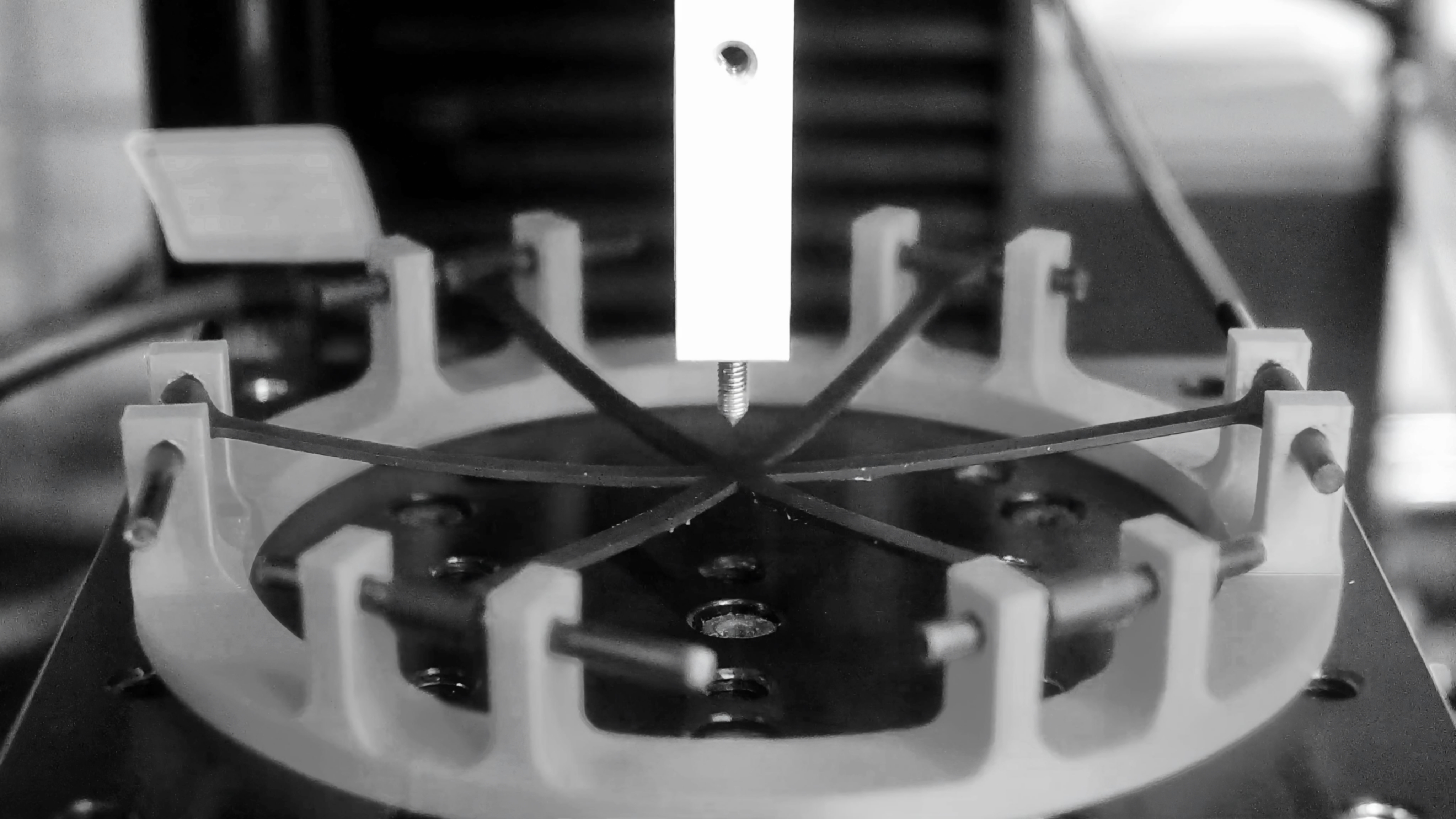}
\caption{}
\label{fig:triplearch_h}
\end{subfigure}

\caption{Deformation of the 3-MFA from FEA (a-d) and experiments (e-h).}
\label{fig:deformation_triplearch}

\end{figure}

 The deformation pathways of the 2- and 3-MFA obtained from FEA are shown in \cref{fig:deformation_doublearcha,fig:deformation_doublearchb,fig:deformation_doublearchc} and \cref{fig:triplearch_a,fig:triplearch_b,fig:triplearch_c,fig:triplearch_d}, respectively. For the 2-MFA, force-free stable configurations are observed in \cref{fig:deformation_doublearcha} and \cref{fig:deformation_doublearchc}, while the configuration in \cref{fig:deformation_doublearchb} corresponds to the midway of its asymmetric deformation pathway (see \cref{fig:Fdelta_n3m0l1}). Similarly, the as-fabricated and toggled force-free stable states of 3-MFA are seen in  \cref{fig:triplearch_a} and \cref{fig:triplearch_d}, respectively. Case 1 deformation occurs between \cref{fig:triplearch_a} and \cref{fig:triplearch_b}. The system then transitions to Case 2, in which all three arches deform asymmetrically, between \cref{fig:triplearch_b} and \cref{fig:triplearch_c}. 

\begin{figure}[!t]
     \centering
     \hfill
     \begin{subfigure}[b]{0.46\textwidth}
         \centering
         \includegraphics[width=\textwidth]{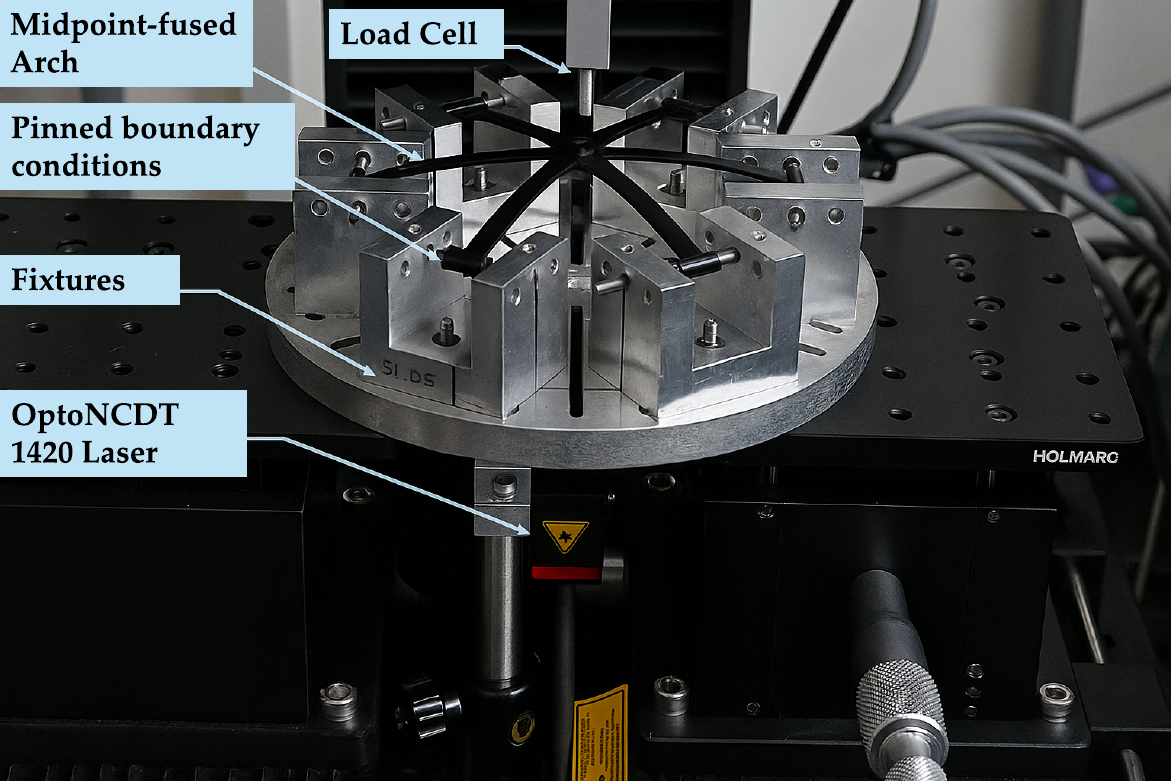}
         \caption{$ $}
         \label{fig:ExpSetup}
     \end{subfigure}
     \hfill
     \begin{subfigure}[b]{0.52\textwidth}
         \centering
         \includegraphics[width=\textwidth]{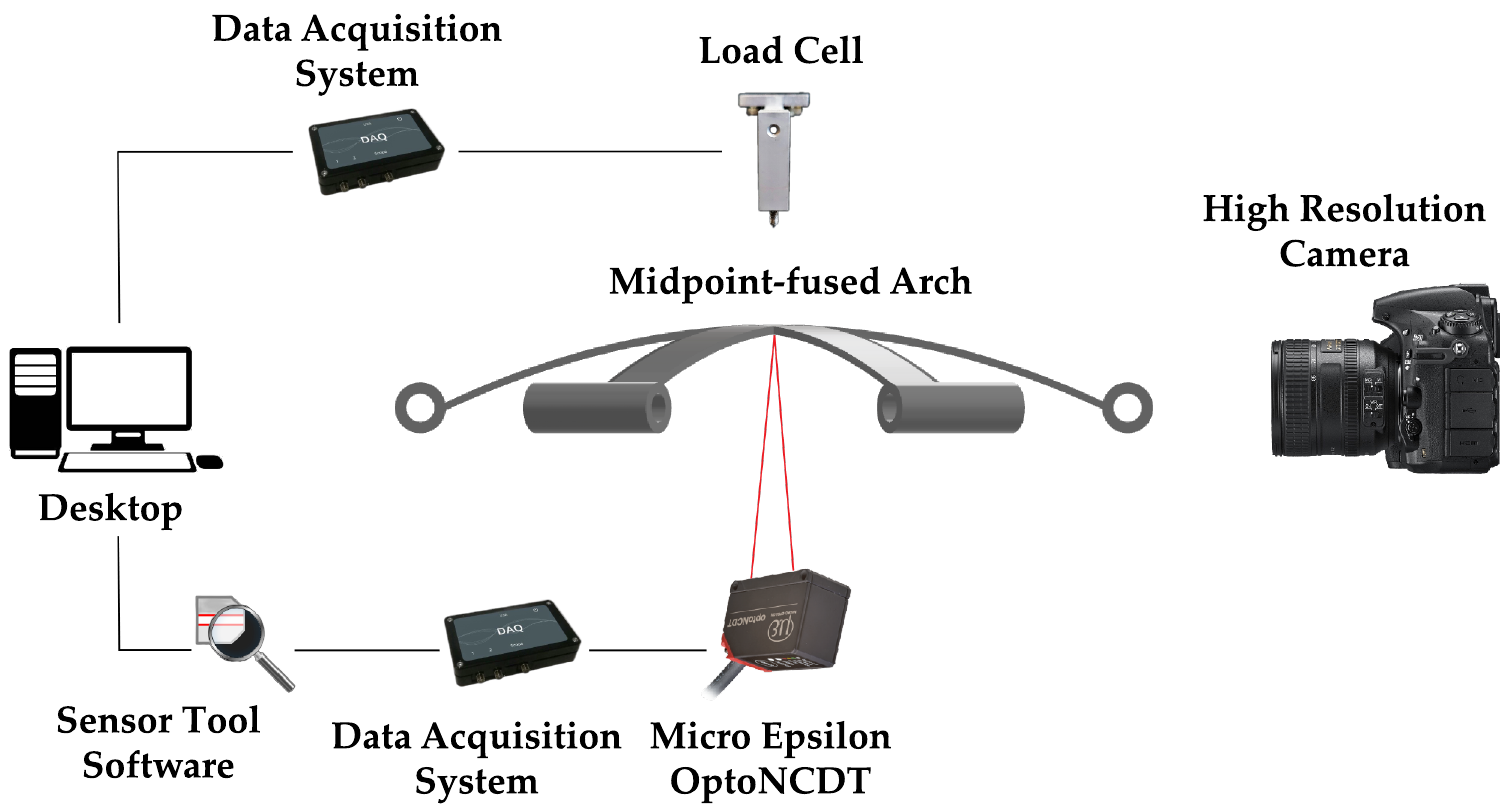}
         \caption{$ $}
         \label{fig:Schematic_expSetup}
     \end{subfigure}
        \caption{(a) Experimental setup used for measuring the force-displacement curve of MFA. (b) Schematic of the experimental setup.}
        \label{fig:exp_setup}
\end{figure}

We compared these key deformation snapshots with results from the experimental setup shown in \cref{fig:ExpSetup}. The setup consists of a base plate with adjustable slots to fix the arch span and fixtures holding carbon-fibre rods that provide pinned-pinned boundary conditions. A load cell mounted on a micro-motion vertical stage measures the load applied at the midpoint by an indenter, while a laser displacement sensor (Micro-Epsilon optoNCDT 1420) below the arch records vertical displacement as it transitions between stable states. A schematic of the setup is shown in \cref{fig:Schematic_expSetup}. The load cell and laser outputs are recorded in real time through a data acquisition system, and a high-resolution camera captures the deformation profiles at key deformed positions. This integrated system enables synchronized force-displacement and visual measurements.

The arches were 3D-printed using Elegoo Standard Black resin, with optimized tessellation and a layer height of 0.03 mm. Material properties of the black resin, determined from standard beam-bending tests on UV-cured samples, are listed in \cref{tab:1}. 

The specimens are printed with the tree support enabled to facilitate their removal, resulting in an optimal surface finish. The post-processing of the specimens is done immediately after the printing. This includes washing of the specimens in the automated wash station containing Isopropyl Alcohol (IPA) of 99.9 percent concentration to remove residual uncured resin. The specimens are then dried completely and then UV-cured at 405 nm for 2 minutes with rotation at the midpoint to ensure uniform exposure. All specimens were stored at room temperature away from ambient light and tested within 12 hours of post-curing to ensure a consistent degree of results across all samples. The experimentally observed deformation pathways for the 2- and 3-MFA, shown in the second rows of \cref{fig:deformation_doublearch,fig:deformation_triplearch}, exhibit strong agreement with both simulations and analytical predictions.

\begin{figure}[!htb]
    \centering
    \begin{subfigure}[t]{0.475\linewidth}
        \centering
        \includegraphics[width=\linewidth]{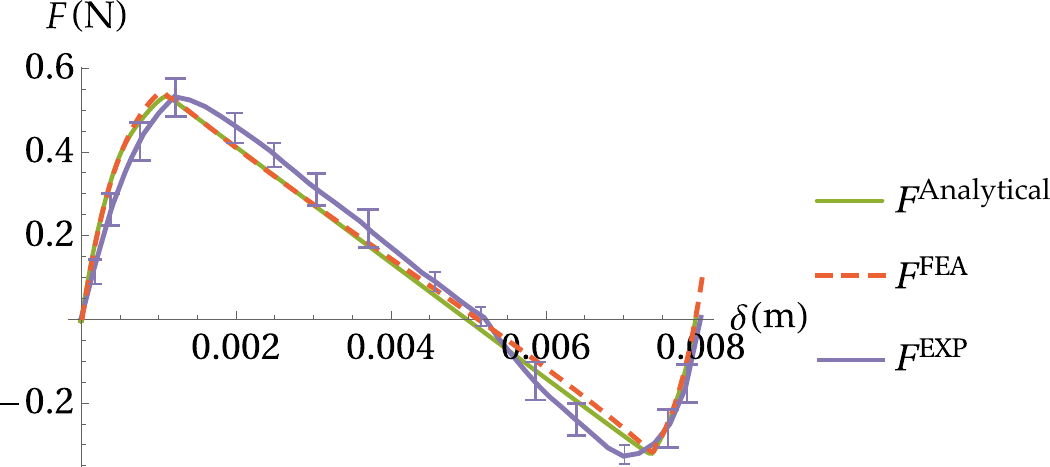}
        \caption{Force-displacement response of 2-MFA (\(n=3, m=0, l=1\)); analytical, FEA, and experiments (\(Q=4, \beta=3, \lambda=25\)).}\label{fig:A4L100T1B3H4_3w_2u_3phi_analytical_fea_EXP}
    \end{subfigure}
    \hfill
    \begin{subfigure}[t]{0.49\linewidth}
        \centering
        \includegraphics[width=\linewidth]{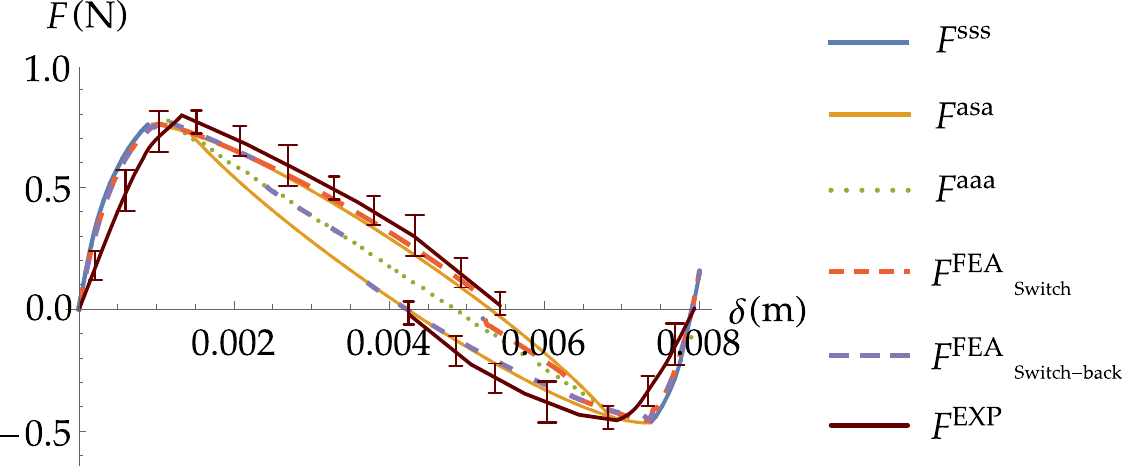}
        \caption{Force-displacement response of 3-MFA (\(n=3, m=0, l=1\)); analytical, FEA, and experiments (\(Q=4, \beta=3, \lambda=25\)).}\label{fig:Fdelta_i3n3m0l1_combined_dim_fea_exp}
    \end{subfigure}
    \caption{Comparison of analytical predictions with FEA and experimental results for 2-MFA and 3-MFA.}
    \label{fig:2MFA_3MFA_comparison}
\end{figure}

Furthermore, the force-displacement responses obtained from FEA and experiments are superimposed on the analytical predictions in \cref{fig:2MFA_3MFA_comparison}. \Cref{fig:A4L100T1B3H4_3w_2u_3phi_analytical_fea_EXP} shows that the analytical results for the 2-MFA agree well with the FEA results and reasonably well with the experimental measurements. The discrepancies observed in the experiments are primarily attributed to the fixtures. In particular, it is difficult to ensure perfectly ideal revolute joints, and to match the arch span precisely with the adjustable slot length of the fixtures. The latter is especially important to avoid introducing pre-stress in the as-fabricated configuration. Note that the first half of the experimental data corresponds to switching from the as-fabricated to the toggled state, while the second half corresponds to switching back.

For the 3-MFA, the comparison is qualitatively similar, as shown in \cref{fig:Fdelta_i3n3m0l1_combined_dim_fea_exp}. The principal difference is the switching from Case 1 to Case 2 observed in the simulations and experiments, which is not captured by the present analytical model, as discussed in \cref{Sec:triplearchescase2}. In addition, for the 3-MFA only partial experimental data could be obtained because the switching and switching-back pathways do not coincide, unlike the 2-MFA case considered here.

\section{Summary}
\label{sec:summary}
 In this work, we presented the mechanics of midpoint-fused arches (MFA) as bistable elastic structures with potential for a range of engineering applications. An analytical model was developed for MFAs with pinned-pinned boundary conditions and a transverse point load applied at the midspan. The formulation accounts for the strain-energy contributions of each constituent arch and enforces kinematic compatibility at the fusion point to couple their deformations. Multiple deformation pathways are then identified by considering different deformation symmetries.

The model assumes shallow, slender arches, within which it provides accurate predictions of MFA mechanics. The fusion region is idealised through kinematic constraints, and the finite overlap of the arches at the fusion point is not explicitly considered. Consequently, for arches with relatively large widths, the model may under-predict the midspan stiffness because the enlarged fusion region alters the local deformation behaviour.

The generality of the framework is demonstrated through detailed analyses of 2-MFA and 3-MFA configurations. Using low-order modal truncations, closed-form expressions are obtained for the symmetric and asymmetric deformation branches, their intersections, and the associated switching behaviour. Higher-mode approximations are then shown to capture the effective hybrid deformation pathway and the force-displacement response, showing good agreement with finite-element simulations and experiments.

\section*{Acknowledgments}

We gratefully acknowledge the funding received from the
 Department of Science and Technology, Government of India
 (SRG/2021/001463) and IIT-H SG-92.
\vspace{6 mm}

\appendix
\section{Details on Kinematics at the Fusion Point}
\label{Appendix:kinematics}
\begin{figure}[!htb]
     \centering
         \includegraphics[width=0.4\textwidth]{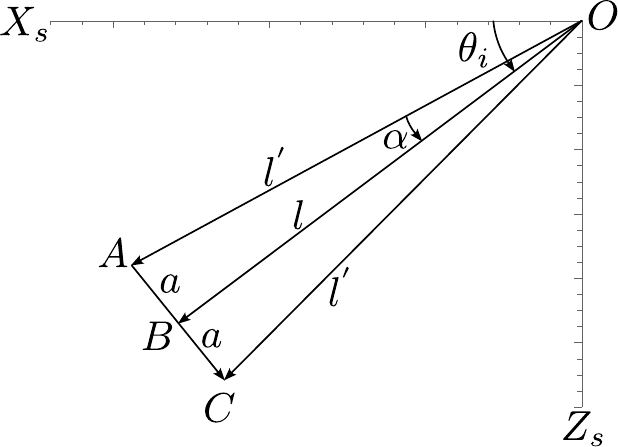}
        \caption{Representation of coordinates of element $dS_i$ along positive $X_{s}$ and $Z_{s}$-axis.}
        \label{fig:kinematic_schematic}
\end{figure}

\setcounter{equation}{0}
\renewcommand{\theequation}{A\arabic{equation}}
As discussed in \cref{sec:Kinematicsatmidpoint}, we derive the kinematic conditions imposed at the fusion point of the MFA. Consider a differential element \(dS_i\) and define its local coordinate axes as shown in \cref{fig:A6_midpatch2}. In the top view given in \cref{fig:kinematic_schematic},  \(O\) denotes the origin and the points \(A\), \(B\), and \(C\) lie on the edge of the element \(dS_i\). The coordinates of the points \(O\), \(A\), \(B\), and \(C\) are
$(0,0,0)$, 
$(l^{\prime} \cos (\theta_i-\alpha),0, l^{\prime} \sin (\theta_i-\alpha)$, $(l \cos \theta_i,0, l \sin \theta_i)$, and 
$(l^{\prime} \cos (\theta_i+\alpha),0, l^{\prime} \sin (\theta_i+\alpha))$, respectively, where \(l^{\prime}=\sqrt{l^2+a^2}\).

To evaluate \(\Phi_i\), i.e., the twist of the \ith\ arch about its own axis, we first compute
\begin{align}
\overrightarrow{A C}=  \overrightarrow{O C}-\overrightarrow{O A}= -2 l^{\prime} \sin \theta_i \sin \alpha \hat{i}+0 \hat{j}+2 l^{\prime} \cos \theta_i \sin \alpha\hat{k}.
\label{Eq:ACvector}
\end{align}
Then the rotated vector \(\overrightarrow{A'C'}\) is obtained by rotating \(\overrightarrow{A C}\) about the \(Z_s\)-axis by an angle \(\omega\), as shown in \cref{fig:A6_midpatch2}:
\begin{align}
\overrightarrow{A'C'}=2 l^{\prime} \sin \theta_i \sin \alpha \cos \omega \hat{i}+2 l^{\prime} \sin \theta_i \sin \alpha \sin \omega \hat{j}+2 l^{\prime} \cos \theta_i \sin \alpha \hat{k}
\end{align}
The angle \(\Phi_i\) between \(\overrightarrow{A C}\) and \(\overrightarrow{A'C'}\) is then,
\begin{align}
 \Phi_i \big|_{X_i = 1/2}=& \cos ^{-1}\left(\frac{\overrightarrow{A C} \cdot \overrightarrow{A'C'}}{|\overrightarrow{A C}||\overrightarrow{A'C'}|}\right) \nonumber\\
=& \cos ^{-1}\left(\sin ^2 \theta_i \cos \omega+\cos ^2 \theta_i\right) \label{eq:appendix_phi}
\end{align}
Now, by assuming that both $\Phi_i \big|_{X_i = 1/2}$ and $\omega$ are small, we can simplify \cref{eq:appendix_phi} as,
\begin{align}
1-\frac{\Phi_i \big|_{X_i = 1/2}^2}{2} &=\sin ^2 \theta_i\left(1-\frac{\omega^2}{2}\right)+\cos ^2 \theta_i, \nonumber\\
\implies \Phi_i \big|_{X_i = 1/2} &=\omega \sin \theta_i.
\end{align}
Similarly, to compute \(\frac{1}{\lambda}\frac{d W_i}{d X_i} \big|_{X_i = 1/2}\), we consider the vectors \(\overrightarrow{O B}\) and \(\overrightarrow{O^{\prime} B^{\prime}}\) and evaluate the angle between them:
\begin{align}
\frac{1}{\lambda} \frac{d W_i}{d X_i} \bigg|_{X_i= 1/2}=&\cos ^{-1}\left(\frac{\overrightarrow{O B} \cdot \overrightarrow{O^{\prime} B^{\prime}}}{|\overrightarrow{O B}||\overrightarrow{O^{\prime} B^{\prime}}|}\right)
=\cos ^{-1}\left(\cos ^2 \theta_i \cos \omega+\sin ^2 \theta_i\right),\\
\implies \frac{1}{\lambda}\frac{d W_i}{d X_i} \bigg|_{X_i = 1/2} & =\omega \cos \theta_i,
\end{align} where,
\begin{align}
\overrightarrow{O B}&=l \cos \theta_i \hat{i}+0 \hat{j}+l \sin \theta_i \hat{k},\\
\overrightarrow{O^{\prime} B^{\prime}}&=l \cos \theta_i \cos \omega \hat{i}+l \cos \theta_i \sin \omega \hat{j}+l \sin \theta_i \hat{k}.
\end{align}

\section{Strain Energy of the \texorpdfstring{$i^\mathrm{th}$}{ith} Arch}
\label{appendix:PEintermsofmodeweights}
\setcounter{equation}{0}
\renewcommand{\theequation}{B\arabic{equation}}

As discussed in \cref{nondimensionalisation}, the non-dimensionalized strain energy due to bending for an individual arch is,
\begin{align}
    S E_{b_i}=\frac{1}{2} \int_0^1\left(\frac{d^2 W_i}{d X_i^2}-\frac{d^2 H_i}{d X_i^2}\right)^2 d X_i+\frac{\beta ^2}{2} \int_0^1\left(\frac{d^2 U_i}{d X_i^2}-\Phi_i \frac{d^2 H_i}{d X_i^2}\right)^2 d X_i \label{SE_{b_i}}.
\end{align}
The as fabricated profile is \(H_i=a_1 \sin(\pi X_i)\) and the deformed  in-plane profile is \(W_i=\sum_{j=1,2,3}^{n} A_{ij} \sin(j\pi X_i)\). The out-of-plane deformation of the \ith arch is expressed as \(U_i=\sum_{j=1,3,5}^{m} B_{i j} \bar{U}_{j}+\sum_{j=2,4,6}^{m} B_{i j} \tilde{U}_{j}\), with  \(\bar{U}_{j}=\frac{1}{2}-\frac{1}{2} \cos [(j+1) \pi X_i]\) and \(\tilde{U}_{j}=\frac{1}{2}\left[1-2 X_i-\cos \left(N_{j} X_i\right)+\frac{2}{N_{j}} \sin \left(N_{j} X_i\right)\right]\). The rotation about the longitudinal axis of the cross-section is \(\Phi_i=\sum_{j=1,3,5}^{l} C_{i j} P_{j}+\sum_{j=2,4,6}^{l} C_{i j} \tilde{P}_{j}\), where \(P_j\), and \(\tilde{P_j}\)  are both defined as \(\sin (j \pi X_i)\) for odd and even entries of \(j\) respectively.

To simplify \cref{SE_{b_i}}, we write \(S E_{b_i}=A+B\), where
$
A=\frac{1}{2} \int_0^1\left(\frac{d^2 W_i}{d X_i^2}-\frac{d^2 H_i}{d X_i^2}\right)^2 d X_i$ and
$B=\frac{\beta ^2}{2} \int_0^1\left(\frac{d^2 U_i}{d X_i^2}-\Phi_i \frac{d^2 H_i}{d X_i^2}\right)^2 d X_i$. Expanding \(A\) yields
\begin{align}
    A&=\frac{1}{2} \int_0^1\left(\frac{d^2 W_i}{d X_i^2}-\frac{d^2 H_i}{d X_i^2}\right)^2 d X_i=\frac{1}{2} \int_0^1\left(\frac{d^2 W_i}{d X_i^2}^2+\frac{d^2 H_i}{d X_i^2}^2-2\frac{d^2 W_i}{d X_i^2}\frac{d^2 H_i}{d X_i^2}\right) d X_i \nonumber\\&
    =\frac{\pi^4}{4}\left(a_1-A_{i1}\right)^2+4 \pi^4 A_{i2}^2+\frac{81}{4} \pi^4 A_{i3}^2+\ldots .+\frac{n^4}{4} \pi^4 A_{in}^2.
\end{align}
Similarly, expanding \(B\) gives,
\begin{align}
    B&=\frac{\beta ^2}{2} \int_0^1\left(\frac{d^2 U_i}{d X_i^2}-\Phi_i \frac{d^2 H_i}{d X_i^2}\right)^2 d X_i \nonumber\\
    &=\frac{\beta ^2}{2} \int_0^1\left(\frac{d^2 U_i}{d X_i^2}^2+\Phi_i^2\frac{d^2 H_i}{d X_i^2}^2-2\Phi_i\frac{d^2 U_i}{d X_i^2}\frac{d^2 H_i}{d X_i^2}\right) d X_i .
\end{align}
For convenience, we denote the three contributions in \(B\)   as \(I_1=\frac{\beta^2}{2}\int_0^1\left(\frac{d^2 U_i}{d X_i^2}\right)^2 d X_i\), \(I_2=\frac{\beta^2}{2}\int_0^1\left(\Phi_i\frac{d^2 H_i}{d X_i^2}\right)^2 d X_i\), and \(I_3=-\frac{\beta^2}{2}\int_0^1\left(2\Phi_i\frac{d^2 U_i}{d X_i^2}\frac{d^2 H_i}{d X_i^2}\right) d X_i\) so that \(B=I_1+I_2+I_3\). We evaluate each term below.
First,
\begin{align}
    I_1&=\frac{\beta^2}{2}\int_0^1\left(\frac{d^2 U_i}{d X_i^2}\right)^2 d X_i=\frac{1}{2}\int_0^1 U_i  ^ {\prime \prime 2} d X_i\nonumber\\&
    =\frac{\beta^2}{2} \int_{0}^{1}\left(\sum_{j=1,3,5}^{m} B_{i j} \bar{U}_{j}^{\prime \prime}+\sum_{j=2,4,6}^{m} B_{i j} \tilde{U}_{j}^{\prime \prime}\right)^{2} d X_i\nonumber\\&
    =\frac{\beta^2}{2} \int_{0}^{1} \sum_{j=1,3,5}^{m} B_{i j}^{2} \bar{U}_{j}^{\prime \prime 2} d X_i+\frac{\beta^2}{2} \int_{0}^{1} \sum_{j=2,4,6}^{m} B_{i j}^{2} \tilde{U}_{j}^{\prime \prime^{2}} d X_i+\beta^2\int_{0}^{1} \sum_{\substack{j=1,3,5 \\
k=2,4,6}}^{m} B_{i j} B_{i k} \bar{U}_{j}^{\prime \prime} \tilde{U}_{k}^{\prime \prime} d X_i\nonumber\\&+\beta^2\int_{0}^{1} \sum_{\substack{j=1,3,5 \\
k=1,3,5}}^{m} B_{i j} B_{i k} \bar{U}_{j}^{\prime \prime} \bar{U}_{k}^{\prime \prime} d X_i+\beta^2\int_{0}^{1} \sum_{\substack{j=2,4,6 \\
k=2,4,6}}^{m} B_{i j} B_{i k} \tilde{U}_{j}^{\prime \prime} \tilde{U}_{k}^{\prime \prime} d X_i
\quad\left\{\begin{array}{c}
\text { where } \\
j \neq k
\end{array}\right\}\nonumber\\&
=\beta^2\sum_{j=1,3,5}^{m} \frac{B_{i j}^{2}}{16}(j+1)^{4} \pi^{4}+\beta^2\sum_{j=2,4,6}^{m} \frac{B_{i j}^{2}}{16} N_{j}^{4}
\end{align}
Next, \(I_2\) is
\begin{align}
    I_2&=\frac{\beta^2}{2}\int_0^1 a_1^2 \left(\Phi_i\frac{d^2 H_i}{d X_i^2}\right)^2 d X_i=\frac{\beta^2}{2}\int_0^1 \Phi_i^2 H_i^{\prime \prime 2} d X_i \nonumber\\&
    =\frac{\beta^2}{2} \int_{0}^{1}\bigg(\sum_{j=1,3,5}^{l} C_{i j} P_{j}+\sum_{j=2,4,6}^{l} C_{i j} \tilde{P}_{j}\bigg)^{2}\left[\sin (\pi X_i)\right]^{\prime \prime 2} d X_i \nonumber\\&
    =\frac{\beta^2}{2} \int_{0}^{1}\bigg(\sum_{j=1,3,5}^{l} C_{i j}^{2} P_{j}^{2}+\sum_{j=2,4,6}^{l} C_{i j}^{2} \tilde{P}_{j}^{2}+2 \sum_{\substack{j=1,3,5 \\
k=2,4,6}}^{l} C_{i j} C_{i k} P_{j} \tilde{P}_{k}+2 \sum_{\substack{j=1,3,5 \\
k=1,3,5}}^{l} C_{i j} C_{i k} P_{j} P_{k}\nonumber\\
& +2 \sum_{\substack{j=2,4,6 \\
k=2,4,6}}^{l} C_{i j} C_{i k} \tilde{P}_{j} \tilde{P}_{k}\bigg)\left[a_{1}^{2} \pi^{4} \sin ^{2}(\pi X_i)\right] d X_i \nonumber\\&
=\pi^{4} a_{1}^{2}\beta^2\biggr[\frac{1}{2} \int_{0}^{1} \sum_{j=1,3,5}^{l} C_{i j}^{2} P_{j}^{2} \sin ^{2}(\pi X_i) d X_i+\frac{1}{2} \int_{0}^{1} \sum_{j=2,4,6}^{l} C_{i j}^{2} \tilde{P}_{j}^{2} \sin ^{2}(\pi X_i) d X_i\nonumber\\&+\int_{0}^{1} \sum_{j=1,3,5}^{l} C_{i j} C_{i k} P_{j} \tilde{P}_{k} \sin ^{2}(\pi X_i) d X_i +\int_{0}^{1} \sum_{\substack{j=1,3,5 \\
k=1,3,5}}^{l} C_{i j} C_{i k} P_{j} P_{k} \sin ^{2}(\pi X_i) d X_i\nonumber\\&+\int_{0}^{1} \sum_{\substack{j=2,4,6 \\
k=2,4,6}}^{l} C_{i j} C_{i k} \tilde{P}_{j} \tilde{P}_{k} \sin ^{2}(\pi X_i) d X_i \biggr] \nonumber\\&
=a_{1}^{2}\beta^2\bigg(\frac{3 \pi^{4}}{16} C_{i 1}^{2}+\frac{\pi^{4}}{8} \sum_{j=2,3,4}^{l} C_{i j}^{2}-\frac{\pi^{4}}{8} \sum_{\substack{j=1,3,5 \\
k=1,3,5}}^{l} C_{i j} C_{i k}-\frac{\pi^{4}}{8} \sum_{\substack{j=2,4,6 \\
k=2,4,6}}^{l} C_{i j} C_{i k}\bigg)
\end{align}
Finally, \(I_3\) becomes
\begin{align}
    I_3&=-\frac{\beta^2}{2}\int_0^1\left(2\Phi_i\frac{d^2 U_i}{d X_i^2}\frac{d^2 H_i}{d X_i^2}\right) d X_i=-\beta^2\int_{0}^{1} U_i^{\prime \prime} \Phi_i H_i^{\prime \prime} d X_i \nonumber\\&
    =a_{1} \beta^2 \pi^{2} {\bigg[\int_{0}^{1} \sum_{\substack{j=1,3,5 \\
k=1,3,5}}^{m, l} B_{i j} C_{i k} \bar{U}_{j}^{\prime \prime} P_{k} \sin (\pi X_i) d X_i+\int_{0}^{1} \sum_{\substack{j=1,3,5 \\ k=2,4,6}}^{m, l} B_{i j} C_{i k} \bar{U}_{j}^{\prime \prime} \tilde{P}_{k} \sin (\pi X_i) d X_i\bigg.} \nonumber\\
& \bigg.+\int_{0}^{1} \sum_{\substack{j=2,4,6 \\
k=1,3,5}}^{m, l} B_{i j} C_{i k} \tilde{U}_{j}^{\prime \prime} P_{k} \sin (\pi X_i) d X_i+\int_{0}^{1} \sum_{\substack{j=2,4,6 \\
k=2,4,6}}^{m, l} B_{i j} C_{i k} \tilde{U}_{j}^{\prime \prime} \tilde{P}_{k} \sin (\pi X_i) d X_i\bigg] \nonumber\\&
=a_{1}\beta^2 \pi^{2} \bigg[\int_{0}^{1} \sum_{\substack{j=1,3,5 \\
k=1,3,5}}^{m, l} B_{i j} C_{i k} \bar{U}_{j}^{\prime \prime} P_{k} \sin (\pi X_i) d X_i+\int_{0}^{1} \sum_{\substack{j=2,4,6 \\
k=2,4,6}}^{m, l} B_{i j} C_{i k} \tilde{U}_{j}^{\prime \prime} \tilde{P}_{k} \sin (\pi X_i) d X_i\bigg] \nonumber\\&
=a_{1} \beta^2 \pi^{2}\Bigg(\sum_{\substack{j=1,3,5 \\
k=1,3,5}}^{m, l} B_{i j} C_{i k} M_{j k}^{*}+\sum_{\substack{j=2,4,6 \\
k=2,4,6}}^{m, l} B_{i j} C_{i k} M_{j k}^{*}\Bigg).
\end{align}
Therefore, the bending strain energy \(S E_{b_i}\) is obtained by combining \(A\) and \(B=I_1+I_2+I_3\), giving,
\begin{align}
    S E_{b}&=\frac{\pi^{4}}{4}\left(a_{1}-A_{i 1}\right)^{2}+\sum_{j=2,3,4}^{n} \frac{j^{4}}{4} \pi^{4} A_{i j}^{2}+\beta^{2}\Bigg[\sum_{j=1,3,5}^{m} \frac{B_{i j}^{2}}{16}(j+1)^{4} \pi^{4}+\sum_{j=2,4,6}^{m} \frac{B_{i j}^{2}}{16} N_{j}^{4}+\Bigg. \nonumber\\
& a_{1}^{2}\Bigg(\frac{3 \pi^{4} C_{i 1}^{2}}{16}+\frac{\pi^{4}}{8} \sum_{j=2,3,4}^{l} C_{i j}^{2}-\frac{\pi^{4}}{8} \sum_{\substack{j=1,3,5 \\
k=1,3,5}}^{l} C_{i j} C_{i k}-\frac{\pi^{4}}{8} \sum_{\substack{j=2,4,6 \\
k=2,4,6}}^{l} C_{i j} C_{i k}\Bigg)+ \nonumber\\
& \Bigg.a_{1} \pi^{2}\Bigg(\sum_{\substack{j=1,3,5 \\
k=1,3,5}}^{m, l} B_{i j} C_{i k} M_{j k}^{*}+\sum_{\substack{j=2,4,6 \\
k=2,4,6}}^{m, l} B_{i j} C_{i k} M_{j k}^{*}\Bigg)\Bigg].
\end{align}
The non-dimensionalized compression strain energy of the \ith arch is
\begin{align}
S E_{c_i} & =\frac{3}{2} Q^2\left\{\int_0^1\left[\left(\frac{d H_i}{d X_i}\right)^2-\left(\frac{d W_i}{d X_i}\right)^2-\left(\frac{d U_i}{d X_i}\right)^2\right] d X_i\right\}^2 \nonumber\\
& =\frac{3}{2} Q^2\left[\frac{\pi^2 a_1^2}{2}-\pi^2\left(\frac{A_{i1}^2}{2}+2 A_{i2}^2+\frac{9}{2} A_{i3}^2+\ldots .+\frac{n^2 A_{in}^2}{2}\right)\right] \nonumber\\ &-\frac{3}{2} Q^2\pi^2\left(\frac{4 B_{i1}^2}{8}+\frac{16 B_{i3}^2}{8}+\frac{36 B_{i5}^2}{8}+\ldots+\frac{(m+1)^2 B_{im}^2}{8}\right)\nonumber\\ &-\frac{3}{2} Q^2\left(\frac{N_2^2 B_{i2}^2}{8}+\frac{N_4^2 B_{i4}^2}{8}+\ldots+\frac{N_j^2 B_{im}^2}{8}\right)^2 \nonumber\\
& =\frac{3}{2} Q^2\left(\frac{\pi^2 a_1^2}{2}-\pi^2 \sum_{j=1,2,3}^{n} \frac{j^2 A_{ij}^2}{2}-\pi^2 \sum_{j=1,3,5}^{m} \frac{(j+1)^2 B_{ij}^2}{8}-\sum_{j=2,4,6}^{m} \frac{N_j^2 B_{ij}^2}{8}\right)^2.
\end{align}
And the non-dimensionalized torsional strain energy of the \ith arch is
\begin{align}
S E_{t_i} & =K \int_0^1\left(\frac{d \Phi_i}{d X_i}\right)^2 d X_i \nonumber\\
& = K \int_0^1 \left(\sum_{j=1,3,5}^{l} C_{ij} P_j + \sum_{j=2,4,6}^{l} C_{ij} \tilde{P}_j \right) \nonumber\\
& =\frac{\pi^2 K}{2} \sum_{j=1,2,3}^{l} j^2 C_{ij}^2.
\end{align}

\bibliographystyle{plain}
\bibliography{sample}

\vskip2pc

\end{document}